\documentclass[aps,pre,notitlepage,10pt,twocolumn, superscriptaddress]{revtex4-2}
\usepackage[utf8]{inputenc}
\usepackage{geometry}
\usepackage{xcolor}

\usepackage{lipsum}
\usepackage{amsthm}
\usepackage{hyperref}
\hypersetup{
    colorlinks,
    linkcolor={red!50!black},
    citecolor={blue!50!black},
    urlcolor={blue!80!black}
}

\usepackage{soul}
\usepackage{graphicx}
\usepackage{amsmath}
\usepackage{wasysym}
\usepackage{amsfonts}
\usepackage{stmaryrd}
\usepackage{subfigure}
\usepackage{listings}
\usepackage{color}
\usepackage[T1]{fontenc}
\usepackage{tabularx}
\usepackage{bm}
\usepackage{enumitem}
\usepackage[normalem]{ulem}

\theoremstyle{definition}
\newtheorem{prop}{Symmetry Proposition}

\DeclareMathOperator{\atanh}{atanh}

\begin{document}
\title{Percolation and criticality of systems with competing interactions on Bethe lattices: limitations and potential strengths of cluster schemes}
\author{Greivin Alfaro Miranda}
\affiliation{Laboratoire de Physique Th\'eorique et Hautes Energies, CNRS-UMR 7589, Sorbonne Universit\'e, 4 Place Jussieu, F-75005 Paris, France}
\author{Mingyuan Zheng}
\email[Author to whom correspondence should be addressed. Electronic mail: ]{mingyuan.zheng@duke.edu}
\affiliation{Advanced Materials Thrust, Function Hub, Hong Kong University of Science and Technology (Guangzhou), Guangzhou 511455, China}
\affiliation{Department of Chemistry, Duke University, Durham, North Carolina 27708, United States}
\author{Patrick Charbonneau}
\affiliation{Department of Physics, Duke University, Durham, North Carolina 27708, United States}
\affiliation{Department of Chemistry, Duke University, Durham, North Carolina 27708, United States}
\author{Antonio Coniglio}
\affiliation{Università degli Studi di Napoli “Federico II”, Physics Department, I-80126 Napoli, Italy}
\author{Leticia F. Cugliandolo}
\affiliation{Laboratoire de Physique Th\'eorique et Hautes Energies, CNRS-UMR 7589, Sorbonne Universit\'e, 4 Place Jussieu, F-75005 Paris, France}
\author{Marco Tarzia}
\affiliation{Laboratoire de Physique Th\'eorique de la Mati\`ere Condens\'ee, CNRS-UMR 7600, Sorbonne Universit\'e, 4 Place Jussieu, F-75005 Paris, France}
\begin{abstract}
The random clusters introduced by Fortuin and Kasteleyn (FK) and analyzed by Coniglio and Klein (CK) for Ising and related models  have led first Swendsen and Wang and then Wolff to formulate remarkably efficient Markov chain Monte Carlo sampling schemes that weaken the critical slowing down. In frustrated models, however, no standard approach has yet been identified to achieve comparable gains in configurational sampling at low frustration---let alone to efficiently sample the strongly frustrated regime. In order to understand why formulating appropriate cluster criteria for frustrated models has thus far been elusive, we here study minimal short-range attractive and long-range repulsive as well as spin-glass models on Bethe lattices. Using a generalization of the CK approach and the cavity-field method, the appropriateness and limitations of FK--CK-type clusters are identified. We find that a standard, constructive cluster scheme is then inoperable, and that the frustration range over which generalized FK--CK clusters are even definable is finite. These results demonstrate the futility of seeking constructive cluster schemes for frustrated systems but leaves open the possibility that alternate approaches could be devised.
\end{abstract}
\date{\today}
\maketitle

\section{Introduction}
In the early 1970s, Kees Fortuin and Piet Kasteleyn (FK) unified the descriptions of Ising, Potts, and percolation models via their formulation of random clusters~\cite{kasteleyn1969phase,fortuin1972random}. That framework has since had a marked impact on stochastic geometry and mathematical physics~\cite{grimmett2006random}. Roughly a decade later, Coniglio and Klein (CK) realized that a geometric characterization of the Ising critical point could be made in terms of clusters that percolate with the Ising critical exponents, thus making these clusters more immediately physically consequent~\cite{coniglio1980clusters,esposito2024}. (That setup was later generalized to a few other models by Robert Edwards and Alan Sokal~\cite{edwards1988generalization}.) Both results were indeed leveraged in the subsequent formulation of novel Markov chain Monte Carlo sampling schemes that could significantly weaken the critical slowing down: the Swendsen--Wang~\cite{swendsen1987nonuniversal,wang1990cluster} and Wolff~\cite{wolff1989collective} algorithms. While the correlation decay time $\tau$ of standard (single-spin) Metropolis sampling grows critically with the equilibrium correlation length $\xi$ as $\tau\sim\xi^z$, with $z\geq2$~\cite{hohenberg1977,halperin2019}, FK--CK cluster--based algorithms do so with $z\ll 2$; see, e.g.,~\cite{newman1999monte,landau2021guide}. The definition of FK--CK clusters was later extended~\cite{coniglio2021correlated} to the antiferromagnetic Ising model~\cite{amitrano1983ab} and to Ising models with any ferromagnetic interaction between any pair of sites~\cite{jan1982study}. Separately, the Swendsen--Wang algorithm has found applications well beyond physics, including in image processing and computer vision~\cite{barbu2005generalizing,barbu2007generalizing}.

In statistical physics proper, however, relatively little progress has since been achieved. One key hurdle is that the FK--CK clusters are not obviously generalizable to models with frustration. For models with short-range attractive and long-range repulsive (SALR) interactions (without quenched disorder), the generalization proposed by Pleimling and Henkel~\cite{pleimling2001anisotropic,henkel2002}, among others, has been met with limited success~\cite{zheng2022weakening}. For spin-glass models (with quenched disorder), the roadblock has motivated the formulation of altogether different cluster schemes, such as the Chayes--Machta--Redner~\cite{chayes1998graphical} and the Houdayer~\cite{houdayer2001cluster} algorithms, although with a similar outcome~\cite{Alfaro2025}.  Irrespective of the type of frustration, FK--CK-like clusters seemingly fail to significantly weaken the critical slowing down even in the weak frustration limit~\cite{fajen2020percolation,zheng2022weakening}. The underlying reason had long remained hazy until some of us recently showed for a specific model with SALR interactions that clusters would need to be constructed using a confounding \emph{negative} probability of including spins with frustrated interactions for the proper clusters to be identified~\cite{zheng2022weakening}. 

It should be emphasized that these families of models are not mere statistical physics curiosities. Lattice SALR interaction models recapitulate the rich phase behavior of microphase formers as varied as diblock copolymers, surfactants, microemulsions, and certain magnetic alloys~\cite{yeomans1988,selke1992,seul1995,charbonneau2022advances}. Their study has been key to disentangling the mesoscale assembly behavior of various ordered~\cite{selke1992} and disordered~\cite{charbonneau2021solution} morphologies. Lattice spin-glass models have a similarly long track record with an even broader array of applications ranging from neural networks to quantum error
correction codes~\cite{mezard1987spin,dotsenko1994introduction,nishimori_statistical_2001,charbonneau2023spin}.
For instance, the connection between the two-dimensional frustrated random-bond Ising model (RBIM) and certain classes of Toric codes has played an important role in determining the error correction threshold of the latter (see Refs.~\cite{Dennis2002, Agrawal2023,Agrawal2024,Lee2025} and references therein). Understanding the root of the failure of standard cluster schemes in these models could therefore have an impact much broader than might first appear.

In this work, we generalize the finding of Ref.~\cite{zheng2022weakening} by presenting a more formal demonstration of the result (Sec.~\ref{sec:theorem}) and by identifying the limitations of FK--CK clusters, however generalized, for three frustrated models on a Bethe lattice. 
The rest of this article is organized as follows. Section~\ref{sec:models} presents the specific models considered and Sec.~\ref{sec:Bethe} their cavity-field method solution~\cite{mezard2009information,semerjian2009exact}. Results are then presented and discussed in Sec.~\ref{sec:discussion}. An extended conclusion follows in Sec.~\ref{sec:conclusion}. 

\section{The models}

Three different frustrated models are considered in this work: a model with isotropic and an another with anisotropic next-nearest-neighbor SALR interactions as well as the diluted random bond Ising model (RBIM) with frustration. These examplars are both canonical members of the SALR interaction (without quenched disorder) and spin-glass (with quenched disorder) classes of models, respectively, and sufficiently simple to be amenable to a complete analysis on Bethe lattices. In this section, their definition and static properties on various lattices is recapitulated; the calculation of their equilibrium properties on Bethe lattices is presented in Sec.~\ref{sec:Bethe}.

\label{sec:models}
\subsection{Isotropic and Anisotropic SALR interaction models}

We consider two types of homogeneous models of Ising spins, $s_i = \pm 1 = \; \uparrow/\downarrow$, with up to next-nearest neighbor interactions. 
The first model has a purely isotropic (iso) Hamiltonian,
\begin{equation}
    \label{eq:Hiso}
    \mathcal{H}_\mathrm{iso}(\{s_i\}) = -J\sum\limits_{\left<i,j\right>}s_i s_j + \kappa J \!\! \sum\limits_{\left<\left<i,j\right>\right>} s_i s_j - h_{\rm ext}\sum\limits_{i} s_i
    \; ,
\end{equation}
with $\langle i,j\rangle$ and $\langle\langle i,j\rangle\rangle$ denoting nearest and next-nearest neighbors, respectively, and the nearest neighbor coupling constant $J > 0$ setting the unit of energy. The second model has an anisotropic Hamiltonian incorporating axial next-nearest neighbor interactions (ANNNI),
\begin{equation}
    \label{eq:HANNNI}
    \mathcal{H}_\mathrm{ANNNI}(\{s_i\}) = -J\sum\limits_{\left<i,j\right>}s_i s_j + \kappa J \!\!\!\! \sum\limits_{\lbrack i,j\rbrack_{\mathrm{axial}}} \!\! s_i s_j - h_{\rm ext}\sum\limits_{i} s_i
    \; ,
\end{equation}
with $[i,j]_\mathrm{axial}$ denoting next-nearest neighbors along the chosen axial direction. Both models either have purely attractive ($\kappa \leq 0$) or SALR ($\kappa > 0$) interactions. (In the rest of this work, the external field of these models is set to zero, $h_{\rm ext}=0$, which endows them with $Z_2$ symmetry, thus simplifying their analysis.) 
The purely attractive regime of both  models straightforwardly presents a paramagnetic-to-ferromagnetic (P-F) phase transition at a finite temperature $T_c(\kappa)$, which belongs to the Ising universality class. Their SALR interaction versions, however, significantly differ.

The ANNNI model with SALR interactions on cubic lattices has been studied with relatively high numerical accuracy using transfer matrices in $d=2$~\cite{hu2021numerical,hu2021resolving}, and with various Monte Carlo simulation schemes and series expansions in $d=3$~\cite{selke1992,charbonneau2022advances}. In both dimensions, at small $\kappa$ the P-F transition remains within the Ising universality class. At large $\kappa$, by contrast, the ground state is modulated along the axial direction. The resulting paramagnetic-to-modulated phase transition is then within the $XY$ universality class. In $d=3$, a finite-temperature tricritical Lifshitz point at $\kappa_\mathrm{L}=0.270$ cleanly separates the two regimes. By contrast, in $d=2$ the details of the low-temperature transition region around $\kappa=1/2$ remain debated. 

The behavior of the model with isotropic SALR interactions is somewhat more complicated. In $d=2$, transfer matrix studies of the SALR interaction regime (or, equivalently, the biaxial next nearest neighbor Ising model, BNNNI) have revealed a phase diagram structurally similar to that of the $d=2$ ANNNI model, albeit with a paramagnetic-to-modulated phase transition that is harder to tease out at high $\kappa$~\cite{hu2021numerical}. In $d=3$, generic field-theoretic arguments suggest that the paramagnetic-to-modulated phase transition is weakly first-order in nature~\cite{charbonneau2021solution}, but the model phenomenology remains incompletely described.  More robust advances have been made for systems on Bethe lattices~\cite{charbonneau2021solution}, in which case the paramagnetic-to-modulated phase transition is clearly within the $XY$ universality class. 

\subsection{Frustrated RBIM}

We also consider the frustrated random-bond Ising model (RBIM) with Hamiltonian
\begin{equation}
\label{eq:HRBIM}
\mathcal H_{\rm RBIM}(\{s_i\}) = -\sum_{\langle i,j\rangle} J_{i j} s_i s_j  -h_{\rm ext}\sum_{i}s_i
\; ,
\end{equation}
where the couplings $J_{ij}$ are taken from a bimodal distribution with probability $\rho$ 
of being ferromagnetic, $J_{ij}=J_0>0$, and probability $1-\rho$ of being anti-ferromagnetic, $J_{ij}= -J_0$. In other words, the couplings are taken at random from the probability distribution:
\begin{equation}
    p(J_{ij}) = \rho \delta(J_{ij} - J_0)+(1-\rho)\delta(J_{ij} + J_0)
    \; ,
    \label{eq:pmJdistribution}
\end{equation}
with $J_0$ setting the unit of energy. Clearly, the standard ferromagnetic Ising model is recovered for $\rho = 1$, and the standard antiferromagnetic Ising model is recovered for $\rho=0$. There also clearly exists a duality: $\rho\leftrightarrow(1-\rho)\,\, \wedge\,\, J_0\leftrightarrow-J_0$.

The RBIM on $d=2$~\cite{cho1997,merz2002,Honecker2001,Honecker2006,Hasenbusch2008a,Hasenbusch2008b,Parisen2009,liu2025, Kanbur2024} and $d=3$~\cite{fajen2020percolation} cubic lattices, in particular, has been extensively studied. For all $d\geq 2$ and $\rho$ close to unity the P-F transition is Ising-like. As $\rho$ decreases, however, dimensional differences emerge. In $d = 3$ the model exhibits a finite-temperature spin-glass phase for $\rho\lesssim0.78$, while in $d = 2$ spin-glass ordering is only present at $T = 0$ (for $\rho \lesssim 0.897$). In mean-field models, such as the infinite-connectivity Sherrington-Kirkpatrick (SK) model as well as comparable models defined on the Bethe lattice, an additional phase emerges: the ferromagnetic-spin-glass (FSG) phase. This intermediate phase between the ferromagnetic and the spin-glass ones exhibits spin-glass behavior while maintaining a non-zero magnetization~\cite{chayes_chayes_sethna_thouless_1986, chayes_sethna_thouless_1988, chayes_sethna_thouless_1990, carlson_chayes_sethna_thouless_1990}.

Interestingly, for all $d \geq 2$, local gauge symmetry gives rise to a Nishimori line \cite{nishimori_internal_1981, nishimori_statistical_2001}  
\begin{eqnarray}
 \label{eq:Nishimori}
    \beta_N J_0= \frac{1}{2} \ln\frac{\rho}{1-\rho},
\end{eqnarray}
where $\beta_N=1/k_B T_N$ is the inverse Nishimori temperature. Along this line certain thermodynamic quantities, such as the internal energy, have closed-form expressions. This line is also invariant under renormalization-group transformations, as is the P-F
transition line. The intersection point of the two
lines therefore gives rise to a multicritical fixed point; this Nishimori point (NP) separates the paramagnetic, ferromagnetic and spin-glass phases in $d \geq 3$. In $d=2$, although no spin glass phase exists, the dynamical (critical) exponent of single spin flip Metropolis Monte Carlo at the Nishimori point has a very high value, $z\simeq 6$, independently of lattice geometry and random bond distribution~\cite{Agrawal2023,Agrawal2024}. Such a high $z$ significantly inhibits configurational sampling (and equilibration) in this regime.

\section{Generalized FK--CK Cluster Scheme}
\label{sec:theorem}

In this section, we generalize the FK--CK random clusters  to the models presented in Sec.~\ref{sec:models}. The general result is that FK--CK clusters in non-frustrated systems  encode thermodynamic correlations~\cite{edwards1988generalization}, 
\begin{equation}
\label{eq:sisj_non}
\langle s_i s_j\rangle = \langle\gamma_{ij}^{\parallel}\rangle
\, , 
\end{equation}
where $\langle s_i s_j\rangle$ is the spin-spin correlation function, and 
$\langle\gamma_{ij}^{\parallel}\rangle$ 
denotes the probability that spins $i$ and $j$ are parallel and belong to a same (random) cluster of the corresponding percolation problem. (Clusters will be defined more explicitly below; for now, the intuitive notion should suffice.) This geometric identification enables updating correlated particles at once, thus accelerating configurational sampling down to the critical temperature, $T_\mathrm{c}$.

In models with frustrated interactions, this relation was later extended by considering both ferromagnetic and antiferromagnetic bonds between neighboring spins~\cite{coniglio1991cluster,cataudella1994critical, Kasai1988, Gandolfi1992, Aizenman2025}
\begin{equation}
\langle s_i s_j\rangle = \langle\gamma_{ij}^{\parallel}\rangle - \langle\gamma_{ij}^{\not\parallel}\rangle,
\label{eq:sisj_frust}
\end{equation}
where $\langle\gamma_{ij}^{\parallel}\rangle$ and $\langle\gamma_{ij}^{\not\parallel}\rangle$ denote the probability that parallel and antiparallel spins $i$ and $j$ belong to the same cluster, respectively. In the frustrated case, the clusters are constructed by joining together neighboring spins such that $J_{ij}s_i s_j >0$. As a result, some of the bonds in a cluster can link neighboring parallel spins and other bonds anti-parallel spins.
Generalized cluster schemes that sum the contributions of parallel and antiparallel spins necessarily overestimate the spin correlations, i.e.,
\begin{equation}
\label{eq:tradclusters}
|\langle s_i s_j\rangle|\leq  \langle\gamma_{ij}^{\parallel}\rangle + \langle\gamma_{ij}^{\not\parallel}\rangle.
\end{equation}
The resulting clusters therefore percolate at temperatures higher than $T_\mathrm{c}$. As a result, these clusters are largely ineffective at configurational sampling in the critical regime, as has been repeatedly demonstrated in numerical simulation~\cite{Leung:1991,deArcangelis1991percolation,cataudella1992percolation,Kalz:2008,Zhu:2015,fajen2020percolation,zheng2022weakening}.

In the rest of this section, we first review the FK--CK construction for the standard Ising model (Sec.~\ref{sec:theorem_Ising}) and then show that the FK--CK formalism can be rigorously generalized to frustrated systems with SALR interactions (Sec.~\ref{sec:theorem_salr}) and to the frustrated RBIM (Sec.~\ref{sec:theorem_rbim})---albeit with some of the bonding ``probabilities'' taking negative values. In contrast to the commonly considered approach corresponding to Eq.~\eqref{eq:tradclusters}, we hence obtain a proper cluster-based geometric description of the generalized relationship for pair correlations.

Before doing so 
it is instructive to recall the difference between the CK and the FK approaches for the Ising model with nearest-neighbor interaction $J$. In the CK formalism, a cluster---also called droplet---is defined as the maximal set of nearest-neighbor parallel spins connected by bonds with probability 
\begin{equation}
\label{eq:pb}
p_B = 1-e^{-2J/kT}
\end{equation}
(The bonds are fictitious; they only define the connectivity and do not change the spin interaction energy.) By contrast, in the FK formalism the introduction of bonds modifies the original spin interaction $J$ into $J= \infty$ with probability $p$ and $J = 0$ with probability $1-p$. This approach leads to the random cluster model with clusters composed of spins connected by infinite interactions. Because of this difference in framing, CK droplets were long seen as unrelated to FK clusters. Only after Swendsen and Wang introduced the cluster dynamics based on the FK formalism was it formally shown that the distribution of CK droplets is the same as that of FK clusters~\cite{coniglio1989exact,coniglio1991cluster}. Although nowadays the CK droplets and the FK clusters are often co-identified, their different origin should nevertheless be kept in mind throughout the demonstration to clarify the argument.

\subsection{Physical clusters for the (unfrustrated) Ising model}
\label{sec:theorem_Ising}
We here review the FK--CK construction for the standard (unfrustrated) Ising model, which leads to the result in Eq.~\eqref{eq:sisj_non}. This demonstration further establishes the notation and highlights the key steps that are then followed for frustrated systems in the following subsections.

Consider a system of Ising spins on a lattice with ferromagnetic nearest-neighbor interactions, for which---up to an irrelevant additive constant---the Hamiltonian can be written as
\begin{equation}
H(\{s_i\}) = - \sum_{\langle i,j \rangle} J (s_i s_j - 1)\, . 
\end{equation} 
The key idea is to replace the original Ising Hamiltonian with an annealed diluted one
\begin{equation}
H'(\{s_i\}) = - \sum_{\langle i,j \rangle} J_{ij}' (s_i s_j - 1) \, ,
\label{eq:annealedH}
\end{equation}
where
\[
J_{ij}' =
\begin{cases}
J' & \text{with probability } p_B \, , \\
0 & \text{with probability } 1 - p_B \, .
\end{cases}
\]
For a given $J'$, the parameter $p_B$ is chosen such that
\begin{equation}
e^{\beta J (s_i s_j - 1)} = p_B \, e^{\beta J'(s_i s_j - 1)} + \left(1 - p_B \right) \, .
\end{equation}
In the limit $\beta J' \rightarrow \infty$, we have $e^{\beta J'(s_i s_j - 1)}=\delta_{s_i, s_j}$, and from Eq.~\eqref{eq:annealedH} we recover Eq.~\eqref{eq:pb}.
As a result, the Boltzmann weight of each spin configuration is
\begin{equation} \label{eq:HpB}
e^{-\beta H(\{s_i\})} = \prod_{\langle i,j \rangle} e^{\beta J (s_i s_j - 1)} = \prod_{\langle i,j \rangle} \left[ p_B \delta_{s_i, s_j} + \left(1 - p_B \right)\right] \, .
\end{equation}
After expanding the products in the above relation, we can write
\begin{equation} \label{eq:WFK-J}
e^{-\beta H(\{s_i\})} = \sum_C W_{\rm FK}(\{s_i\}, C) \, ,
\end{equation}
where
\begin{eqnarray} \label{eq:WFK1}
\nonumber
W_{\rm FK}(\{s_i\}, C) &=& \prod_{\langle i,j\rangle \in C} p_B \delta_{s_i, s_j} \prod_{\langle i,j\rangle \not \in C} (1 - p_B) \\
&=& p_B^{|C|} (1 - p_B)^{|A|} \prod_{\langle i,j\rangle \in C} \delta_{s_i, s_j}  \, .
\end{eqnarray}
Here, $C$ is a subset of all the bonds that correspond to a specific configuration of the interactions $J'$, such that the bonds with $J' = \infty$ belong to the cluster configuration $C$, and the subset of bonds with $J' = 0$ defines $A$, with $|C| + |A| = |E|$ for $E$ the set of all bonds. In other words, $W_{\rm FK}(\{s_i\}, C)$ is the statistical weight of a spin configuration $\{s_i\}$ with the set of interactions $\{J_{ij}'\}$ in the diluted model with $|C|$ edges interacting with infinite strength and all other edges interacting with zero strength. The Kronecker delta indicates that two spins connected by an infinite interaction strength must be in the same state. Therefore, the cluster configuration $C$ can be decomposed into clusters of parallel spins connected by interactions of infinite strength. 

\begin{figure}
    \centering
    \includegraphics[width=\linewidth]{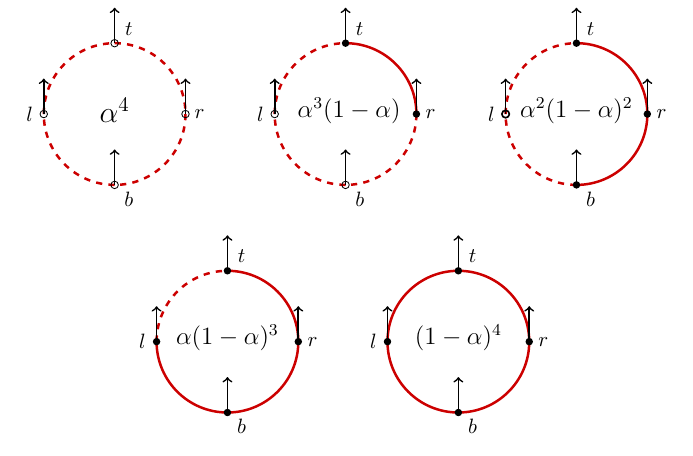}
    \caption{Chain of four Ising spins with ferromagnetic interactions, $J_{ij}=J>0$, under periodic boundary conditions, in its minimal energy spin configuration $(s_t, s_r, s_b, s_l) = (\uparrow, \uparrow, \uparrow, \uparrow)$. All possible clusters are shown, with zero to four links (solid red lines), thus identifying spins as being either part (filled circle) or not (unfilled circle) of a cluster. Cluster multiplicities are 1, 4, 6, 4, and 1, respectively, and FK--CK cluster weights $W_\mathrm{FK}$ (center) are expressed for bonding probability $\alpha=e^{-2\beta J}\in(0,1]$.
    }
    \label{fig:clusters-example-FM}
\end{figure}

The partition function $Z$ can then be obtained by summing over all spin configurations with their Boltzmann weight given by Eq.~\eqref{eq:WFK-J}. Because each disconnected 
cluster in $C$ gives a factor of 2, one then gets
\begin{equation}
Z = \sum_C 2^{N_C} p_B^{|C|} (1 - p_B)^{|A|} \, ,
\end{equation}
where $N_C$ is the number of clusters in $C$. Put differently, the FK--CK formalism gives a partition function, $Z=\sum_C W(C)$, whose structure is equivalent to that of a correlated bond percolation model,
\begin{equation} \label{eq:WCperc}
\begin{aligned}
W(C) & = \sum_{\{s_i\}} W_{\rm FK}(\{s_i\}, C) 
\\
& 
= 2^{N_C}  p_B^{|C|} (1 - p_B)^{|A|}   \, ,
\end{aligned}
\end{equation}
which coincides with the weight of the random bond percolation except for the extra factor $2^{N_C}$.
All clusters and their weights for the spin configuration $(s_t, s_r, s_b, s_l) = (\uparrow, \uparrow, \uparrow, \uparrow)$ are shown in Fig.~\ref{fig:clusters-example-FM}.

From Eqs.~\eqref{eq:WFK-J} and~\eqref{eq:WFK1}, it follows that
\begin{equation} \label{eq:corr}
\langle s_i s_j \rangle = \langle \gamma_{ij}^{\parallel} \rangle_W \, ,
\end{equation}
where $\gamma_{ij}^\parallel (C) = 1$ if $i$ and $j$ belong to the same cluster $C$ and $0$ otherwise, $\langle \cdots \rangle$ is the thermodynamic average with the Boltzmann weights and $\langle \cdots \rangle_W$ is the average over bond configurations in the bond correlated percolation with weights given by Eq.~\eqref{eq:WCperc}.  This result is obtained by taking the sum over all spin configurations, using the weight in Eq.~\eqref{eq:WFK1}, which gives
\begin{equation}
s_is_j=
\begin{cases}
     1 & \textrm{if $i$ an $j$ belong to the same cluster}\\
     0& \textrm{otherwise}.
\end{cases}    
\end{equation}

\subsection{Generalization to SALR interaction models}
\label{sec:theorem_salr}

Having established the FK--CK framework for unfrustrated systems, we now extend it to models with SALR interactions. 
The key conceptual step is recognizing that the mathematical formalism remains intact even when NNN interactions are antiferromagnetic, thus making some bonding ``probabilities'' negative. In other words, although the equality between spin correlations and cluster connectivity still holds exactly, the constructive sampling interpretation is lost.

Consider a model with isotropic SALR interactions. (The extension to the anisotropic version is straightforward.) Again, apart from an irrelevant additive constant (and in the absence of external magnetic field), the Hamiltonian of the isotropic model in Eq.~\eqref{eq:Hiso} can be rewritten as
\begin{equation}
H_{\rm iso}(\{s_i\}) = - J_1 \sum_{\langle i,j \rangle} (s_i s_j - 1) - J_2 \sum_{\langle \langle i,j \rangle \rangle} (s_i s_j - 1) \, ,
\end{equation}
with $J_1 = J >0$ and $J_2 = - \kappa J < 0$. For any pair of interacting spins we define
\begin{equation}
e^{\beta J_\nu(s_i s_j - 1)} = p_B^{(\nu)} \delta_{s_i s_j} + \left(1 - p_B^{(\nu)} \right) \, ,
\end{equation}
with a generalization of Eq.~\eqref{eq:pb},
\begin{equation}
\label{eq:NNN_CKFK_definition}
p_B^{(\nu)} = 1 - e^{-2 \beta J_\nu}, \quad \nu = 1, 2 \, .
\end{equation}
We can then generalize Eq.~\eqref{eq:WFK-J} as 
\begin{equation}
e^{-\beta H_{\rm iso}(\{s_i\})} = \sum_C W_{\rm FK}(\{s_i\},  
C_1, C_2) \, ,
\end{equation}
where
\begin{equation}
\begin{aligned}
&W_{\rm FK}(\{s_i\}, 
C_1, C_2, ) = \left( p_B^{(1)} \right) ^{|C_1|} \left(1 - p_B^{(1)} \right)^{|A_1|} \\
& \qquad \times \left(p_B^{(2)}\right)^{|C_2|} \left(1 - p_B^{(2)} \right)^{|A_2|}\!\!\! \prod_{\langle i,j \rangle \in C_1} \!\!\!\delta_{s_i s_j}\!\!\!\!\!\! \prod_{\langle \langle i,j \rangle \rangle \in C_2} \!\!\!\delta_{s_i s_j} \, .
\end{aligned}
\end{equation}
Here, $C_1$ and $C_2$ are the cluster 
configurations of the interactions $J_1$ (of type 1) on NN bonds  and $J_2$ (of type 2) on NNN bonds, respectively.

The partition function is then
\begin{equation}
\begin{aligned}
Z & = \sum_{C_1} \sum_{C_2} \left( p_B^{(1)} \right)^{|C_1|} \left(1 - p_B^{(1)} \right)^{|A_1|} \\
& \qquad \qquad \times \left(p_B^{(2)} \right)^{|C_2|} \left(1 - p_B^{(2)} \right)^{|A_2|} 2^{N(C_1, C_2)} \, ,
\end{aligned}
\end{equation}
where $N(C_1, C_2)$ is the number of clusters with the combined bond configuration. Therefore, the weight of the cluster configuration $(C_1,C_2)$ in the associated correlated bond percolation problem is 
\begin{equation}
\begin{aligned}
W(C_1, C_2) & = \left( p_B^{(1)} \right)^{|C_1|} \left(1 - p_B^{(1)} \right)^{|A_1|} \\
& \qquad  \times \left(p_B^{(2)} \right)^{|C_2|} \left(1 - p_B^{(2)} \right)^{|A_2|} 2^{N(C_1, C_2)} \, ,
\end{aligned}
\end{equation}
and, as for the unfrustrated model,
\begin{equation}
\langle s_i s_j \rangle = \langle \gamma_{ij}^{\parallel} \rangle_W \, ,
\end{equation}
with $\gamma_{ij}^{\parallel} = 1$ if $i$ and $j$ belong to the same cluster, and $0$ otherwise. This result is independent of the sign of the coupling constants, and in particular holds in the SALR interaction case, i.e., $J_1>0$ and $J_2<0$. For this model, however, $p_B^{(2)}$ is negative and hence $1-p_B^{(2)}>1$. Therefore, rewriting the Boltzmann weights can no longer be interpreted in terms of probabilities.

\subsection{Generalization to the frustrated RBIM}
\label{sec:theorem_rbim}
We also consider the specific case of the random bond Ising model (RBIM) that introduces frustration through the presence of quenched disorder, as described in Eq.~\eqref{eq:HRBIM}. The FK--CK generalization nevertheless proceeds analogously to the SALR interaction case. The crucial result is that for any fixed disorder realization, the spin-spin correlation function equals the cluster connectivity probability, and this equality survives averaging over the quenched disorder. This establishes that the generalized FK--CK clusters properly encode thermodynamic correlations in the presence of \emph{both} frustration and disorder. 

Once more, apart from an irrelevant additive constant (and in the absence of external magnetic field), the Hamiltonian in Eq.~\eqref{eq:HRBIM} can be rewritten  as
\begin{equation}
H_{\rm RBIM} (\{s_i\}) = - \sum_{\langle i,j \rangle} J_{ij} (s_i s_j - 1)\, .
\end{equation}
We then replace this expression with an annealed diluted Hamiltonian
\begin{equation}
H'(\{s_i\}) = - \sum_{\langle i,j \rangle} J'_{ij}(s_i s_j - 1) \, ,
\label{eq:annealed_HRBIM}
\end{equation}
where
\[
J'_{ij} =
\begin{cases}
J' & \text{with probability } p_B^{(ij)} \, , \\
0 & \text{with probability } 1 - p_B^{(ij)} \, .
\end{cases}
\]
For a fixed $J'$, $p_B^{(ij)}$ is chosen such that
\begin{equation}
e^{\beta J_{ij} (s_i s_j - 1)} = p_B^{(ij)} \, e^{\beta J'(s_i s_j - 1)} + \left(1 - p_B^{(ij)} \right)
\end{equation}
for each bond. 
In the limit $J' \rightarrow \infty$, we have $e^{\beta J'(s_i s_j - 1)}=\delta_{s_i, s_j}$, and $p_B^{(ij)}$ is given by
\begin{equation}
p_B^{(ij)} = 1 - e^{-2 \beta J_{ij}} \, .
\label{eq:FKCK_definition_rbim}
\end{equation}
(As anticipated, for $J_{ij}<0$ the parameter $p_B^{(ij)}$ is negative; the formal construction can nevertheless be continued.)
Consequently, the Boltzmann weight is
\begin{equation}
e^{-\beta H_{\rm RBIM}(\{s_i\})} = \prod_{\langle i,j \rangle} \left[ p_B^{(ij)} \delta_{s_i, s_j} + \left(1 - p_B^{(ij)} \right)\right] \, ,
\end{equation}
and hence we can write
\begin{equation} \label{eq:WFK}
e^{-\beta H_{\rm RBIM}(\{s_i\})} = \sum_C W_{\rm FK}(\{s_i\}, \{J_{ij} \}, C) \, ,
\end{equation}
where
\begin{eqnarray}
&&
W_{\rm FK}(\{s_i\}, \{J_{ij} \}, C) = 
\nonumber\\
[5pt]
&& 
\qquad\qquad
\prod_{\langle i,j\rangle \in C} p_B^{(ij)} \delta_{s_i, s_j} 
\prod_{\langle i,j\rangle \not \in C} (1 - p_B^{(ij)}) \, .
\end{eqnarray}
Here, $C$ is a subset of all the bonds that correspond to a specific configuration of the interactions $J'_{ij}$, such that the bonds with $J' = \infty$ belong to the cluster configuration $C$, and the subset of bonds with $J' = 0$ defines $A$, with $|C| + |A| = |E|$ for $E$ the set of all bonds.
Note that we have here included the dependence on the full realization of the random couplings $\{J_{ij}\}$ in the definition of the FK statistical weights, because the quenched disorder over bond types make their ``probabilities'' differ for each system realization.
 
The partition function $Z$ can then be obtained by summing over all spin configurations,
\begin{equation}
Z = \sum_C 2^{N_C} \!\!\! \prod_{\langle i,j\rangle \in C} p_B^{(ij)} \prod_{\langle i,j\rangle \not \in C} (1 - p_B^{(ij)})  \, ,
\end{equation}
where $N_C$ is the number of clusters in $C$. 
Therefore, the FK--CK formalism gives a partition function, $Z=\sum_C W(C)$, whose structure is equivalent---albeit, as anticipated above, with some negative $p_B$---to that of a correlated bond percolation model, 
\begin{equation}
\begin{aligned}
W(C) & = \sum_{\{s_i\}} W_{\rm FK}(\{s_i\}, \{J_{ij} \}, C) \\
& = 2^{N_C} \!\!\! \prod_{\langle i,j\rangle \in C} p_B^{(ij)} \prod_{\langle i,j\rangle \not \in C} (1 - p_B^{(ij)})  \, .
\end{aligned}
\end{equation}
It follows that
\begin{equation}
\langle s_i s_j \rangle = \langle \gamma_{ij}^{\parallel} \rangle_W \, ,
\end{equation}
where $\gamma_{ij}^\parallel (C) = 1$ 
if $i$ and $j$ belong to the same cluster, and $0$ otherwise. We emphasize that this equality holds when averaging over Boltzmann weights and averaging over bond configurations, for {\it any} fixed disorder realization $\{J_{ij} \}$. Consequently, the equality must hold after averaging over the quenched disorder as well. A concrete example is discussed in the concluding Sec.~\ref{sec:conclusion}. 

\subsection{Cluster definition remarks}
In all the above cases, the equality between the spin–spin correlation function and $\langle \gamma_{ij}^\parallel \rangle_W$ follows from the FK--CK clusters ensuring that the Boltzmann weight of every spin configuration  coincides with the statistical weight of the corresponding random-bond percolation model defined by the measure $W(C)$. This property crucially implies that, even if one constructs another cluster model for which clusters percolate exactly at the Ising critical point and exhibit the same critical exponents---as for the $\alpha$-parameter cluster model described in Secs.~\ref{sec:iso_revCK} and~\ref{sec:alpha_cluster_rbim}---the equality between the Boltzmann weight of the spin configurations and the statistical weight of the cluster model no longer holds. Hence, nothing guarantees that throughout the phase diagram, the spin–spin correlation function is equal to the percolation correlation function. Consequently, an algorithm based on such a cluster model is not expected to weaken the critical slowing down with any significance, as was observed in Ref.~\cite{zheng2022weakening} for a specific SALR interaction model. We are tempted to conjecture that this equality is a necessary condition for weakening the critical slowing down in general, but cannot formally demonstrate it at this point. 

\section{Exact solution of the models on the Bethe lattice}
\label{sec:Bethe}

The Bethe lattice provides an ideal testing ground for cluster schemes because it admits exact solutions via the cavity method~\cite{mezard1987spin, mezard2009information}. In this section, we solve the three models presented in Sec.~\ref{sec:models} on this lattice. 
The analysis of each model follows a common structure. We first derive recursive equations for the local spin configuration probabilities, $\eta$, which encode all thermodynamic properties such as spin correlations and the system free energies. We then develop analogous recursive equations for the cluster percolation probabilities---denoted $P$ for site probabilities and $\pi$ for the cavity form, in which the influence of a given neighbor is omitted to construct the recursion. These equations characterize the geometric properties of clusters, in particular whether spins belong to the infinite percolating cluster. These two sets of equations are structurally similar but serve different purposes. The configuration probabilities $\eta$ determine when the system undergoes thermodynamic phase transitions (e.g., P-F), while the percolation probabilities $P$ and $\pi$ determine whether clusters span the entire system. Comparing the temperatures at which the thermodynamic and the percolation transitions occur reveals whether a given cluster definition properly captures the physics of the critical regime. The technical derivations presented here can be solved iteratively for given parameter values; their solutions and physical implications are discussed in Sec.~\ref{sec:discussion}.

\subsection{Isotropic SALR interaction model}
\label{sec:iso}

\begin{figure*}[t]
\centering
\includegraphics[scale=0.5]{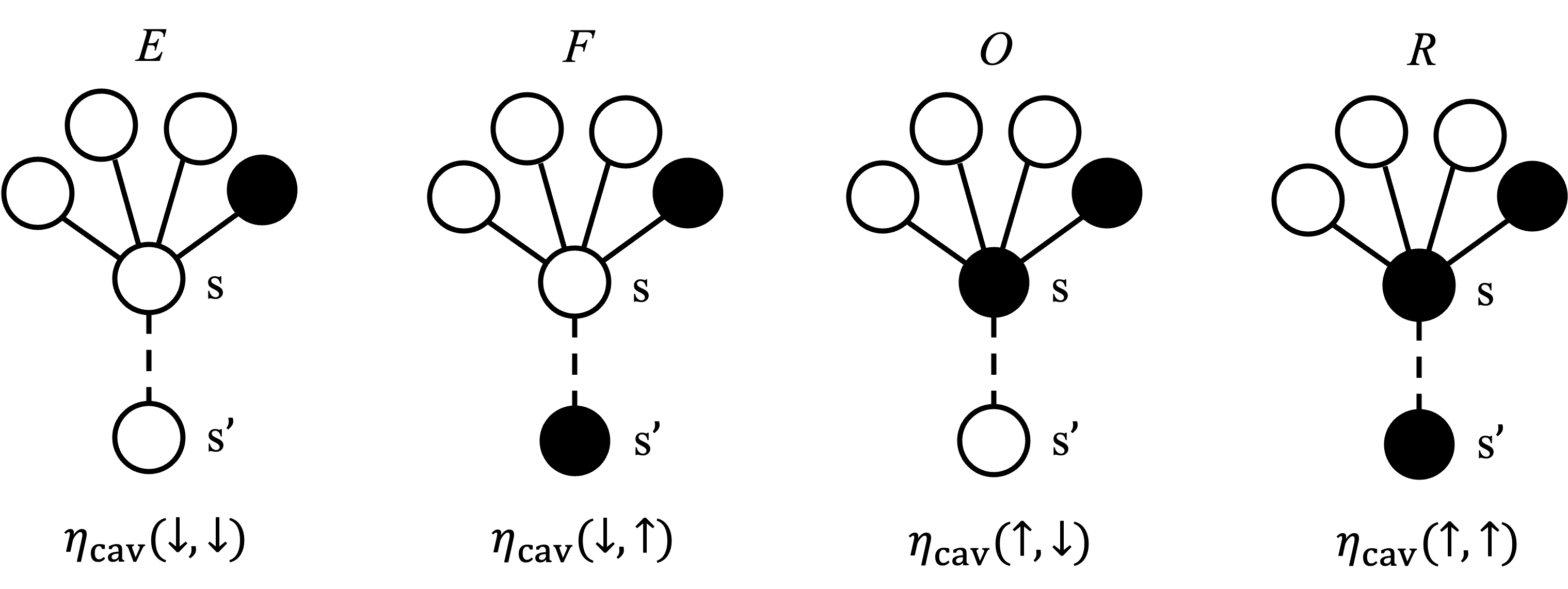}
\caption{Configurations with up (black) and down (white) spins corresponding to the various cavity fields of the isotropic SALR interaction model on a Bethe lattice with connectivity 
$c+1=5$.
$\eta_\mathrm{cav}(s, s')$ is the (local) configuration probability for the current site $s$ and the backward (cavity) site  $s'$. The upper labels $E, \, F, \, O, \, R$ equal 
$\eta_\mathrm{cav}(s, s')$ in each of the four cases and are used as short-hand notation
for the latter in the text. Here and in what follows we will denote with an up arrow, $\uparrow \ $, the 
spins pointing up and with downarrow, $\downarrow \ $, the spins pointing down.} 
\label{fig:bethe_configs}
\end{figure*}

For the isotropic SALR interaction model of Eq.~\eqref{eq:Hiso}, the recursive equations of the thermodynamic properties as well as the percolation behavior of two different cluster criteria are considered. Because the phase behavior of the model is non-trivial even at relatively small $\kappa$, we also obtain the linear expansion in $\kappa$. To facilitate the numerical work, we additionally analyze the linear stability of the solutions. 

\subsubsection{Cavity field and recursion relations}
\label{sec:isingmodel}

For a Bethe lattice with connectivity $c+1$, the cavity method provides the (local) configuration probabilities for the current site $s$ and the backward (or cavity) site  $s'$~\cite{semerjian2009exact} 
\begin{align}
\label{eq:spin_configs_iso}
    &\eta_\mathrm{cav}(s, s') = Z_\mathrm{cav}^{-1} \sum\limits_{l=0}^{c} \binom{c}{l}[\eta_\mathrm{cav}(\uparrow,s)]^l [\eta_\mathrm{cav}(\downarrow,s)]^{c-l} \\
    &\times e^{\beta J s(2l-c)}e^{-\beta \kappa J s'(2l-c)}e^{-\beta \kappa J\frac{l(l-1)+(c-l)(c-l-1)-2l(c-l)}{2}},\nonumber
\end{align}
where $Z_\mathrm{cav}$ is the normalization factor of these probabilities, $\sum_{s,s'}\eta_\mathrm{cav}(s,s')=1$, and  the spin density
\begin{align}
\label{eq:spin_density_iso}
    &\eta(s) = Z_{\mathrm{site}}^{-1} \sum\limits_{l=0}^{c+1} \binom{c+1}{l}[\eta_\mathrm{cav}(\uparrow,s)]^l [\eta_\mathrm{cav}(\downarrow,s)]^{c+1-l}\\
    &\times e^{\beta J s(2l-c-1)} e^{-\beta \kappa J\frac{l(l-1)+(c+1-l)(c-l)-2l(c+1-l)}{2}},\nonumber
\end{align}
where $Z_{\mathrm{site}}$ is the normalization factor, such that $\eta(\uparrow)+\eta(\downarrow)=1$. The free energy per site, which is obtained by averaging over all possible choices of removed sites and removed edges~\cite{charbonneau2021solution}, is 
\begin{equation}
\label{eq:salr_bf}
    \beta f = \beta f_{\mathrm{site}} - \frac{c+1}{2} \beta f_{\mathrm{link}}
\end{equation}
where $\beta  f_{\mathrm{site}} = -\ln Z_{\mathrm{site}}\nonumber$ and
\begin{equation}
\beta f_{\mathrm{link}} = -\ln{(E^2e^{\beta J} + R^2e^{\beta J} + 2F\, O \, e^{-\beta J})}, 
\end{equation}
using the short-hand notation $E/F/O/R$ to encode the different spin configuration probabilities $\eta_\mathrm{cav}(s,s')$ (see Fig.~\ref{fig:bethe_configs}). 
\vspace{0.25cm}
\begin{prop}
The $Z_2$ symmetry of the isotropic SALR interaction model provides the following identities:
\label{prop:sym1}
\begin{enumerate}[font=\upshape,label=(\alph*)]
\item for $\kappa = 0$,
\begin{align}
E &= F = Z_\mathrm{cav}^{-1} \ (E e^{\beta J} + O e^{-\beta J})^c \ , \label{eq:spin_iso_kappa0_1}\\
O &=R = Z_\mathrm{cav}^{-1} \ (E e^{-\beta J} + O e^{\beta J})^c \ ;\label{eq:spin_iso_kappa0_2}
\end{align}
\item for any $\kappa$ within the paramagnetic phase, $E=R$ and $F=O$;
\item for any $\kappa$ within the paramagnetic phase, the net magnetization is zero with $\eta(\uparrow)=\eta(\downarrow)=1/2$, and hence $E+F=O+R=1/2$.
\end{enumerate}
\end{prop}

\subsubsection{Linear expansion around $\kappa=0$}
\label{sec:spin_config_ls}
For the special case $\kappa=0$, closed-form expressions of the cavity field equations for the paramagnetic phase can be obtained as in Eqs.~\eqref{eq:spin_iso_kappa0_1}-\eqref{eq:spin_iso_kappa0_2}. 
For $\kappa\neq0$, however, these equations need to be solved iteratively. In order to validate these results, we here also consider corrections linear in $\kappa$ to the configuration probabilities, 
\begin{equation}
    \eta_\mathrm{cav}(s,s') = \eta_\mathrm{cav}^0(s,s') + \kappa J \left.\frac{\partial\eta_\mathrm{cav}(s,s')}{\partial \kappa J}\right|_{\kappa=0},
\end{equation}
which entails computing 
\begin{equation}
\label{eq:linear1}
    \frac{\partial\eta_\mathrm{cav}(s,s')}{\partial \kappa J}=Z_\mathrm{cav}^{-1}\frac{\partial f_\mathrm{cav}(s,s')}{\partial \kappa J} - \frac{f_\mathrm{cav}(s,s')}{Z_\mathrm{cav}^2}\frac{\partial Z_\mathrm{cav}}{\partial \kappa J}.
\end{equation}
To streamline the presentation, we separately consider the derivatives of the summation, $f_\mathrm{cav}(s,s')=\eta_\mathrm{cav}(s,s')Z_\mathrm{cav}$, and of the normalization factor, $Z_\mathrm{cav}$. From Eq.~\eqref{eq:spin_configs_iso}, we have 
\begin{widetext}
\begin{equation}
\label{eq:linear2}
\begin{split}
    \left.\frac{\partial f_\mathrm{cav}(s,s')}{\partial \kappa J}\right|_{\kappa=0}
    &=\sum\limits_{l=0}^{c} \binom{c}{l}[\eta_\mathrm{cav}^0(\uparrow,s)]^l [\eta_\mathrm{cav}^0(\downarrow,s)]^{c-l} e^{\beta J s(2l-c)} (-\beta) \left[s'(2l-c) + \frac{l(l-1)+(c-l)(c-3l-1)}{2}\right]\\
    &+ \sum\limits_{l=0}^{c-1} \binom{c}{l}[\eta_\mathrm{cav}^0(\uparrow,s)]^l [\eta_\mathrm{cav}^0(\downarrow,s)]^{c-l-1} e^{\beta J s(2l-c)}(c-l)\frac{\partial\eta_\mathrm{cav}(\downarrow,s)}{\partial \kappa J}\\
    &+ \sum\limits_{l=1}^{c} \binom{c}{l}[\eta_\mathrm{cav}^0(\uparrow,s)]^{l-1} [\eta_\mathrm{cav}^0(\downarrow,s)]^{c-l} e^{\beta J s(2l-c)}l\frac{\partial\eta_\mathrm{cav}(\uparrow,s)}{\partial \kappa J},
\end{split}
\end{equation}
\end{widetext}
where $\eta_\mathrm{cav}^0(s,s')$ denotes the corresponding configuration probabilities at $\kappa=0$, and 
\newpage
\begin{equation}
\label{eq:linear3}
    \frac{\partial Z_\mathrm{cav}}{\partial \kappa J}= \sum\limits_{s,s'} \frac{\partial f_\mathrm{cav}(s,s')}{\partial \kappa J}.
\end{equation}

Equations~\eqref{eq:linear1}-\eqref{eq:linear3} provide recursive relations with which the four derivatives $\partial_{\kappa J}\eta_\mathrm{cav}(s,s')$ can be directly obtained. Various simplifying relations can then be invoked. For conciseness, we write the coefficients of the three terms in Eq.~\eqref{eq:linear2} as $\widehat{\phi}_i$, where $i\in \{0,1,2\}$ is the order of the term, and $\phi \in \{E, F, O, R \}$ denotes the configuration of $f_\mathrm{cav}(s,s')$, as in Fig.~\ref{fig:bethe_configs}.

\begin{itemize}[leftmargin=*]
\item  Because the \textit{coefficients} of the second and third terms are independent of $s'$, we get that $\hat{E}_i=\hat{F}_i$ and $\hat{O}_i=\hat{R}_i$ for $i=1$ or $2$.

\item From Prop.~\ref{prop:sym1}(a) and (b) for the paramagnetic phase, we have that each $\eta^0_\mathrm{cav}(s,s')=1/4$, and hence
\begin{align}
\hat{E}_1&=\hat{F}_1=\hat{O}_2=\hat{R}_2=c\left(\frac{\cosh{(\beta J)}}{2}\right)^{c-1}e^{\beta J},\\
\hat{E}_2&=\hat{F}_2=\hat{O}_1=\hat{R}_1=c\left(\frac{\cosh{(\beta J)}}{2}\right)^{c-1}e^{-\beta J},\\
\hat{E}_0&=\hat{R}_0,\\
\hat{F}_0&=\hat{O}_0.
\end{align} 

\item From Prop.~\ref{prop:sym1}(b) and (c) for the paramagnetic phase, we have 
\begin{equation}
\label{eq:linear_symmetry}
    \dfrac{\partial E}{\partial \kappa J} = -\dfrac{\partial F}{\partial \kappa J} = -\dfrac{\partial O}{\partial \kappa J} = \dfrac{\partial R}{\partial \kappa J}.
\end{equation}
\end{itemize}

In short-hand notation, Eq.~\eqref{eq:linear3} then becomes
\begin{widetext}
\begin{equation}
\begin{split}
\label{eq:linear_dz}
    \frac{\partial Z_\mathrm{cav}}{\partial \kappa J} 
    &= (\hat{E}_0+\hat{F}_0+\hat{O}_0+\hat{R}_0) + (\hat{E}_1+\hat{F}_1)\frac{\partial E}{\partial \kappa J} + (\hat{O}_1+\hat{R}_1)\frac{\partial F}{\partial \kappa J} + (\hat{E}_2+\hat{F}_2)\frac{\partial O}{\partial \kappa J} + (\hat{O}_2+\hat{R}_2)\frac{\partial R}{\partial \kappa J}.
\end{split}
\end{equation}
\end{widetext}
Applying Eqs.~\eqref{eq:linear2} and \eqref{eq:linear_dz} to Eq.~\eqref{eq:linear1}, gives four equations for the four variants of $\partial_{\kappa J}\eta_\mathrm{cav}(s,s')$, hence reducing the problem to an exercise in linear algebra. From Eq.~\eqref{eq:linear_symmetry}, we have that the various $\partial_{\kappa J}\eta_\mathrm{cav}(s,s')$ for the paramagnetic phase are interchangeable, which we denote $\partial_{\kappa J} E$. Using the aforementioned symmetry relations, Eq.~\eqref{eq:linear1} at $\kappa = 0$ becomes
\begin{equation}
    \left.\frac{\partial E}{\partial \kappa J}\right|_{\kappa=0} = \frac{\hat{E}_0-\hat{F}_0}{2 Z_\mathrm{cav}},
\end{equation}
which, as expected, only depends on the (inverse) temperature $\beta J$ and on connectivity $c+1$. To linear order in $\kappa$, the configuration probabilities  are then 
\begin{equation}
\begin{split}
    \eta_\mathrm{cav}(s,s') 
    &= \frac{1}{4} + (2\delta_{s,s'}-1)\kappa J \left.\frac{\partial E}{\partial \kappa J}\right|_{\kappa=0}.
\end{split}
\end{equation}

\subsubsection{Linear stability analysis}
\label{sec:spin_config_lsa}

To sidestep some of the numerical difficulties associated with determining $T_c$ from the free energy, we analyze the linear stability of the recursive expressions in Eq.~\eqref{eq:spin_configs_iso}. For small fluctuations around a fixed state, we can express the cavity field as
\begin{equation}
    \bm{\eta_\mathrm{cav}(s,s')}
    = \bm{\eta_\mathrm{cav}^\mathrm{fix}(s,s')} + \mathcal{J} d\bm{ \eta_\mathrm{cav}(s,s')}.
\end{equation}
In short-hand notation, this matrix equation becomes
\begin{equation}
\left(\begin{array}{c}
E  \\
F  \\
O  \\
R 
\end{array}\right)=
\left(\begin{array}{c}
E^\mathrm{fix}  \\
F^\mathrm{fix}  \\
O^\mathrm{fix}  \\
R^\mathrm{fix} 
\end{array}\right)
+ \mathcal{J}
\left(\begin{array}{c}
dE  \\
dF  \\
dO  \\
dR 
\end{array}\right),
\end{equation}
where ${\mathcal J}$ is the $4\times 4$ Jacobian matrix of derivatives of Eq.~\eqref{eq:spin_configs_iso} with respect to each configuration variable 
\begin{equation}
    \mathcal{J}_{ij} = Z_\mathrm{cav}^{-1}\frac{\partial f_\mathrm{cav}(s_i,s_i')}{\partial \eta_\mathrm{cav}(s_j,s_j')}
    -\frac{\eta_\mathrm{cav}(s_i,s_i')}{Z_\mathrm{cav}}\frac{\partial Z_\mathrm{cav}}{\partial \eta_\mathrm{cav}(s_j,s_j')}
    \label{eq:J_salr_entry}
\end{equation}
with $f_\mathrm{cav}(s,s')$ defined as in Sec.~\ref{sec:spin_config_ls}.
For conciseness, in the rest of this subsection we denote $f_\mathrm{cav}(s,s')$ and $Z_\mathrm{cav}$ as $f_{\phi}$ ($\phi \in \{E, F, O, R \}$) and $Z$, respectively, thus yielding
\begin{widetext}
\begin{equation}
\label{eq:lsa_matrix}
\mathcal{J} = 
\left(\begin{array}{cccc}
\frac{1}{Z}\frac{\partial f_E}{\partial E}-\frac{E}{Z}\frac{\partial Z}{\partial E}  
& -\frac{E}{Z}\frac{\partial Z}{\partial F} 
& \frac{1}{Z}\frac{\partial f_E}{\partial O}-\frac{E}{Z}\frac{\partial Z}{\partial O}
& -\frac{E}{Z}\frac{\partial Z}{\partial R} 
\\
[5pt]
\frac{1}{Z}\frac{\partial f_F}{\partial E}-\frac{F}{Z}\frac{\partial Z}{\partial E}  
& -\frac{F}{Z}\frac{\partial Z}{\partial F} 
& \frac{1}{Z}\frac{\partial f_F}{\partial O}-\frac{F}{Z}\frac{\partial Z}{\partial O}
& -\frac{F}{Z}\frac{\partial Z}{\partial R} 
\\
[5pt]
-\frac{O}{Z}\frac{\partial Z}{\partial E} 
& \frac{1}{Z}\frac{\partial f_O}{\partial F}-\frac{O}{Z}\frac{\partial Z}{\partial F} 
& -\frac{O}{Z}\frac{\partial Z}{\partial O}  
& \frac{1}{Z}\frac{\partial f_O}{\partial R}-\frac{O}{Z}\frac{\partial Z}{\partial R} 
\\ 
[5pt]
-\frac{R}{Z}\frac{\partial Z}{\partial E} 
& \frac{1}{Z}\frac{\partial f_R}{\partial F}-\frac{R}{Z}\frac{\partial Z}{\partial F} 
& -\frac{R}{Z}\frac{\partial Z}{\partial O}  
& \frac{1}{Z}\frac{\partial f_R}{\partial R}-\frac{R}{Z}\frac{\partial Z}{\partial R} 
\\ 
\end{array}\right).
\end{equation}
\end{widetext}

For a given set of converged configurational probabilities---as obtained from Eq.~\eqref{eq:spin_configs_iso}---this matrix and its eigenvalues can be computed. The leading eigenvalue, $\lambda_\mathrm{max}$ determines the stability of a state. For  $\lambda_\mathrm{max}<1$, repeatedly iterating the cavity equations leads to convergence, and hence the corresponding state is equilibrated; for $\lambda_\mathrm{max} > 1$, divergence ensues, and hence that state is unstable; for $\lambda_\mathrm{max} = 1$, the state is marginally stable.  
\begin{figure}[h!]
\centering
\includegraphics[scale=0.45,trim={0.8cm 0 0.1cm 0.2cm},clip]{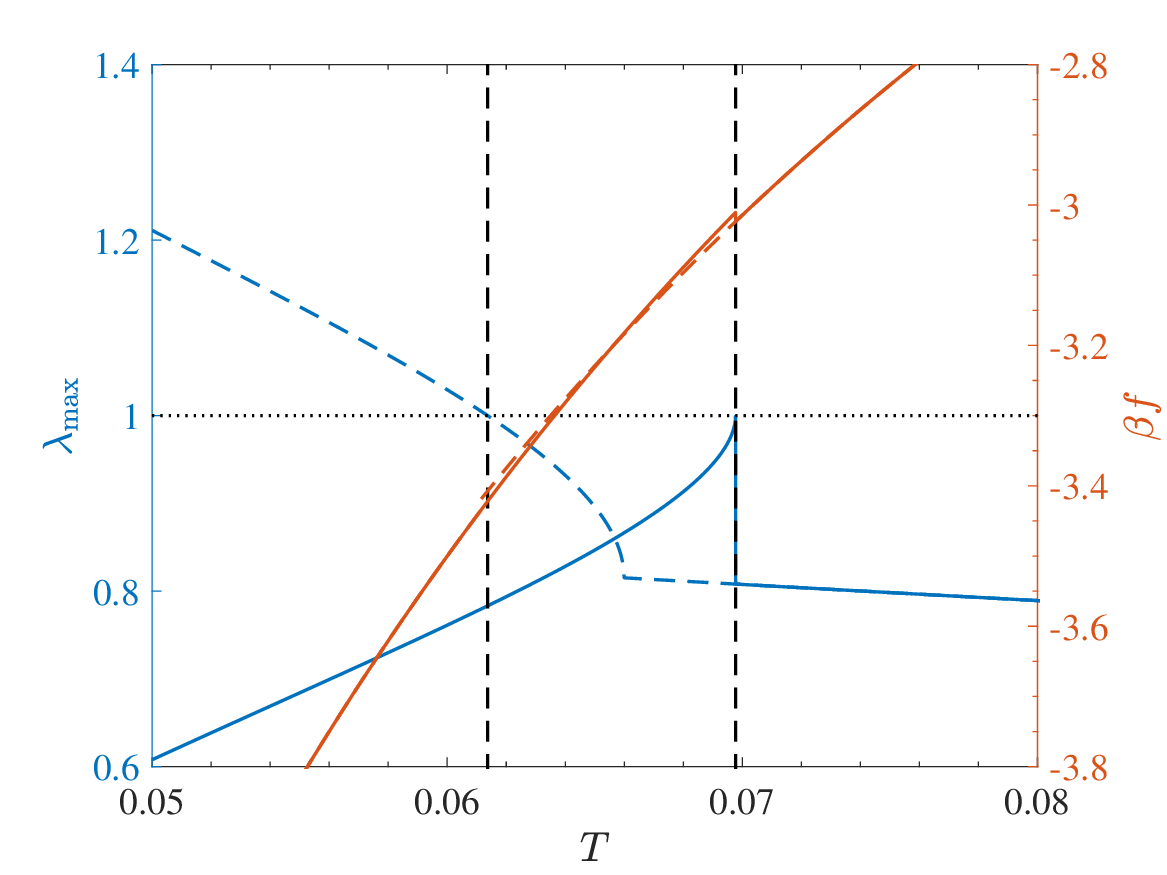}
\caption{Free energy (right axis, red) and leading eigenvalue (left axis, blue) of the isotropic SARL model for $\kappa=0.22$ on the $c+1=3$ Bethe lattice. The hysteresis of the heating (solid line) and cooling (dashed line) curves hints at the presence of a (weakly) first-order P-F phase transition---as suggested in Ref.~\cite{charbonneau2021solution}. The end of the metastability range of these curves, $T_{\mathrm{heat}}=0.06977$  and $T_{\mathrm{cool}}=0.06137$ (dashed vertical lines), coincides with  $\lambda_\mathrm{max} = 1$ (horizontal dotted line) for the complete expression in Eq.~\eqref{eq:lsa_matrix} (solid blue line) and its homogeneous reduction in Eq.~\eqref{eq:homogeneouslambdamax} (dashed blue line), respectively. These results bound the thermodynamic transition temperature, $T_c=0.06560$, where the free energy curves cross.} 
\label{fig:salr_eigs}
\end{figure}

The leading eigenvalue of the Jacobian matrix can be obtained numerically for any configuration, whether in the paramagnetic or the ferromagnetic phase. For the paramagnetic phase, leveraging symmetry further simplifies the analysis. In particular, Prop.~\ref{prop:sym1}(b) yields:
\begin{align}
& 
\frac{\partial f_E}{\partial E} = \frac{\partial f_R}{\partial R}, \,\,\, \quad
\frac{\partial f_E}{\partial O} = \frac{\partial f_R}{\partial F}, \,\,\, \quad
\frac{\partial f_F}{\partial E} = \frac{\partial f_O}{\partial R}, \,\,\,
\nonumber\\
&
\frac{\partial f_F}{\partial O} = \frac{\partial f_O}{\partial F}, \,\,\,\quad
\frac{\partial Z}{\partial E} = \frac{\partial Z}{\partial R}, \,\,\, \! \quad\quad
\frac{\partial Z}{\partial F} = \frac{\partial Z}{\partial O}.
\end{align}
The Jacobian matrix structure is therefore
\[\mathcal{J} = \left(\begin{array}{cccc}
a & b & c & d \\
e & f & g & h \\
h & g & f & e \\
d & c & b & a
\end{array}\right),\]
for which the four eigenvectors can be organized in two families of the form  $\mathbf{x}_1 = (u,v,-v,-u)^T$ and $\mathbf{x}_2 = (u,v,v,u)^T$, each associated with a pair of distinct eigenvalues and associated parameters $u$ and $v$.
As a result, for the homogeneous phase all four eigenvalues of $\mathcal{J}$  can be extracted without actually diagonalizing the $4 \times 4$ matrix. In particular, because the eigenvector associated with the leading eigenvalue is of the form $\mathbf{x}_1$, the eigen-equation $\mathcal{J}\mathbf{x}_1=\lambda \mathbf{x}_1$ gives
\begin{align}
\label{eq:lsa_config_eig}
   \dfrac{u}{v}=\dfrac{ \dfrac{u}{v}\dfrac{\partial f_E}{\partial E} - \dfrac{\partial f_E}{\partial O}}{\dfrac{u}{v}\dfrac{\partial f_F}{\partial E} - \dfrac{\partial f_F}{\partial O}}.
\end{align}
Treating the ratio $u/v$ as a single variable leaves a simple quadratic equation to solve. The resulting eigenvalue is (in the original notation) 
\begin{equation}
\label{eq:homogeneouslambdamax}
    \lambda = Z_{\mathrm{cav}}^{-1}\left(\frac{\partial f_\mathrm{cav}(\downarrow,\downarrow)}{\partial \eta_\mathrm{cav}(\downarrow,\downarrow)} - \frac{v}{u}\frac{\partial f_\mathrm{cav}(\downarrow,\downarrow)}{\partial \eta_\mathrm{cav}(\uparrow,\downarrow)}\right),
\end{equation}
where $\lambda_\mathrm{max}$ is determined by taking $u/v$ as the greater of the (positive) solutions of Eq.~\eqref{eq:lsa_config_eig}.

Although this strategy enforces homogeneity even when such homogeneity is pathologically unstable---such as at low temperatures---the onset of that instability helps identify phase transitions. For an Ising-like transition, for instance, the largest eigenvalue of the complete $4\times 4$ matrix in Eq.~\eqref{eq:lsa_matrix} as well as the eigenvalue of the simplified form in Eq.~\eqref{eq:homogeneouslambdamax} give numerically consistent estimates of the critical temperature, $T_c$. The symmetry-informed linear stability analysis is particularly effective at accelerating convergence around the Lifshitz point, whereat the numerical diagonalization of $\mathcal{J}$ is unstable.

Figure~\ref{fig:salr_eigs} illustrates yet another way in which the linear stability analysis can help detect phase transitions. In the vicinity of a (weakly) first-order P-F transition (see Sec.~\ref{sec:discussion} for details), the leading eigenvalue obtained from two different approaches to the transition behave differently. For the full $4\times4$ matrix (with configurational probabilities initialized in the ferromagnetic phase), the leading eigenvalue increases with temperature and reaches $\lambda_\mathrm{max} = 1$ at a temperature identified as $T_\mathrm{heat}$, where 
the metastable ferromagnetic solution is lost. 
By contrast, for the homogeneously reduced expression, the leading eigenvalue 
extends to low temperatures but crosses unity at $T_\mathrm{cool}$, denoting the limit of 
metastability of the paramagnetic phase. These two estimates concisely bracket the free energy discontinuity calculated from Eq.~\eqref{eq:salr_bf}. Because that discontinuity  presents but a faint numerical signal, analyzing leading eigenvalues provides useful and robust bounds to the actual transition temperature.

\subsubsection{Percolation of the $\alpha$-parameter cluster model}
\label{sec:iso_revCK}
A na\"ive extension  of the nearest-neighbor FK--CK cluster scheme to systems with next-nearest neighbor interactions entails rescaling the bond probability in Eq.~\eqref{eq:pb} as
\begin{equation}
\label{eq:alphaparameterdef}
    p_B^\mathrm{rev}(\alpha) = 1 - \exp{(-2\beta J\alpha)},
\end{equation}
with tuning parameter $\alpha$. So as to account in an effective sense for next-nearest neighbor interactions in this $\alpha$-parameter cluster model, we expect $\alpha\geq1$ for models with purely attractive interactions and $\alpha<1$ for those with SALR interactions. Prior simulations of $d=2$ lattices have shown that tuning $p_B^\mathrm{rev}$ such that $T_p=T_c$ for this $\alpha$-parameter cluster model improves the sampling efficiency of the associated Monte Carlo cluster algorithm for small to moderately-sized systems, but not so for larger ones~\cite{zheng2022weakening}. As discussed in Sec.~\ref{sec:theorem}, the resulting clusters do not properly capture structural correlations, thus explaining the breakdown. Here, an analysis of this cluster model on the Bethe lattice provides additional physical insights into the mismatch.

In order to determine $T_p$ for the $\alpha$-parameter cluster model, we follow the strategy of Ref.~\onlinecite{charbonneau2021solution} of designing recursive relations to  calculate the \textit{percolation probability} that a spin belongs or not to the percolating cluster $C_\infty$. For simplicity and without loss of generality, we consider only clusters of up spins. (Results for clusters of down spins are the same, by symmetry.) We then define the probabilities that spin $i$ points up and belongs or not to $C_\infty$, which we denote $\theta_i=1$ or 0, respectively, in terms of the cavity field
\begin{align}
\label{eq:ck_pr_def}
    \pi_R &\equiv \textrm{Pr} (s_i=\; \uparrow \wedge \; s_j= \; \uparrow \wedge \; \theta_i=1) , \\ 
    q_R &\equiv \textrm{Pr}(s_i=\; \uparrow \wedge \;  s_j= \; \uparrow \wedge \;  \theta_i=0), \end{align}
where $\pi_R + q_R = R$. Taking the bonding probability into account, we define the auxiliary quantities
\begin{equation}
\label{eq:revised_pr}
        \hat{\pi} = \pi_R \, p_{B}^\mathrm{rev}, \qquad \hat{q} = q_R + (1-p_{B}^\mathrm{rev})\pi_R,
\end{equation}
which are the probabilities that an up spin is or is not connected to the parallel and already percolating neighbor spin, respectively. Recursive expressions for these quantities can be obtained
\begin{widetext}
\begin{align}
\label{eq:ck_pr}
    \pi_R &= Z_R^{-1} \sum\limits_{l=1}^{c} \binom{c}{l}e^{\beta J(2l-c)} e^{-\beta \kappa J(2l-c)} F^{c-l} \sum\limits_{k=1}^{l}\binom{l}{k}\hat{\pi}^k \, \hat{q}^{l-k} e^{-\beta \kappa J\frac{l(l-1)+(c-l)(c-1-l)-2l(c-l)}{2}},\\
    q_R &= Z_R^{-1} \sum\limits_{l=0}^{c} \binom{c}{l}e^{\beta J(2l-c)} e^{-\beta \kappa J(2l-c)} F^{c-l} \hat{q}^l e^{-\beta \kappa J\frac{l(l-1)+(c-l)(c-1-l)-2l(c-l)}{2}},
\end{align}
where $Z_R$ is such that $\pi_R + q_R = R$. Given converged $\pi_R$ and $q_R$, the percolation probability $P$ and the probability of non-percolated up spin $Q$ are
\begin{align}
\label{eq:salr_nnn_p-bis}
    P &= Z_\mathrm{site}^{-1} \sum\limits_{l=1}^{c+1} \binom{c+1}{l}e^{\beta J(2l-c-1)} e^{-\beta \kappa J\frac{l(l-1)+(c+1-l)(c-l)-2l(c+1-l)}{2}} F^{c+1-l} \sum\limits_{k=1}^{l}\binom{l}{k}\hat{\pi}^k \, \hat{q}^{l-k},\nonumber\\
    Q &= Z_\mathrm{site}^{-1} \sum\limits_{l=0}^{c+1} \binom{c+1}{l}e^{\beta J(2l-c-1)} e^{-\beta \kappa J\frac{l(l-1)+(c+1-l)(c-l)-2l(c+1-l)}{2}} F^{c+1-l} \hat{q}^l,
\end{align}
\end{widetext}
where $Z_\mathrm{site}$ is such that $P + Q = \eta(\uparrow)$.

As validation of the above scheme, Appendix~\ref{sec:appendix_iso_revCK} solves for $\alpha$ as a function of connectivity $c+1$ in the regime linear in $\kappa$ around $\kappa = 0$. That analysis also reveals that the change of $p_B^\mathrm{rev}$ around $\kappa = 0$ is continuous and smooth for attractive and repulsive next-nearest-neighbor interactions. 

\subsubsection{Percolation of generalized FK--CK clusters}
\label{sec:iso_nnn}

For the generalized FK--CK cluster scheme, we define as in Eq.~\eqref{eq:NNN_CKFK_definition} the bond probabilities between nearest and next-nearest neighbors, $p_1=1-e^{-2\beta J}$ and $p_2=1-e^{-2\beta\kappa J}$, respectively. We also define variables that describe the probabilities of different local configurations and percolation status:
\begin{itemize}
    \item $\pi_i(s_j, \theta_j)$ corresponds to the current site $i$ having $s_i= \; \uparrow$ and being part of the percolated cluster, $\theta_i=1$, \emph{and} its backward site $j$ being in state $(s_j, \theta_j)$; 
    \item $q_i(s_j, \theta_j)$ corresponds to the current site $i$ having $s_i= \; \uparrow$ and not being part 
    of the percolated cluster, $\theta_i=0$, \emph{and} the backward site $j$ being in state $(s_j, \theta_j)$;
    \item $w_i(s_j, \theta_j)$ corresponds to the current site $i$ having $s_i= \; \downarrow$ and not being part of the percolated cluster, $\theta_i=0$, \emph{and} the cavity site $j$ being in state $(s_j, \theta_j)$.
\end{itemize}   
As in Sec.~\ref{sec:iso_revCK}, we only consider clusters consisting of up spins. In this case, the backward site can take one of three possible states, $(s_j, \theta_j) = (\uparrow,1)$, $(\uparrow,0)$, or $(\downarrow,0)$. As a result, the recursive relations for a set of nine variables can be written (with indices $i$ and $j$ neglected for generality) as Eqs.~\eqref{eq:SALRperc1}--\eqref{eq:SALRperc9} in Appendix~\ref{sec:appendix_iso_perc}. 

The approach to obtain these equations can be understood by first considering the $d=1$ chain (or, equivalently, the Bethe lattice with $c+1=2$). We can then more straightforwardly examine how to calculate the percolation probability involving both nearest and next-nearest neighbor bonds. The recursive equations for $d=1$ are obtained in Eqs.~\eqref{eq:perc_1d_1}--\eqref{eq:perc_1d_9} below by considering three consecutive spins \includegraphics[scale=0.4,trim={0cm 0.4cm 0cm 0cm},clip]{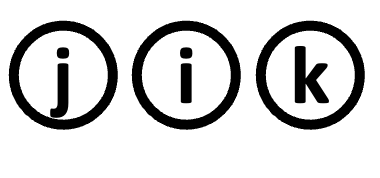}. The solid or empty circle denotes whether the spin is in the percolated cluster or not, respectively. 
The recursive relation is $\pi_i(s_j, \theta_j)$ (or $q_i(s_j, \theta_j)$ or $w_i(s_j, \theta_j)$) expressed as a function of $\pi_k(s_i, \theta_i)$ (or $q_k(s_i, \theta_i)$ or $w_k(s_i, \theta_i)$). Since the edge between $i$ and $k$ is recovered while that of $i$ and $j$ is removed through the cavity method, we here consider the possibility of a nearest-neighbor bond between $i$ and $k$ and a next-nearest-neighbor bond between $j$ and $k$. If a new bond is formed and therefore the spin joins the percolating cluster, then it is denoted as a shaded circle.

For the configuration ($\uparrow, \uparrow$) (denoted $E$ in Fig.~\ref{fig:bethe_configs}), one has $\eta_\mathrm{cav}(\uparrow,\uparrow) = \pi(\uparrow,1)+\pi(\uparrow,0)+q(\uparrow,1)+q(\uparrow,0)$ with
\begin{equation}
\label{eq:perc_1d_1}
\begin{tabular}{@{}p{6.5cm}@{\quad}p{2cm}@{}}
\begin{minipage}[t]{7cm}
$\pi_i(\uparrow, 1) Z_\mathrm{cav} = e^{\beta J(1 - \kappa)} \pi_k(\uparrow, 1) p_2$
\end{minipage} & 
\begin{minipage}[t]{2cm}
\centering
\includegraphics[width=1.5cm,trim={0cm 0.35cm 0cm 0cm},clip]{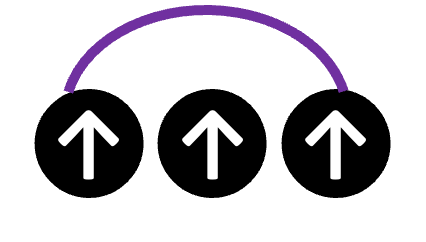}
\end{minipage} \\[0.em]

\begin{minipage}[t]{7cm}
$\phantom{\pi_i(\uparrow, 1) Z_\mathrm{cav}} + e^{\beta J(1 - \kappa)} \pi_k(\uparrow, 0) p_1 p_2$
\end{minipage} & 
\begin{minipage}[t]{2cm}
\centering
\includegraphics[width=1.5cm,trim={0cm 0.35cm 0cm 0cm},clip]{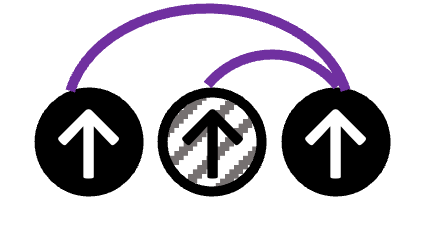}
\end{minipage} \\[0.em]

\begin{minipage}[t]{7cm}
$\phantom{\pi_i(\uparrow, 1) Z_\mathrm{cav}} + e^{\beta J(1 - \kappa)} q_k(\uparrow, 1) p_1 p_2$
\end{minipage} & 
\begin{minipage}[t]{2cm}
\centering
\includegraphics[width=1.5cm,trim={0cm 0.35cm 0cm 0cm},clip]{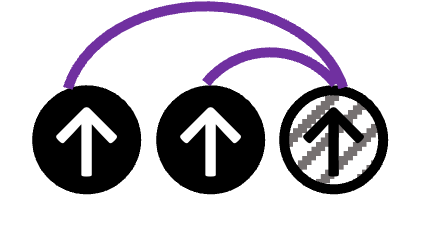}
\end{minipage}
\end{tabular}
\end{equation}

\begin{equation}
\label{eq:perc_1d_2}
\begin{tabular}{@{}p{6.5cm}@{\quad}p{2cm}@{}}
\begin{minipage}[t]{7cm}
$\pi_i(\uparrow, 0) Z_\mathrm{cav} = e^{\beta J(1 - \kappa)} \pi_k(\uparrow, 1) (1-p_2)$
\end{minipage} & 
\begin{minipage}[t]{2cm}
\centering
\includegraphics[width=1.5cm,trim={0cm 0.35cm 0cm 0cm},clip]{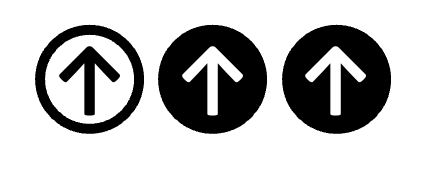}
\end{minipage} \\[0.em]

\begin{minipage}[t]{7cm}
$\phantom{\pi_i(\uparrow, 1) Z_\mathrm{cav}} + e^{\beta J(1 - \kappa)} \pi_k(\uparrow, 0) p_1 (1-p_2)$
\end{minipage} & 
\begin{minipage}[t]{2cm}
\centering
\includegraphics[width=1.5cm,trim={0cm 0.35cm 0cm 0cm},clip]{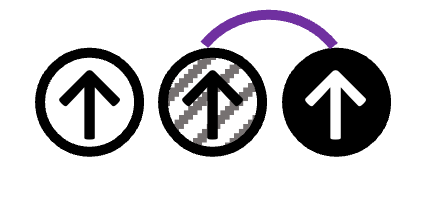}
\end{minipage} \\[0.em]

\begin{minipage}[t]{7cm}
$\phantom{\pi_i(\uparrow, 1) Z_\mathrm{cav}} + e^{\beta J(1 - \kappa)} q_k(\uparrow, 1) p_1(1-p_2)$
\end{minipage} & 
\begin{minipage}[t]{2cm}
\centering
\includegraphics[width=1.5cm,trim={0cm 0.35cm 0cm 0cm},clip]{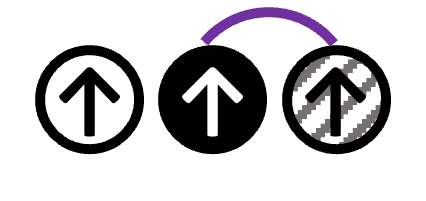}
\end{minipage} \\[0.em]

\begin{minipage}[t]{7cm}
$\phantom{\pi_i(\uparrow, 1) Z_\mathrm{cav}} + e^{\beta J(1 - \kappa)} q_k(\uparrow, 1) (1-p_1)$
\end{minipage} & 
\begin{minipage}[t]{2cm}
\centering
\includegraphics[width=1.5cm,trim={0cm 0.35cm 0cm 0cm},clip]{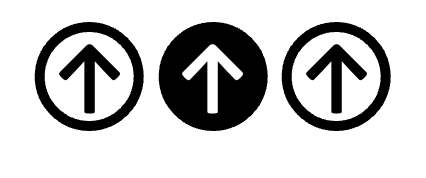}
\end{minipage} \\[0.em]

\begin{minipage}[t]{7cm}
$\phantom{\pi_i(\uparrow, 1) Z_\mathrm{cav}} + e^{-\beta J(1 - \kappa)} w_k(\uparrow, 1)$
\end{minipage} & 
\begin{minipage}[t]{2cm}
\centering
\includegraphics[width=1.5cm,trim={0cm 0.35cm 0cm 0cm},clip]{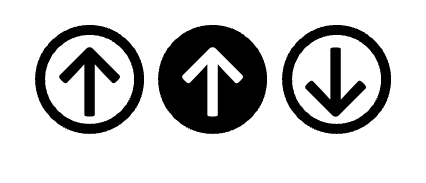}
\end{minipage}
\end{tabular}
\end{equation}

\begin{equation}
\label{eq:perc_1d_3}
\begin{tabular}{@{}p{6.5cm}@{\quad}p{2cm}@{}}
\begin{minipage}[t]{7cm}
$q_i(\uparrow, 1) Z_\mathrm{cav} = e^{\beta J(1 - \kappa)} \pi_k(\uparrow, 0) (1-p_1)p_2$
\end{minipage} & 
\begin{minipage}[t]{2cm}
\centering
\includegraphics[width=1.5cm,trim={0cm 0.35cm 0cm 0cm},clip]{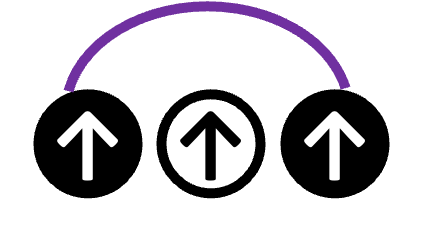}
\end{minipage}
\end{tabular}
\end{equation}

\begin{equation}
\label{eq:perc_1d_4}
\begin{tabular}{@{}p{6.5cm}@{\quad}p{2cm}@{}}
\begin{minipage}[t]{7cm}
$q_i(\uparrow, 0) Z_\mathrm{cav} = e^{\beta J(1 - \kappa)} \pi_k(\uparrow, 0)(1-p_1)(1-p_2)$
\end{minipage} & 
\begin{minipage}[t]{2cm}
\centering
\includegraphics[width=1.5cm,trim={0cm 0.35cm 0cm 0cm},clip]{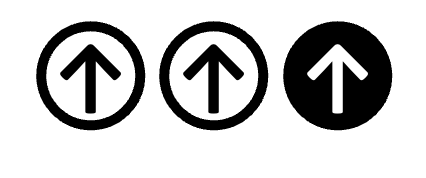}
\end{minipage} \\[0.em]

\begin{minipage}[t]{7cm}
$\phantom{\pi_i(\uparrow, 1) Z_\mathrm{cav}} + e^{\beta J(1 - \kappa)} q_k(\uparrow, 0)$
\end{minipage} & 
\begin{minipage}[t]{2cm}
\centering
\includegraphics[width=1.5cm,trim={0cm 0.35cm 0cm 0cm},clip]{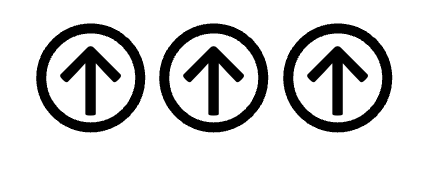}
\end{minipage} \\[0.em]

\begin{minipage}[t]{7cm}
$\phantom{\pi_i(\uparrow, 1) Z_\mathrm{cav}} + e^{-\beta J(1 - \kappa)} w_k(\uparrow, 0)$
\end{minipage} & 
\begin{minipage}[t]{2cm}
\centering
\includegraphics[width=1.5cm,trim={0cm 0.35cm 0cm 0cm},clip]{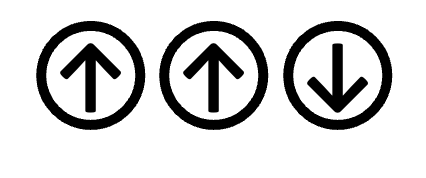}
\end{minipage}
\end{tabular}
\end{equation}
For the configuration ($\uparrow, \downarrow$) (denoted $O$ in Fig.~\ref{fig:bethe_configs}), one has 
$\eta_\mathrm{cav}(\uparrow,\downarrow) = \pi(\downarrow,0)+q(\downarrow,0)$ with
\begin{equation}
\label{eq:perc_1d_5}
\begin{tabular}{@{}p{6.5cm}@{\quad}p{2cm}@{}}
\begin{minipage}[t]{7cm}
$\pi_i(\downarrow, 0) Z_\mathrm{cav} = e^{\beta J(1 + \kappa)} \pi_k(\uparrow, 1)$
\end{minipage} & 
\begin{minipage}[t]{2cm}
\centering
\includegraphics[width=1.5cm,trim={0cm 0.35cm 0cm 0cm},clip]{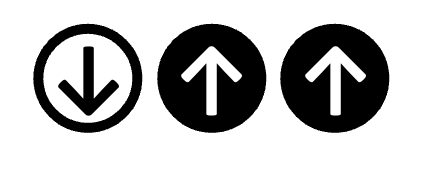}
\end{minipage} \\[0.em]

\begin{minipage}[t]{7cm}
$\phantom{\pi_i(\uparrow, 1) Z_\mathrm{cav}} + e^{\beta J(1 + \kappa)} \pi_k(\uparrow, 0) p_1$
\end{minipage} & 
\begin{minipage}[t]{2cm}
\centering
\includegraphics[width=1.5cm,trim={0cm 0.35cm 0cm 0cm},clip]{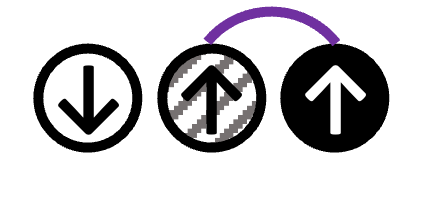}
\end{minipage} \\[0.em]

\begin{minipage}[t]{7cm}
$\phantom{\pi_i(\uparrow, 1) Z_\mathrm{cav}} + e^{\beta J(1 + \kappa)} q_k(\uparrow, 1)$
\end{minipage} & 
\begin{minipage}[t]{2cm}
\centering
\includegraphics[width=1.5cm,trim={0cm 0.35cm 0cm 0cm},clip]{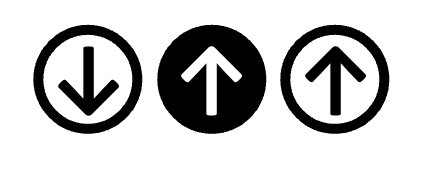}
\end{minipage} \\[0.em]

\begin{minipage}[t]{7cm}
$\phantom{\pi_i(\uparrow, 1) Z_\mathrm{cav}} + e^{-\beta J(1 + \kappa)} w_k(\uparrow, 1)$
\end{minipage} & 
\begin{minipage}[t]{2cm}
\centering
\includegraphics[width=1.5cm,trim={0cm 0.35cm 0cm 0cm},clip]{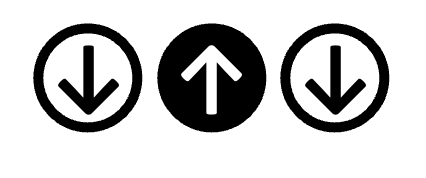}
\end{minipage}
\end{tabular}
\end{equation}

\begin{equation}
\label{eq:perc_1d_6}
\begin{tabular}{@{}p{6.5cm}@{\quad}p{2cm}@{}}
\begin{minipage}[t]{7cm}
$q_i(\downarrow, 0) Z_\mathrm{cav} = e^{\beta J(1 + \kappa)} \pi_k(\uparrow, 0) (1-p_1)$
\end{minipage} & 
\begin{minipage}[t]{2cm}
\centering
\includegraphics[width=1.5cm,trim={0cm 0.35cm 0cm 0cm},clip]{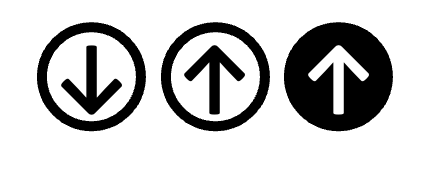}
\end{minipage} \\[0.em]

\begin{minipage}[t]{7cm}
$\phantom{\pi_i(\uparrow, 1) Z_\mathrm{cav}} + e^{\beta J(1 + \kappa)} q_k(\uparrow, 0)$
\end{minipage} & 
\begin{minipage}[t]{2cm}
\centering
\includegraphics[width=1.5cm,trim={0cm 0.35cm 0cm 0cm},clip]{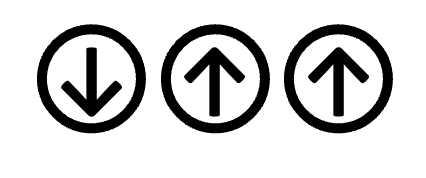}
\end{minipage} \\[0.em]

\begin{minipage}[t]{7cm}
$\phantom{\pi_i(\uparrow, 1) Z_\mathrm{cav}} + e^{-\beta J(1 + \kappa)} w_k(\uparrow, 0)$
\end{minipage} & 
\begin{minipage}[t]{2cm}
\centering
\includegraphics[width=1.5cm,trim={0cm 0.35cm 0cm 0cm},clip]{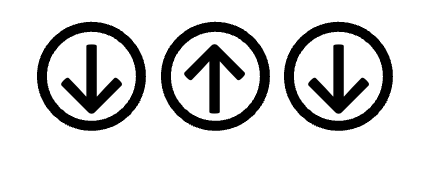}
\end{minipage}
\end{tabular}
\end{equation}
For the configuration ($\downarrow, \uparrow$) (denoted $F$ in Fig.~\ref{fig:bethe_configs}), one has 
$\eta_\mathrm{cav}(\downarrow,\uparrow) = w(\uparrow,1)+w(\uparrow,0)$ with
\begin{equation}
\label{eq:perc_1d_7}
\begin{tabular}{@{}p{6.5cm}@{\quad}p{2cm}@{}}
\begin{minipage}[t]{7cm}
$w_i(\uparrow, 1) Z_\mathrm{cav} = e^{-\beta J(1 + \kappa)} \pi_k(\downarrow, 0) p_2$
\end{minipage} & 
\begin{minipage}[t]{2cm}
\centering
\includegraphics[width=1.5cm,trim={0cm 0.35cm 0cm 0cm},clip]{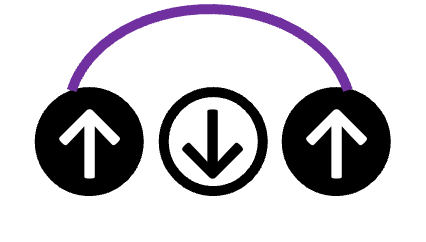}
\end{minipage}
\end{tabular}
\end{equation}

\begin{equation}
\label{eq:perc_1d_8}
\begin{tabular}{@{}p{6.5cm}@{\quad}p{2cm}@{}}
\begin{minipage}[t]{7cm}
$w_i(\uparrow, 0) Z_\mathrm{cav} = e^{-\beta J(1+\kappa)} \pi_k(\downarrow, 0) (1-p_2)$
\end{minipage} & 
\begin{minipage}[t]{2cm}
\centering
\includegraphics[width=1.5cm,trim={0cm 0.35cm 0cm 0cm},clip]{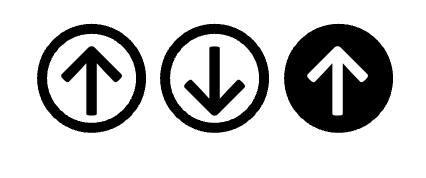}
\end{minipage} \\[0.em]

\begin{minipage}[t]{7cm}
$\phantom{\pi_i(\uparrow, 1) Z_\mathrm{cav}} + e^{-\beta J(1 + \kappa)} q_k(\downarrow, 0)$
\end{minipage} & 
\begin{minipage}[t]{2cm}
\centering
\includegraphics[width=1.5cm,trim={0cm 0.35cm 0cm 0cm},clip]{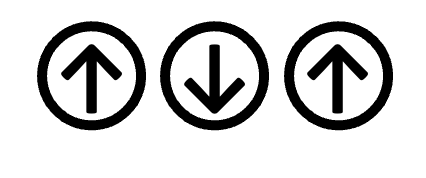}
\end{minipage} \\[0.em]

\begin{minipage}[t]{7cm}
$\phantom{\pi_i(\uparrow, 1) Z_\mathrm{cav}} + e^{\beta J(1 + \kappa)} w_k(\downarrow, 0)$
\end{minipage} & 
\begin{minipage}[t]{2cm}
\centering
\includegraphics[width=1.5cm,trim={0cm 0.35cm 0cm 0cm},clip]{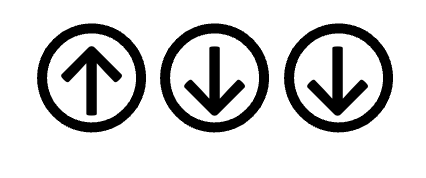}
\end{minipage}
\end{tabular}
\end{equation}
Finally, for the configuration ($\downarrow, \downarrow$) (denoted $E$ in Fig.~\ref{fig:bethe_configs}), one has 
$\eta_\mathrm{cav}(\downarrow,\downarrow) = w(\downarrow,0)$ with
\begin{equation}
\label{eq:perc_1d_9}
\begin{tabular}{@{}p{6.5cm}@{\quad}p{2cm}@{}}
\begin{minipage}[t]{7cm}
$w_i(\downarrow, 0) Z_\mathrm{cav} = e^{-\beta J(1-\kappa)} \pi_k(\downarrow, 0)$
\end{minipage} & 
\begin{minipage}[t]{2cm}
\centering
\includegraphics[width=1.5cm,trim={0cm 0.35cm 0cm 0cm},clip]{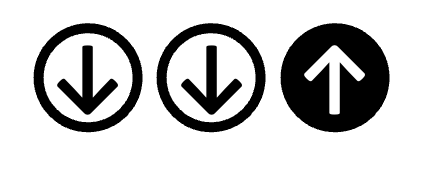}
\end{minipage} \\[0.em]

\begin{minipage}[t]{7cm}
$\phantom{\pi_i(\uparrow, 1) Z_\mathrm{cav}} + e^{-\beta J(1 - \kappa)} q_k(\downarrow, 0)$
\end{minipage} & 
\begin{minipage}[t]{2cm}
\centering
\includegraphics[width=1.5cm,trim={0cm 0.35cm 0cm 0cm},clip]{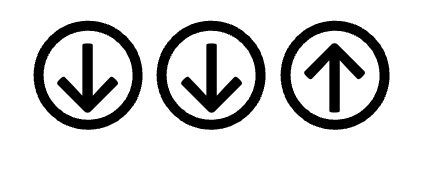}
\end{minipage} \\[0.em]

\begin{minipage}[t]{7cm}
$\phantom{\pi_i(\uparrow, 1) Z_\mathrm{cav}} + e^{\beta J(1 - \kappa)} w_k(\downarrow, 0)$
\end{minipage} & 
\begin{minipage}[t]{2cm}
\centering
\includegraphics[width=1.5cm,trim={0cm 0.35cm 0cm 0cm},clip]{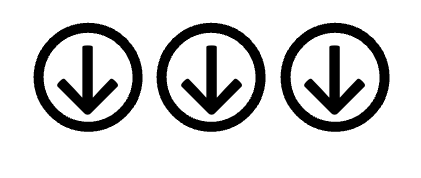}
\end{minipage}
\end{tabular}
\end{equation}

In order to calculate the (non)percolation probability of an up spin, we consider the additional case of the node $i$ being the cavity site of both nodes $j$ and $k$, and we recover the complete $d=1$ chain by adding the two prior edges. For convenience, we calculate the nonpercolation probability $Q$---recalling that the percolation probability $P$ can be obtained from $P+Q = \eta(\uparrow)$:
\begin{equation}
\label{eq:perc_1d_10}
\begin{tabular}{@{}p{6.5cm}@{\quad}p{2cm}@{}}
\begin{minipage}[t]{7cm}
$Q Z_\mathrm{perc} = e^{\beta J(-2-\kappa)} w(\uparrow, 0)^2$
\end{minipage} & 
\begin{minipage}[t]{2cm}
\centering
\includegraphics[width=1.5cm,trim={0cm 0.35cm 0cm 0cm},clip]{Figures_Mingyuan/percolation_configs_1d_2/q_down_0_3.png}
\end{minipage} \\[0.em]

\begin{minipage}[t]{7cm}
$\phantom{Q Z_\mathrm{perc}} + 2e^{\beta \kappa J} w(\uparrow, 0)\pi(\uparrow, 0)(1-p_1)$
\end{minipage} & 
\begin{minipage}[t]{2cm}
\centering
\includegraphics[width=1.5cm,trim={0cm 0.35cm 0cm 0cm},clip]{Figures_Mingyuan/percolation_configs_1d_2/q_down_0_1.png}
\end{minipage} \\[0.em]

\begin{minipage}[t]{7cm}
$\phantom{Q Z_\mathrm{perc}} + 2e^{\beta \kappa J} w(\uparrow, 0)q(\uparrow, 0)$
\end{minipage} & 
\begin{minipage}[t]{2cm}
\centering
\includegraphics[width=1.5cm,trim={0cm 0.35cm 0cm 0cm},clip]{Figures_Mingyuan/percolation_configs_1d_2/q_down_0_2.png}
\end{minipage} \\[0.em]

\begin{minipage}[t]{7cm}
$\phantom{Q Z_\mathrm{perc}} + e^{\beta J(2 - \kappa)} \pi(\uparrow, 0)^2(1-p_1)^2$
\end{minipage} & 
\begin{minipage}[t]{2cm}
\centering
\includegraphics[width=1.5cm,trim={0cm 0.35cm 0cm 0cm},clip]{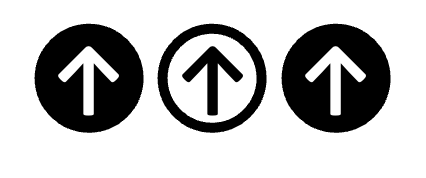}
\end{minipage} \\[0.em]

\begin{minipage}[t]{7cm}
$\phantom{Q Z_\mathrm{perc}} + e^{\beta J(2 - \kappa)} q(\uparrow, 0)^2$
\end{minipage} & 
\begin{minipage}[t]{2cm}
\centering
\includegraphics[width=1.5cm,trim={0cm 0.35cm 0cm 0cm},clip]{Figures_Mingyuan/percolation_configs_1d_2/q_up_0_2.png}
\end{minipage} \\[0.em]

\begin{minipage}[t]{7cm}
$\phantom{Q Z_\mathrm{perc}} + 2e^{\beta J(2 - \kappa)} \pi(\uparrow, 0)q(\uparrow, 0)\phi$
\end{minipage} & 
\begin{minipage}[t]{2cm}
\centering
\includegraphics[width=1.5cm,trim={0cm 0.35cm 0cm 0cm},clip]{Figures_Mingyuan/percolation_configs_1d_2/q_up_0_1.png}
\end{minipage}
\end{tabular}
\end{equation}
where $\phi = (1-p_1)^2p_2 + (1-p_1)(1-p_2)$, corresponding to the cases that the percolated site $k$ is connected to site $j$ by a NNN bond or not, respectively. Although this percolation probability is trivial in $d=1$ because percolation is not possible for $T>0$, the approach can be directly adopted for the ANNNI model (which has a SALR interaction chain along one direction) and extended to the isotropic SALR interaction model.

In order to validate the above solution of the generalized FK--CK cluster scheme, we here check that the spin-spin correlation $\langle s_i s_j\rangle$ and the probability of two parallel spins being part of the same cluster $\langle\gamma_{ij}^\parallel\rangle$ are indeed equivalent for the $d=1$ chain (see Eq.~\eqref{eq:sisj_non} and Sec.~\ref{sec:theorem_salr}). The former is retrieved using the transfer matrix scheme described in Ref.~\cite{zheng2022weakening}, while the latter is obtained from the recursive Eqs.~\eqref{eq:perc_1d_1}--\eqref{eq:perc_1d_9} for $c+1=2$ (see Appendix~\ref{sec:appendix_tm}). As can be seen in Fig.~\ref{fig:corr_1d}, for both positive and negative $\kappa$ we have $\langle s_i s_j\rangle = \langle\gamma_{ij}^\parallel\rangle$. In particular, the two quantities are characterized by a same correlation length, $\xi$, extracted from fitting an exponential form to the numerical results. In other words, the thermodynamic and geometric properties for the generalized FK--CK cluster scheme match perfectly. 

\begin{figure}[h!]
\centering
\includegraphics[scale=0.48,trim={0.2cm 0 0.5cm 0.2cm},clip]{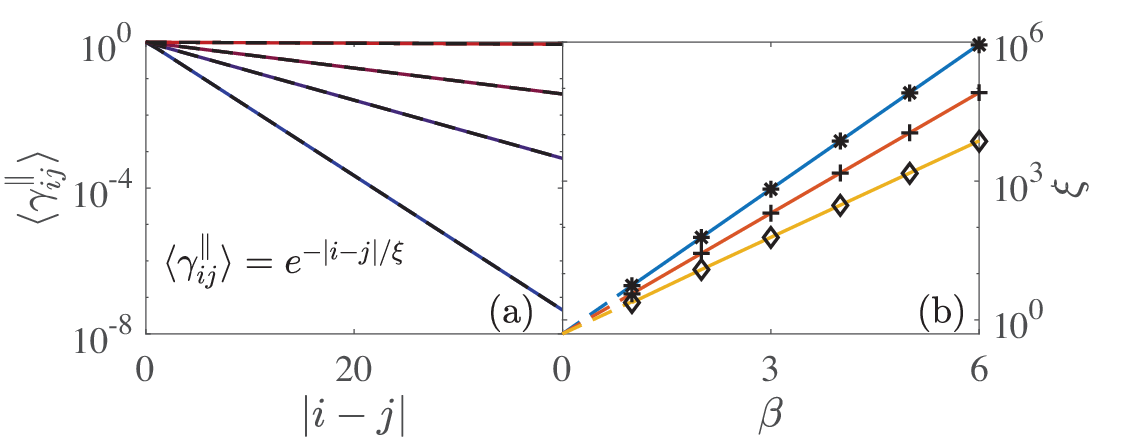}
\caption{(a) Probability that spins $i$ and $j$ are parallel and part of a same cluster, $\langle\gamma_{ij}^\parallel\rangle$, for the case $d=1$, as determined from the generalized FK--CK cluster scheme (see Eqs.~\eqref{eq:perc_1d_1}--\eqref{eq:perc_1d_9} and Appendix~\ref{sec:appendix_tm}) for $\kappa = 0.1$ and $\beta = 1, 1.5, 2$, and $4$ (colored lines, from bottom to top). The decay of the numerical results is well described by an exponential from, $\langle\gamma^\parallel_{ij}\rangle=e^{-\lvert i-j\rvert/\xi}$ (dashed black lines) with fitted correlation length $\xi$. (b) The correlation length results from $\langle\gamma_{ij}^\parallel\rangle$ (markers) match those from $\langle s_i s_j\rangle$ (lines) for $\kappa = 0.1, 0$ and $-0.1$ (from bottom to top) in Ref.~\cite{zheng2022weakening}. The lengths correspond at all temperatures and are consistent with the scaling $\xi=e^{2\beta J(1-2\kappa)}/2$, here extrapolated to $\beta\rightarrow0$ (dashed lines).} 
\label{fig:corr_1d}
\end{figure}

For the Bethe lattice more generally the same nine variables describe the percolation probability of the cavity field, with the caveat that now more than one site $k$ is connected to the current site $i$ due to the tree-like structure of the lattice. As a result, many more cases of NNN bond(s) have to be considered in the recursive relations. In this case, we define auxiliary functions to describe the probabilities of different cases forming NNN bond(s), assuming the current site $i$ has $l$ nearest-neighbor sites with up spin (excluding the cavity site).

(i) Let $\Phi(k,m)$ be the probability that $m$ NN sites ($\uparrow,0$) of site $i$ are connected to $k$ NN sites ($\uparrow,1$) of site $i$ through one or more NNN bonds. Notice that the $m$ sites ($\uparrow,0$) may be connected to each other. Then, any of these sites connected to the sites ($\uparrow,1$) would result in the percolation of all these sites. A recursive formula of the conditional probability is thereby obtained,
\begin{widetext}
\begin{equation}
    \Phi(k,m) = \left\{
    \begin{array}{l}
    1, \qquad \mbox{for $m = 0$,} \\
    [5pt]
    0, \qquad \mbox{for $m > 0$ and $k = 0$,} \\
    [5pt]
    \sum\limits_{i=0}^{m-1}\binom{m}{i}[1-(1-p_2)^k]^{m-i}(1-p_2)^{ki}\Phi(m-i,i), \qquad \mbox{for $m > 0$ and $k > 0$.}
\end{array}
\right .
\end{equation}
The probability that only $m$ out of $l-k$ nearest-neighbor sites ($\uparrow,0$) of the site $i$ become percolated through one or more NNN bonds to the $k$ nearest-neighbor sites ($\uparrow,1$) of site $i$ is therefore
\begin{equation}
    \binom{l-k}{m}\Phi(k,m)(1-p_2)^{(k+m)(l-k-m)}.
\end{equation}

(ii) Let $\Psi(k,m)$ be the probability that $m$ nearest-neighbor sites ($\uparrow,0$) of site $i$ become percolated through either one or more NNN bonds to $k$ nearest-neighbor sites ($\uparrow,1$) of site $i$, or one direct NN bond to the current site $i$ which is  already percolated through other branches. The connections within the sites ($\uparrow,0$) are also considered,
\begin{equation}
    \Psi(k,m) = \left\{
    \begin{array}{l}
    1, \qquad \mbox{for $m = 0$,} \\
    [5pt]
    \sum\limits_{i=0}^{m-1}\binom{m}{i}[1-(1-p_1)(1-p_2)^k]^{m-i}[(1-p_1)(1-p_2)^k]^i \Phi(m-i,i), \qquad \mbox{for $m > 0$.}
\end{array}
\right .
\end{equation}
\end{widetext}
The probability that only $m$ out of $l-k$ nearest-neighbor sites ($\uparrow,0$) of site $i$ become percolated through the aforementioned two types of bonds is therefore
\begin{equation}
    \binom{l-k}{m}\Psi(k,m)[(1-p_1)(1-p_2)^{k+m}]^{l-k-m}.
\end{equation}

(iii) Let $\Theta(k,m,n)$ be the probability that the current site $i$ with ($\uparrow,0$) is connected to its percolated nearest-neighbor sites directly (through one NN bond) or indirectly (through one NN bond plus one or more NNN bonds), and $m+n$ nearest-neighbor sites ($\uparrow,0$) of site $i$ become percolated either through NNN bonds or NN bonds or both. In order to take all possible cases into account, we classify the $m+n$ nearest-neighbor sites ($\uparrow,0$) of site $i$ into two types---$m$ sites are connected to $k$ nearest-neighbor sites ($\uparrow,1$) of site $i$ through one or more NNN bonds, and $n$ sites do \emph{not} have such bonds but they are (directly or indirectly) connected to the current site $i$. It then gives

\begin{widetext}
\begin{equation}
    \Theta(k,m,n) = \left\{
    \begin{array}{l}
    0, \qquad \mbox{for $k = 0$,} \\
    \Phi(k,m)[1-(1-p_1)^{k+m}](1-p_2)^{(k+m)n}\cdot \sum\limits_{i=0}^{n-1}\binom{n}{i}\Phi(n-i,i)p_1^{n-i} (1-p_1)^i, \qquad \mbox{for $k > 0$.}
\end{array}
\right .
\end{equation}
The probability that only $m+n$ out of $l-k$ nearest-neighbor sites ($\uparrow,0$) of site $i$ become percolated in this case is
\begin{equation}
    \binom{l-k}{m}\binom{l-k-m}{n}\Theta(k,m,n)[(1-p_1)(1-p_2)^{k+m+n}]^{l-k-m-n}.
\end{equation}
The complete recursive expressions for the nine variables are given in Appendix~\ref{sec:appendix_iso_perc}.

Overall (non)percolation probabilities also ensure $P + Q = \eta(\uparrow)$,
\begin{equation}
\label{eq:iso_perc_Q}
\begin{split}
    P &= Z_\mathrm{site}^{-1}\left\{\sum_{l=0}^{c+1}\sum_{k=0}^{l}\sum_{a+b+d>0}\binom{c+1}{l}\binom{l}{k}e^{\beta J (2l-c-1)} x' \cdot w(\uparrow,0)^{c+1-l-a} w(\uparrow,1)^a q(\uparrow,0)^{l-k-b} q(\uparrow,1)^b \pi(\uparrow,0)^{k-d} \pi(\uparrow,1)^d\right.\\
    &\left.+\sum\limits_{l=0}^{c+1}\binom{c+1}{l}e^{\beta J (2l-c-1)} x' \cdot w(\uparrow,0)^{c+1-l} \sum\limits_{k=0}^{l}\binom{l}{k}q(\uparrow,0)^{l-k}\pi(\uparrow,0)^{k}
    \sum\limits_{m=0}^{l-k} \binom{l-k}{m}\Phi(k,m)[1-(1-p_1)^{k+m}]\right\}\\
    Q &= Z_\mathrm{site}^{-1} \sum\limits_{l=0}^{c+1}\binom{c+1}{l}e^{\beta J (2l-c-1)} x' \cdot w(\uparrow,0)^{c+1-l} \sum\limits_{k=0}^{l}\binom{l}{k}q(\uparrow,0)^{l-k}\pi(\uparrow,0)^{k} \sum\limits_{m=0}^{l-k} \binom{l-k}{m}\Phi(k,m)(1-p_1)^{k+m}.
\end{split}
\end{equation}
where we defined $x' = e^{-\beta\kappa\frac{l(l-1)+(c+1-l)(c-3l)}{2}}$ for conciseness.
\end{widetext}

Note, however, that the above recursive equations miss one percolation case. In $\pi_i(s_j, \theta_j)$ (or $q_i(s_j, \theta_j)$ or $w_i(s_j, \theta_j)$), the percolation states of the backward site $j$ only depend on its possible NNN bonds to other nearest-neighbor sites $k$ of site $i$. (Because the edge between $i$ and $j$ is removed in the cavity field method, the NN bond between site $i$ and $j$ is not considered.) We should therefore consider cases for which sites $k$ become percolated in the recursive equations: 1) site $k$ has the status $\pi_{k}(s_i, \theta_i)$; 2) unpercolated site $k$ forms a NN bond to a percolated site $i$; 3) unpercolated site $k$ forms NNN bond(s) to other percolated nearest-neighbor sites of site $i$; 4) site $k$ has the status $q_{k}(\uparrow, 0)$, while site $i$ becomes percolated through other branch(es) and has NNN bonds to the nearest-neighbor site(s) of site $k$ in the previous shell, and that site(s) form NN bond(s) to site $k$. Because an additional level of the tree-like graph then needs to be considered, this last case leads to a marked increase in algebraic complexity. (It could be computed using the strategies of models with up to third-nearest-neighbor interactions in Ref.~\cite{charbonneau2021solution}.) Because we estimate this contribution to the percolation probability  in the regime considered to be at most $10^{-4}$, which is of the same order as the numerical tolerance used to identify the point at which the percolation probability $P$ vanishes---and hence the percolation temperature $T_p$---we here safely neglect it. 

Note that the percolation line could also be obtained by linear stability analysis, but the approach leads to fairly complicated algebraic expressions for all nine variables, and is therefore not particularly helpful in practice.

\subsection{ANNNI Model}
\begin{figure*}[t]
\centering
\includegraphics[scale=0.5]{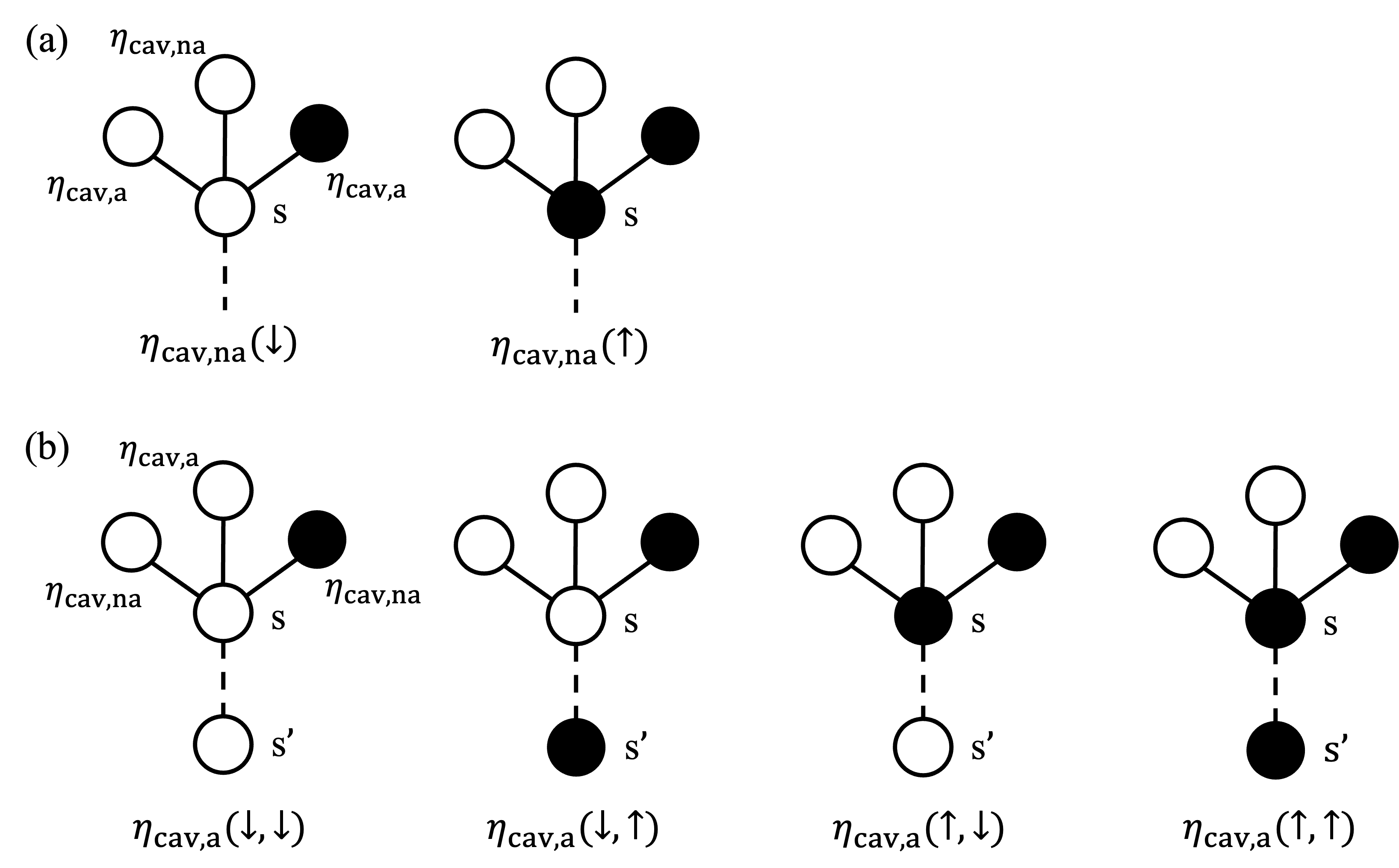}
\caption{Configurations with up (black) and down (white) spins corresponding to the cavity fields of ANNNI models defined in (a) non-axial and (b) axial directions on a Bethe lattice with $c+1=4$.} 
\label{fig:annni_configs}
\end{figure*}
The analysis of the ANNNI model given by Eq.~\eqref{eq:HANNNI} shares various similarities with the isotropic SALR interaction case.
The linear expansion and the linear stability analysis, however, are then much more involved. We therefore only consider the recursive equations for  thermodynamic properties as well as for the percolation properties of FK--CK clusters. 

\subsubsection{Cavity fields and recursion relations}

For the ANNNI model, a special consideration must be made for the axial direction (see Fig.~\ref{fig:annni_configs}). We denote $\eta_\mathrm{cav, a}(s, s')$ the configuration probabilities in the axial direction, where $s$ stands for the current site, $s'$ for the backward (or cavity) site, and $\eta_\mathrm{cav,na}(s)$ denotes sites in non-axial directions. We then have
\begin{widetext}
\begin{equation}
\label{eq:spin_configs_a}
\begin{split}
    \eta_\mathrm{cav, a}(s, s') &= {Z_\mathrm{cav,a}}^{-1} \sum\limits_{l=0}^{c-1} \binom{c-1}{l}[\eta_\mathrm{cav,na}(\uparrow)]^l [\eta_\mathrm{cav,na}(\downarrow)]^{c-1-l} e^{\beta J s(2l-c+1)}\\
    &\qquad\qquad\quad \times (\eta_\mathrm{cav,a}(\uparrow, s) e^{\beta J(s-\kappa s')} + \eta_\mathrm{cav,a}(\downarrow, s) e^{-\beta J(s-\kappa s')}),
\end{split}
\end{equation}
where $Z_\mathrm{cav,a}$ is the normalization factor, such that $\sum\limits_{s,s'}\eta_\mathrm{cav,a}(s,s')=1$, and 
\begin{eqnarray}
    \eta_\mathrm{cav,na}(s) &=& {Z_\mathrm{cav,na}}^{-1} \sum\limits_{l=0}^{c-2} \binom{c-2}{l}[\eta_\mathrm{cav,na}(\uparrow)]^l [\eta_\mathrm{cav,na}(\downarrow)]^{c-1-l} e^{\beta J s(2l-c+2)} 
    \nonumber\\ 
    &&
    \qquad\quad \times \lbrack (\eta_\mathrm{cav,a}(\uparrow, s) e^{\beta Js})^2 e^{\beta\kappa J} + (\eta_\mathrm{cav,a}(\downarrow, s) e^{-\beta Js})^2 e^{\beta\kappa J} + 2 \eta_\mathrm{cav,a}(\uparrow, s)\eta_\mathrm{cav,a}(\downarrow, s) e^{-\beta\kappa J}\rbrack,
    \label{eq:spin_configs_na}
\end{eqnarray}
where $Z_\mathrm{cav,na}$ is the normalization factor such that $\sum\limits_{s}\eta_\mathrm{cav,na}(s)=1$. The density is then
\begin{eqnarray}
\label{eq:annni_density}
    \eta(s) &=& Z_\mathrm{site}^{-1} \sum\limits_{l=0}^{c-1} \binom{c-1}{l}[\eta_\mathrm{cav,na}(\uparrow)]^l [\eta_\mathrm{cav,na}(\downarrow)]^{c-1-l} e^{\beta J s(2l-c+1)} 
    \nonumber\\
    &&
    \qquad\quad\times \lbrack (\eta_\mathrm{cav,a}(\uparrow, s) e^{\beta Js})^2 e^{\beta\kappa J} + (\eta_\mathrm{cav,a}(\downarrow, s) e^{-\beta Js})^2 e^{\beta\kappa J} + 2 \eta_\mathrm{cav,a}(\uparrow, s)\eta_\mathrm{cav,a}(\downarrow, s) e^{-\beta\kappa J}\rbrack,
\end{eqnarray}
where $Z_\mathrm{site}$ is such that $\eta(\uparrow)+\eta(\downarrow)=1$, and the free energy per site is
\begin{equation}
    \beta f = \beta f_{\mathrm{site}} - \beta f_\mathrm{link,a} - \frac{c-1}{2} \beta f_\mathrm{link, na},
\end{equation}
where $\beta f_{\mathrm{site}} = -\ln Z_{\mathrm{site}}$ and 
\begin{eqnarray}
    \beta f_\mathrm{link,a} &=& -\ln{\lbrack (\eta_\mathrm{cav,a}(\uparrow, \uparrow))^2 e^{\beta J} + (\eta_\mathrm{cav,a}(\downarrow, \downarrow))^2 e^{\beta J} + 2 \eta_\mathrm{cav,a}(\uparrow, \downarrow)\eta_\mathrm{cav,a}(\downarrow, \uparrow) e^{-\beta J}\rbrack},
\\
[5pt]
    \beta f_\mathrm{link, na} &=& -\ln{\lbrack (\eta_\mathrm{cav, na}(\uparrow))^2 e^{\beta J} + (\eta_\mathrm{cav, na}(\downarrow))^2 e^{\beta J} + 2 \eta_\mathrm{cav, na}(\uparrow)\eta_\mathrm{cav, na}(\downarrow) e^{-\beta J}\rbrack}.
\end{eqnarray}
\end{widetext}

\subsubsection{Percolation of generalized FK--CK clusters}
\label{sec:annni_nnn}
In order to calculate the percolation probability through the cavity field method, we consider the two bond probabilities for nearest neighbor and next-nearest neighbor $p_1$ and $p_2$ as in Sec.~\ref{sec:iso_nnn}. 
In the axial direction, the probability distributions $\pi_k(s_j, \theta_j)$, $q_k(s_j, \theta_j)$ and $w_k(s_j, \theta_j)$ have  meanings similar to those in Sec.~\ref{sec:iso_nnn}. In the non-axial directions, we define $\pi$ as the probability that the cavity site points up and belongs to percolated cluster, and $q$ as the probability that the cavity site points up and does \emph{not} belong to percolated cluster. As for the isotropic case, we only consider clusters consisting of up spins (the case of clusters of down spins is symmetric), and therefore down spins naturally do not belong to the percolated cluster, \textit{i.e.} $(s_j, \theta_j) = (\downarrow,0)$. As a result, the recursive relations among the $(9+2)$ variables (the indices $j$ and $k$ can be neglected for generality) are extensions based on the generalized FK--CK criterion for the $d=1$ chain (Sec.~\ref{sec:iso_nnn}) in the axial direction and on the nearest-neighbor FK--CK criterion (Sec.~\ref{sec:iso_revCK}) in non-axial directions:
\begin{widetext}
\begin{equation}
\begin{split}
    \eta_\mathrm{cav,a}(\uparrow, \uparrow) &= \pi(\uparrow,1)+\pi(\uparrow,0)+q(\uparrow,1)+q(\uparrow,0),\\
    \eta_\mathrm{cav,a}(\downarrow, \uparrow) &= w(\uparrow,1)+w(\uparrow,0),\\
    \eta_\mathrm{cav,a}(\uparrow, \downarrow) &= \pi(\downarrow,0)+q(\downarrow,0),\\
    \eta_\mathrm{cav,a}(\downarrow, \downarrow) &= w(\downarrow,0),\\
    \eta_\mathrm{cav, na}(\uparrow) &= \pi+q.
\end{split}
\end{equation}
For conciseness, we define 
\begin{equation}
    Y(s) = \sum\limits_{l=0}^{c-1} \binom{c-1}{l}\eta_\mathrm{cav,na}(\uparrow)^l \eta_\mathrm{cav,na}(\downarrow)^{c-1-l} e^{\beta J s(2l-c+1)},
\end{equation}
\begin{equation}
    Y_p = \sum\limits_{l=1}^{c-1} \binom{c-1}{l}e^{\beta J s(2l-c+1)}\eta_\mathrm{cav,na}(\downarrow)^{c-1-l}\sum\limits_{k=1}^{l}\hat{q}^{l-k}\hat{\pi}^{k},
\end{equation}
\begin{equation}
    Y_q = \sum\limits_{l=0}^{c-1} \binom{c-1}{l}e^{\beta J (2l-c+1)}\eta_\mathrm{cav,na}(\downarrow)^{c-1-l}\hat{q}^l,
\end{equation}
to describe the contribution of non-axial neighbors, where $\hat{\pi}=p_1\pi$, $\hat{q}=q+(1-p_1)\pi$.
For the configuration $\eta_\mathrm{cav,a}(\uparrow, \uparrow) = \pi(\uparrow,1)+\pi(\uparrow,0)+q(\uparrow,1)+q(\uparrow,0)$, with
\begin{equation}
\begin{split}
    \pi(\uparrow,1) &= {Z_\mathrm{cav,a}}^{-1}e^{\beta J(1+\kappa)}\lbrace \lbrack \left(\pi(\uparrow,1)+(\pi(\uparrow,0)+q(\uparrow,1))p_1\right \rbrack p_2 Y(\uparrow) 
    +\lbrack (\pi(\uparrow,0)(1-p_1)+q(\uparrow,0)p_1\rbrack p_2 Y_p \rbrace,
\end{split}
\end{equation}
\begin{equation}
\begin{split}
    q(\uparrow,1) &= {Z_\mathrm{cav,a}}^{-1}e^{\beta J(1+\kappa)} \pi(\uparrow,0)(1-p_1)p_2 Y_q
\end{split}
\end{equation}
\begin{equation}
\begin{split}
    \pi(\uparrow,0) &= {Z_\mathrm{cav,a}}^{-1}\lbrace  e^{\beta J(1+\kappa)}\lbrack\left(\pi(\uparrow,1)+(\pi(\uparrow,0)+q(\uparrow,1))p_1\right)(1-p_2) + q(\uparrow,1)(1-p_1)\rbrack Y(\uparrow)\\
    &+ e^{\beta J(1+\kappa)}\lbrack (\pi(\uparrow,0)(1-p_1)(1-p_2)+q(\uparrow,0)(1-p_1+p_1-p_1p_2)\rbrack Y_p \\
    &+ e^{-\beta J(1+\kappa)}\lbrack w(\uparrow,1)Y(\uparrow)+w(\uparrow,0)Y_p\rbrack\rbrace,
\end{split}
\end{equation}
\begin{equation}
\begin{split}
    q(\uparrow,0) &= {Z_\mathrm{cav,a}}^{-1}\lbrace e^{\beta J(1+\kappa)}\lbrack \pi(\uparrow,0)(1-p_1)(1-p_2) + q(\uparrow,0)\rbrack + e^{-\beta J(1+\kappa)} w(\uparrow,0)\rbrace Y_q
\end{split}
\end{equation}
For the configuration $\eta_\mathrm{cav,a}(\downarrow, \uparrow) = w(\uparrow,1)+w(\uparrow,0)$, with
\begin{equation}
\begin{split}
    w(\uparrow,1) &= {Z_\mathrm{cav,a}}^{-1} e^{\beta J(-1+\kappa)}\pi(\downarrow,0)p_2 Y(\downarrow),
\end{split}
\end{equation}
\begin{equation}
\begin{split}
    w(\uparrow,0) &= {Z_\mathrm{cav,a}}^{-1}\lbrace e^{\beta J(-1+\kappa)}\lbrack \pi(\downarrow,0)(1-p_2) + q(\downarrow,0)\rbrack + e^{-\beta J(-1+\kappa)} w(\downarrow,0)\rbrace Y(\downarrow).
\end{split}
\end{equation}
For the configuration $\eta_\mathrm{cav,a}(\uparrow, \downarrow) = \pi(\downarrow,0)+q(\downarrow,0)$, with
\begin{equation}
\begin{split}
    \pi(\downarrow,0) &= {Z_\mathrm{cav,a}}^{-1}\lbrace \lbrack e^{\beta J(1-\kappa)}\left(\pi(\uparrow,1)+\pi(\uparrow,0)p_1+q(\uparrow,1)\right) + e^{-\beta J(1-\kappa)}w(\uparrow,1)\rbrack Y(\uparrow)\\
    &+\lbrack e^{\beta J(1-\kappa)}(\pi(\uparrow,0)(1-p_1)+q(\uparrow,0))+e^{-\beta J(1-\kappa)}w(\uparrow,0)\rbrack Y_p \rbrace,
\end{split}
\end{equation}
\begin{equation}
\begin{split}
    q(\downarrow,0) &= {Z_\mathrm{cav,a}}^{-1}\lbrace e^{\beta J(1-\kappa)}\lbrack \pi(\uparrow,0)(1-p_1)+q(\uparrow,0)\rbrack + e^{-\beta J(1-\kappa)}q(\downarrow,0) \rbrace Y_q
\end{split}
\end{equation}
For the configuration $\eta_\mathrm{cav,a}(\downarrow, \downarrow) = w(\downarrow,0)$, with
\begin{equation}
\begin{split}
    w(\downarrow,0) &= {Z_\mathrm{cav,a}}^{-1}\lbrace e^{-\beta J(1+\kappa)}\lbrack \pi(\downarrow,0) + q(\downarrow,0)\rbrack + e^{\beta J(1+\kappa)} w(\downarrow,0)\rbrace Y(\downarrow).
\end{split}
\end{equation}
In the non-axial direction, we have $\eta_\mathrm{cav, na}(\uparrow) = \pi+q$, 
\begin{equation}
\begin{split}
    q &= {Z_\mathrm{cav,na}}^{-1}\sum\limits_{l=0}^{c-2} \binom{c-2}{l}e^{\beta J (2l-c+2)}\eta_\mathrm{cav,na}(\downarrow)^{c-2-l}\hat{q}^l \lbrace e^{\beta(-2J+\kappa)}w(\uparrow,0)^2 + 2e^{-2\beta\kappa}w(\uparrow,0)\lbrack \pi(\uparrow,0)(1-p_1)+q(\uparrow,0)\rbrack\\ 
    &+ e^{\beta J(2+\kappa)} \lbrack \pi(\uparrow,0)^2(1-p_1)^2 + q(\uparrow,0)^2 + 2\pi(\uparrow,0)q(\uparrow,0)((1-p_1)(1-p_2)+(1-p_1)^2p_2)\rbrack\rbrace.
\end{split}
\end{equation}
As a result, the percolation probability $P$ can be obtained from the relation $P + Q = \eta(\uparrow)$ and
\begin{equation}
\begin{split}
    Q &= Z_\mathrm{site}^{-1}\sum\limits_{l=0}^{c-1} \binom{c-1}{l}e^{\beta J (2l-c+2)}\eta_\mathrm{cav,na}(\downarrow)^{c-1-l}\hat{q}^l \lbrace e^{\beta J(-2+\kappa)}w(\uparrow,0)^2 + 2e^{-2\beta\kappa J}w(\uparrow,0)\lbrack \pi(\uparrow,0)(1-p_1)+q(\uparrow,0)\rbrack\\ 
    &+ e^{\beta J(2+\kappa)} \lbrack \pi(\uparrow,0)^2(1-p_1)^2 + q(\uparrow,0)^2 + 2\pi(\uparrow,0)q(\uparrow,0)((1-p_1)(1-p_2)+(1-p_1)^2p_2)\rbrack\rbrace,
\end{split}
\end{equation}
where $Z_\mathrm{site}$ is the same as for Eq.~\eqref{eq:annni_density}.
\end{widetext}
\onecolumngrid
\clearpage
\twocolumngrid

\subsection{Frustrated RBIM Model}

The RBIM defined in Eq.~\eqref{eq:HRBIM} only involves nearest-neighbor interactions, which in some ways simplifies the analysis relative to that of models with SALR interactions. The presence of quenched disorder, 
however,
breaks the translational invariance of the lattice. As a result, the recursive equations for the percolation probability get rapidly more involved with growing connectivity. We therefore restrict most of our analysis to the case $c+1=3$. The presence of quenched disorder also makes the linear expansion around the standard Ising model too onerous to consider.

\subsubsection{Cavity field and recursion equations}
\begin{figure}[h!]
    \centering
    \includegraphics[width=0.85\linewidth]{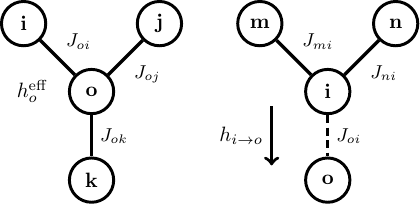}
    \caption{Indexing scheme for a generic frustrated RBIM configuration on a Bethe lattice with connectivity $ c+1 = 3 $. (Left) The central site $ o $ is subjected to an effective field $ h_o^{\rm eff} $, arising from interactions with its three nearest neighbors $ i, j, $ and $ k $, via couplings $ J_{oi}, J_{oj}, $ and $ J_{ok} $, respectively. (Right) The contribution of site $ i $ to $ h_o^{\rm eff} $, denoted $ h_{i \to o} $, reflects the influence of its own neighboring sites $ m $ and $ n $, computed in the absence of site $ o $.}
    \label{fig:RBIM_configuration}
\end{figure}

Because translational invariance is broken in this model, explicit site indexing is restored. We denote the current---or central---site with $o$, and label its $c+1 = 3 $ nearest neighbors as $i$, $j$ and $k$ (see Fig.~\ref{fig:RBIM_configuration}). The marginal probability of a microscopic spin configuration at the central site $o$ can be parametrized in terms of an effective local field $h_o^{\rm eff}$ as 
\begin{equation}
\eta_{o}(s_o) = \frac{e^{\beta h^{\rm eff}_{o} s_o}}{2 \cosh(\beta h_{o}^{\rm eff})}.
\label{eq:conf_probability_parametrization}
\end{equation}  
This effective field encapsulates the influence of neighboring spins on site $o$. Alternatively, this marginal probability can be expressed explicitly in terms of interactions with neighboring sites,
\begin{equation}
\eta_{o}(s_o) = \frac{e^{\beta h_{\rm ext} s_o}}{Z_{\rm site}} \prod_{l \in \partial o} \sum_{s_l} e^{\beta J_{ol} s_o s_l} \eta_{l \to o}(s_l)\;,
\label{eq:conf_probability_RBIM}
\end{equation}
where $Z_{\rm site}$ is the normalization factor, such that $\eta_o(\uparrow) + \eta_o(\downarrow) = 1$. The quantity $\eta_{l \to o }(s_l)$ is the cavity marginal configuration probability for the site $l \in \{ i,j,k\}$, computed in the absence of site $o$ (see Fig.~\ref{fig:RBIM_configuration}), where the contribution from site $i$ depends on the remaining neighbors $\{m,n\} = \partial i \setminus o$. This cavity configuration probability satisfies a recursive equation of the same form
\begin{equation}
\eta_{l\to r}(s_l) = \frac{e^{\beta h_{\rm ext} s_l}}{Z_{\rm cav}} \prod_{p \in \partial l \setminus r} \sum_{s_p} e^{\beta J_{lp} s_l s_p} \eta_{p \to l}(s_p)\;,
\label{eq:conf_probability_cav}
\end{equation} 
for any generic site $l$, with $r$ denoting the cavity (or backward) site. The factor $Z_{\rm cav}$ is the normalization factor, such that $\eta_{l \to r}(\uparrow) + \eta_{l \to r}(\downarrow) = 1$. Analogously to Eq.~\eqref{eq:conf_probability_parametrization}, the cavity configuration probability can be parametrized using a cavity field $h_{l \to r}$
\begin{equation}
\eta_{l \to r}(s_l) = \frac{e^{\beta h_{l \to r} s_l}}{2 \cosh(\beta h_{l \to r})}.
\label{eq:conf_probability_parametrization_cavity}
\end{equation} 
With both effective and cavity configuration probabilities expressed in this form, we can derive equivalent expressions for their corresponding fields. The effective field at site $o$ is then
\begin{equation}
    h^{\rm eff}_o = h_{\rm ext} + \frac{1}{\beta} \sum_{l \in \partial o}  \atanh(\tanh{\beta J_{ol}} \tanh{\beta h_{l \to o}}),
    \label{eq:effective_field} 
\end{equation}
while the cavity field $h_{l \to r}$, which captures the influence on site $l$ in the absence of site $r$, satisfies the recursion
\begin{equation}
    h_{l \to r} = h_{\rm ext}+ \frac{1}{\beta}\sum_{p\in\partial l \setminus r} \atanh(\tanh{\beta J_{lp}} \tanh{\beta h_{p \to l}}).
    \label{eq:cavity_fields}
\end{equation} 
Note that because of the disorder in the couplings $\{ J_{0l}\}$ and $\{J_{lp} \}$ whenever $\rho \neq 1$, both effective and cavity fields are themselves random variables. The stationary distribution for the cavity fields in Eq.~\eqref{eq:cavity_fields} is therefore obtained with arbitrary precision  using the population dynamics algorithm \cite{mezard2009information}.  Numerically, this scheme involves initializing a random sample---or population---of size $\mathcal M$ of the cavity fields $\{ h_{l \to r}\}$ and iteratively updating each member of this population according to Eq.~\eqref{eq:cavity_fields}. Each member undergoes $\mathcal{N}$ such updates until the entire population converges to a stationary probability distribution. Once the distribution of the cavity fields is known, the effective fields defined in Eq.~\eqref{eq:effective_field} can be computed. Knowing the distribution of effective fields, $\{ h^{\rm eff}_o \}$, enables the calculation of all relevant thermodynamic quantities. In particular, the average magnetization density is
\begin{equation}
m \equiv [\langle s_o\rangle] = [\tanh(\beta h_o^{\rm eff})],
\label{eq:magnetization}
\end{equation}
where $\langle \cdots \rangle$ denotes the thermal average with respect to the Boltzmann--Gibbs distribution, and $[\cdots]$ denotes the average over disorder realizations. In this context, averaging over disorder is equivalent to averaging over the distribution of effective fields $\{ h_o^{\rm eff} \}$.

\subsubsection{Phase diagram}
\label{sec:rbim_phase}
Interestingly, for $h_\mathrm{ext}=0$, closed-form analytical expressions for the critical temperatures of the P-F and paramagnetic-to-spin-glass (P-SG) transitions can be obtained~\cite{dotsenko1994introduction, nishimori_statistical_2001}
\begin{equation}
    \begin{aligned}
        & \frac{J_0}{T_{c}} = \atanh\Bigg(\frac{1}{c(2\rho-1)}\Bigg) \; , 
        \\ 
        & \frac{J_0}{T_{\rm P-SG}} = \atanh\Bigg(\frac{1}{\sqrt{c}}\Bigg) \; ,
    \end{aligned}
    \label{eq:critical_temps_RBIM}
\end{equation}
respectively. These expressions are calculated from the linear stability of the first two moments of the cavity fields in Eq.~\eqref{eq:cavity_fields} around zero. In other words, they correspond to the temperatures for which
\begin{equation}
[h_{l \to r}] = 0 \qquad\text{and}\qquad  [h_{l \to r}^2] = 0,
\end{equation}
respectively.  The critical dilution at which these critical temperatures coincide is
\begin{equation}
    \rho_\star = \frac{1}{2}\left( 1+\frac{1}{\sqrt{c}}\right) \;,
\end{equation} 
which gives rise to a multicritical point separating a regime in which the system undergoes a P-F transition ($\rho$ close to 1) from another one in which the ferromagnetic phase is replaced by a phase with spin-glass ordering 
($\rho$ close to 1/2). The three phases meet at this point. Coming from $T$ large, the P-F transition line corresponds to the temperature at which the system develops a non-zero magnetization, given by Eq.~\eqref{eq:magnetization}. By contrast, along the P-SG transition line, the spontaneous magnetization remains zero, but the system exhibits a non-zero Edwards--Anderson (EA) order parameter,
\begin{equation}
    q_{\rm EA} \equiv [\langle s_o \rangle^2] = [\tanh^2(\beta h_{o}^{\rm eff})].
\end{equation} 
A phase with both magnetization and EA order parameter being non-zero can be identified, denoted here as the ferromagnetic-spin-glass (FSG). The transition lines between SG-FSG and FSG-F can be estimated numerically, at the replica-symmetric level, using population dynamics. Coming from the spin-glass phase, the critical line SG-FSG is estimated at the onset of finite magnetization, while the FSG-F transition line is determined from the stability of the EA order parameter, measured as 
\begin{equation}
    \delta q_{\rm EA} = |q_{\rm EA}-q_{ab}|,
    \label{eq:stabilityofRS}
\end{equation} 
where $q_{ab}$ is the overlap between replicas $a$ and $b$, i.e.,
\begin{equation}
    q_{ab} = [\langle s_o^{(a)} \rangle \langle s_o^{(b)} \rangle] = [m^{(a)}m^{(b)}]\;,
\end{equation} 
and $\delta q_{\rm EA} = 0$ in the purely ferromagnetic phase.
Numerically, this corresponds to two populations of the cavity fields being updated simultaneously, with different initial conditions. A crucial point is that both populations $a$ and $b$ are evolved concomitantly in the population dynamics algorithm~\cite{Pagnani2003}. Put differently, at each selected site, $p \in \partial l \setminus r$, we draw the same random bonds $J_{lp}$ for both populations. The EA order parameter $q_{\rm EA}$ is then calculated using any of the two replicas $a$ or $b$, the choice being immaterial as they should produce the same result. For the specific case $T/J_0 = 0$, the cavity equations must be modified~\cite{Mzard2003}. Therefore, for $T/J_0 = 0$, we have extracted the corresponding known results for the SG-FSG and FSG-F transition lines from Ref.~\cite{Kwon1988}, $\rho = 0.86950(3)$ and $\rho =0.91665(5)$, respectively. Finally, the Nishimori line is given by Eq.~\eqref{eq:Nishimori}---independently of the Bethe lattice connectivity---and goes through the multicritical point $(\rho_\star, T_{\rm SG})$, as expected (see Sec.~\ref{sec:models}).

\subsubsection{Percolation of FK--CK clusters}
\label{sec:FKCK-RBIM}
To determine the percolation probabilities in this model, we build on the above notation. We denote $P^{(n)}_{o}$ the probability that site $o$ belongs to the percolating cluster, with $n$ of its nearest-neighbor spins pointing up, and $Q^{(n)}_{o}$ the complementary probability that site $o$ does \textit{not} belong to the percolating cluster, with $n$ of its nearest-neighbor spins pointing up. As in previous sections, we consider the percolation transition of positive magnetized domains, and hence that site $o$ belongs to the infinite percolating cluster the spin in $o$ must point up, i.e., $s_o = \;\uparrow$. Compactly, we write these conditions as 
\begin{equation}
    \begin{aligned}
        P_{o}^{(n)} \equiv \textrm{Pr} \left\{s_o =\;\uparrow \, \wedge \;\,  o \in \mathcal{C_\infty} \, \wedge \, \sum_{l \in \partial o} \delta_{s_l, \uparrow} = n\right\} \;, \\
        Q_{o}^{(n)} \equiv \textrm{Pr} \left\{s_o =\;\uparrow \, \wedge \;\,  o \notin \mathcal{C_\infty} \, \wedge \, \sum_{l \in \partial o} \delta_{s_l, \uparrow} = n \right\} \;.
    \end{aligned}
    \label{eq:ProbDefinitions}
\end{equation}
Clearly, $n \in \{ 0, 1, \ldots , c+1\} $. Summing over all possible $n$ we obtain the overall probabilities that the central site $o$ belongs or does not belong to the percolating cluster, $P_o$ or $Q_o$, respectively. These last two probabilities are related to the configuration probability defined in Eq.~\eqref{eq:conf_probability_RBIM} via
\begin{equation}
    \eta_{o}(\uparrow) = P_o + Q_o,
    \label{eq:normalization_Zcav}
\end{equation} 
letting us parametrize all our expressions in terms of the percolation probability, $P_o$, and the spin up configuration probability $\eta_{o}(\uparrow)$.
\begin{widetext}
Using the labels defined in Fig. \ref{fig:RBIM_configuration}, the constituent equations for $n = 0,1,2,3$, are: 
\begin{eqnarray}
    P^{(0)}_{o} 
    &=& 
    0 \; ,
    \label{eq:P0FKCK}
\\
[5pt]
    P^{(1)}_o &=& \dfrac{e^{\beta h_{\rm ext}}}{Z_{\rm{site}}} \sum_{(x,y,z) \in C_3} e^{\beta(J_{ix} - J_{iy} - J_{iz})} \pi_{x \to o} p_B^{(ox)} \left[1-\eta_{y\to o}(\uparrow) \right] \left[1-\eta_{z \to o}(\uparrow) \right]  \;,
\label{eq:P1FKCK}
\\
[5pt]
    P_o^{(2)} & = & \frac{e^{\beta h_{\rm ext}}}{Z_{\rm site}} 
\sum_{(x,y,z) \in C_3} 
e^{\beta (J_{ix} + J_{iy} - J_{iz})} [1 - \eta_{z \to o}(\uparrow)] \times \\ \nonumber
&& \qquad \qquad \qquad \qquad \left\{
\pi_{x \to o} \, p_B^{(ox)} [\eta_{y \to o}(\uparrow) - \pi_{y \to o}] + 
\pi_{y \to o} \, p_B^{(oy)} [\eta_{x \to o}(\uparrow) - \pi_{x \to o}] + 
\pi_{x \to o} \pi_{y \to o} \Phi_{xy}^{(o)}
\right\}
\label{eq:P2FKCK}
\\
[5pt]
P_o^{(3)} & = & \frac{e^{\beta (J_{oi} + J_{oj} + J_{ok} + h_{\rm ext})}}{Z_{\rm site}} \Bigg\{ 
 \sum_{(x,y,z) \in C_3} 
\pi_{x \to o}\, p_B^{(ox)} [\eta_{y \to o} (\uparrow) - \pi_{y \to o}][\eta_{z \to o}(\uparrow) - \pi_{z \to o}] \\ [3pt] \nonumber
&& \qquad \qquad \qquad \qquad  \qquad 
+  \sum_{(x,y,z) \in C_3} 
\pi_{x \to o} \pi_{y \to o} \, [\eta_{z \to o}(\uparrow) - \pi_{z \to o}] \Phi_{xy}^{(o)} 
+  \pi_{i \to o} \pi_{j \to o} \pi_{k \to o} \Psi_{ijk}^{(o)} \Bigg\}
\label{eq:P3FKCK}
\end{eqnarray}
with 
\begin{equation}
\begin{array}{c}
\Phi_{xy}^{(o)} = 
p_B^{(ox)} p_B^{(oy)} + 
\left(1 - p_B^{(ox)}\right) p_B^{(oy)} + 
p_B^{(ox)} \left(1 - p_B^{(oy)} \right) \\[1.5ex]
\Psi_{ijk}^{(o)} = 
\sum_{(x,y,z) \in C_3} \left[
p_B^{(ox)} \left(1 - p_B^{(oy)}\right) \left(1 - p_B^{(oz)}\right) + 
p_B^{(ox)} p_B^{(oy)} \left(1 - p_B^{(oz)}\right)
\right] + 
p_B^{(oi)} p_B^{(oj)} p_B^{(ok)},
\end{array}
\end{equation}
\end{widetext}
where we used the cyclic permutation group of order 3, $C_3 = \{(i,j,k), (j,k,i), (k,i,j)\}$, to consider compactly all possible combinations of sites with up spins. The overall percolation probability is given by the sum over all constituent contributions
\begin{equation}
    P_o = \sum_{n = 0}^{c+1} P_{o}^{(n)}.
    \label{eq:OverallPercolationProb}
\end{equation}
The factor \( Z_{\rm site} \) is such that \( \eta(\uparrow) = P_o + Q_o \). The cavity quantities appearing in Eqs.~\eqref{eq:P0FKCK}--\eqref{eq:P3FKCK}, denoted \( \pi_{x \to o} \), represent the probability that a site $x\, (= i, j, k)$ belongs to the percolating cluster in the absence of site \( o \), and its complement \( q_{x \to o} \) corresponds to the probability that \( x \) does \emph{not} belong to the percolating cluster in the absence of \( o \). The latter quantities, $q_{x 
\to o}$, do not appear explicitly in the preceding equations because the condition $\eta_{x \to o}(\uparrow) = \pi_{x \to o} + q_{x \to o}$ has already been imposed.
These equations are constructed considering $c = 2$ neighboring spins, reflecting the absence of site $o$. As with the overall percolation probability, the cavity percolation probability---and its complement---is decomposed into constituent terms corresponding to each possible number $n$ of neighboring up spins,
\begin{equation}
    \pi_{x \to o} = \sum_{n = 0}^{c} \pi_{x \to o}^{(n)}.
\end{equation}
Analogous to the definitions in Eq.~\eqref{eq:ProbDefinitions}, the constituent cavity percolation probabilities are formally defined as:
\begin{equation}
    \begin{aligned}
        \pi_{x \to o}^{(n)} &\equiv \textrm{Pr}\left\{ s_x = \uparrow \, \wedge \;\, x \in \mathcal{C}_\infty \, \wedge \; \sum_{l \in \partial x \setminus o} \delta_{s_l, \uparrow} = n\right\}, \\
        q_{x \to o}^{(n)} &\equiv \textrm{Pr}\left\{ s_x = \uparrow \, \wedge \;\, x \notin \mathcal{C}_\infty \, \wedge \; \sum_{l \in \partial x \setminus o} \delta_{s_l, \uparrow} = n\right\},
    \end{aligned}
\end{equation} 
where in this case $n\in\{0, 1, 2\}$. The explicit form of these equations, following the labeling used in the right-hand side of Fig.~\ref{fig:RBIM_configuration}, is
\begin{widetext}
\begin{eqnarray}
    \pi^{(0)}_{i \to o} 
    &=& 
    0 \; ,
    \label{eq:pi0FKCK}
\\
[5pt]
    \pi^{(1)}_{i \to o}  &=& \dfrac{e^{\beta h_{\rm ext}}}{Z_{\rm{cav}}} \Big\{ e^{\beta(J_{im} - J_{in})} \pi_{m \rightarrow i} \,p_B^{(im)} \left[1-\eta_{n \rightarrow i}(\uparrow) \right] + e^{-\beta(J_{im} - J_{in})} \pi_{n \rightarrow i} \,p^{(in)}_B \left[1-\eta_{m \to i}(\uparrow)\right] \Big\} \;,
\label{eq:pi1FKCK}
\\
[5pt]
    \pi^{(2)}_{i \to o} &=& 
    \dfrac{e^{\beta h_{\rm ext} + \beta(J_{im} + J_{in})}}{Z_{\rm cav}} 
    \Big\{ \pi_{m \to i}\, p^{(im)}_B \left[\eta_{n \to i}(\uparrow)-\pi_{n \to i} \right]  + 
    \pi_{n \to i} \,p_B^{(in)} \left[\eta_{m \to i}(\uparrow)- \pi_{m \to i} \right] \,  
    \nonumber\\ 
   [3pt]
   &&  
   \qquad\qquad\qquad\qquad
   +  \pi_{m \to i}\,\pi_{n \to i} \left[ p_B^{(im)}\left(1-p_B^{(in)}\right) + p_B^{(in)}\left(1-p_B^{(im)}\right) + p_B^{(im)} \,p_B^{(in)}\right]\Big\} \; , 
\label{eq:pi2FKCK}
\end{eqnarray} 
where  $Z_{\rm cav}$ is such that $\pi_{x \to o} + q_{x \to o}=\eta_{x \to o}(\uparrow)$. The most explicit form of the cavity percolation probability, using $p_B^{(im)} = 1 - e^{-2 \beta J_{im}}$, is then
\begin{equation}
\pi_{i \to o} = \eta_{i \to o}(\uparrow) \frac{ \pi_{n \to i} \left( e^{2 \beta J_{in}} - 1 \right)  \left[\eta_{m \to i}(\uparrow) \left( e^{2 \beta J_{im}} - 1 \right) +1\right]
+\pi_{m \to i} \left( e^{2 \beta J_{im}} - 1 \right) \left[\left( e^{2 \beta J_{in}} - 1 \right)
   (\eta_{n \to i}(\uparrow)-\pi_{n \to i})+1\right]}{\left[\eta_{m \to i}(\uparrow) \left( e^{2 \beta J_{im}} - 1 \right)+1\right] \left[\eta_{n \to i}(\uparrow) \left( e^{2 \beta J_{in}} - 1 \right)+1\right]} \;.
   \label{eq:Probc2}
\end{equation}
\end{widetext}
For the standard Ising model, Eq.~\eqref{eq:Probc2} can simply be solved by iteration, because the coupling constants are all equal, $J_{im} = J > 0 \;\forall\, (im)$. However, introducing antiferromagnetic bonds ($J_{im} < 0$) makes the quantity $\pi_{i \to o}$ depend explicitly on the specific realization of $J_{im}$. Because these bonds are quenched random variables, $\pi_{i \to o}$ becomes a quenched random variable itself. We again solve these equations using the population dynamics algorithm~\cite{mezard2009information}, this time the population being the cavity percolation probabilities $\pi_{i \to o}$. For this model, we omit the computation of the effective percolation probabilities, $P_{o}$. The solution of Eq.~\eqref{eq:OverallPercolationProb} is more complex and is unnecessary to study the critical properties of the system. Because $P_0 = 0 \Leftrightarrow \pi_{x \to 0} = 0$, the cavity percolation probabilities also serve as order parameters for the percolation transition. Both the average cavity and the effective ($P \equiv [P_o]$) percolation probabilities suffice.
In practice, we numerically compute $\pi \equiv [\pi_{x \to o}]$ as the average over a population of cavity percolation probabilities once their distribution reaches a stationary state.

\subsubsection{Percolation of the $\alpha$-parameter cluster model}
\label{sec:alpha_cluster_rbim}
As for the isotropic SALR interaction model, an $\alpha$-parameter cluster model can be defined for the frustrated RBIM. In this case, we adjust the bond probabilities defined over satisfied links---neighboring sites $(ij)$ that satisfy the condition $J_{ij} s_i s_j > 0$. For such links, a bond is placed with probability $p_B^{(ij)} = 1 - e^{-2\beta |J_{ij}|}$, which is strictly positive. Because under this formulation the percolation temperature does not coincide with the thermodynamic transition (see Sec.~\ref{sec:theorem}), we introduce a parameter in the bond probability to enforce alignment as in Eq.~\eqref{eq:alphaparameterdef},
\begin{equation}
    p_B^{(ij)}(\alpha) = 1 - e^{-\beta \alpha (J_{ij} s_i s_j + |J_{ij}|)} \;,
    \label{eq:modified-FKCK}
\end{equation} 
with $\alpha$ such that $T_p(\alpha) = T_c$. Equations~\eqref{eq:pi0FKCK}-\eqref{eq:pi2FKCK} can be updated accordingly, and we denote  $\pi(\alpha)$ the resulting average cavity percolation probability. 

\section{Results and discussion}
\begin{figure*}
    \centering
    \includegraphics[width=0.48\linewidth]{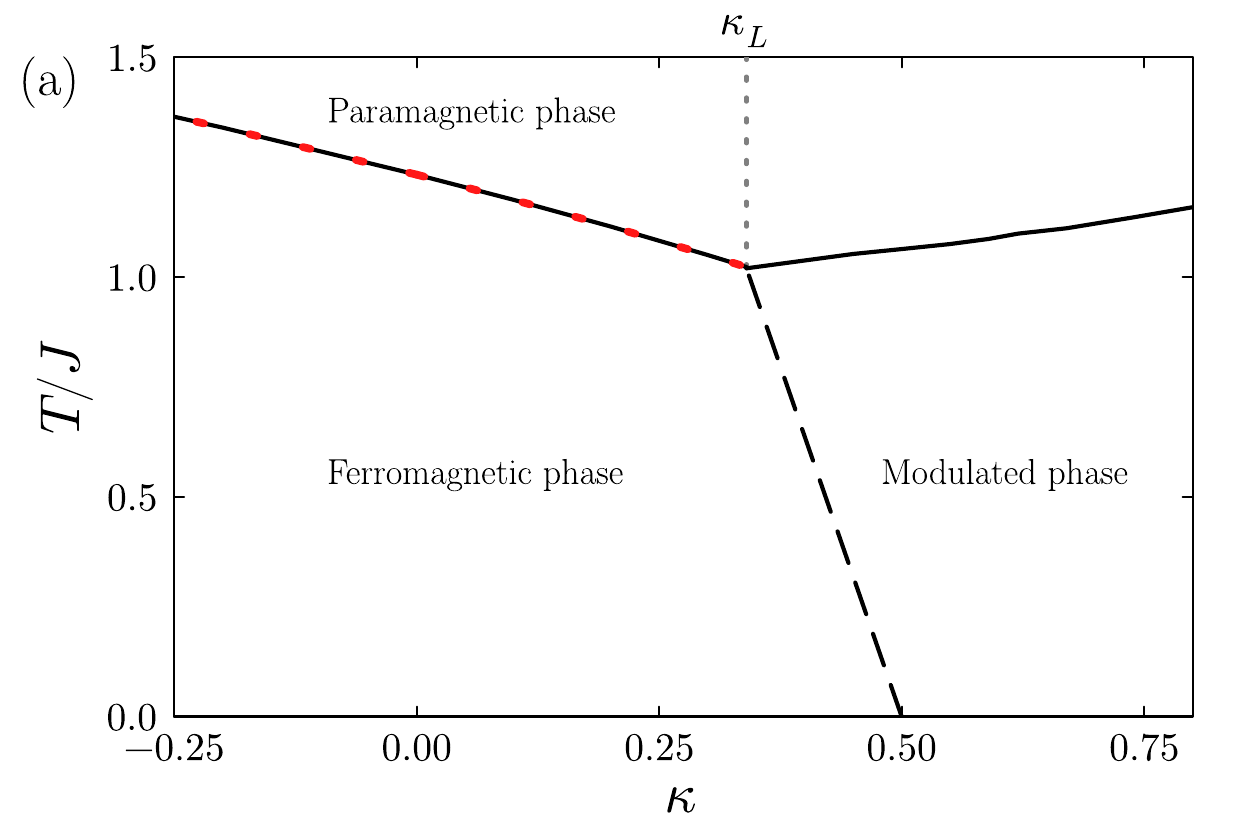}
    \includegraphics[width=0.48\linewidth]{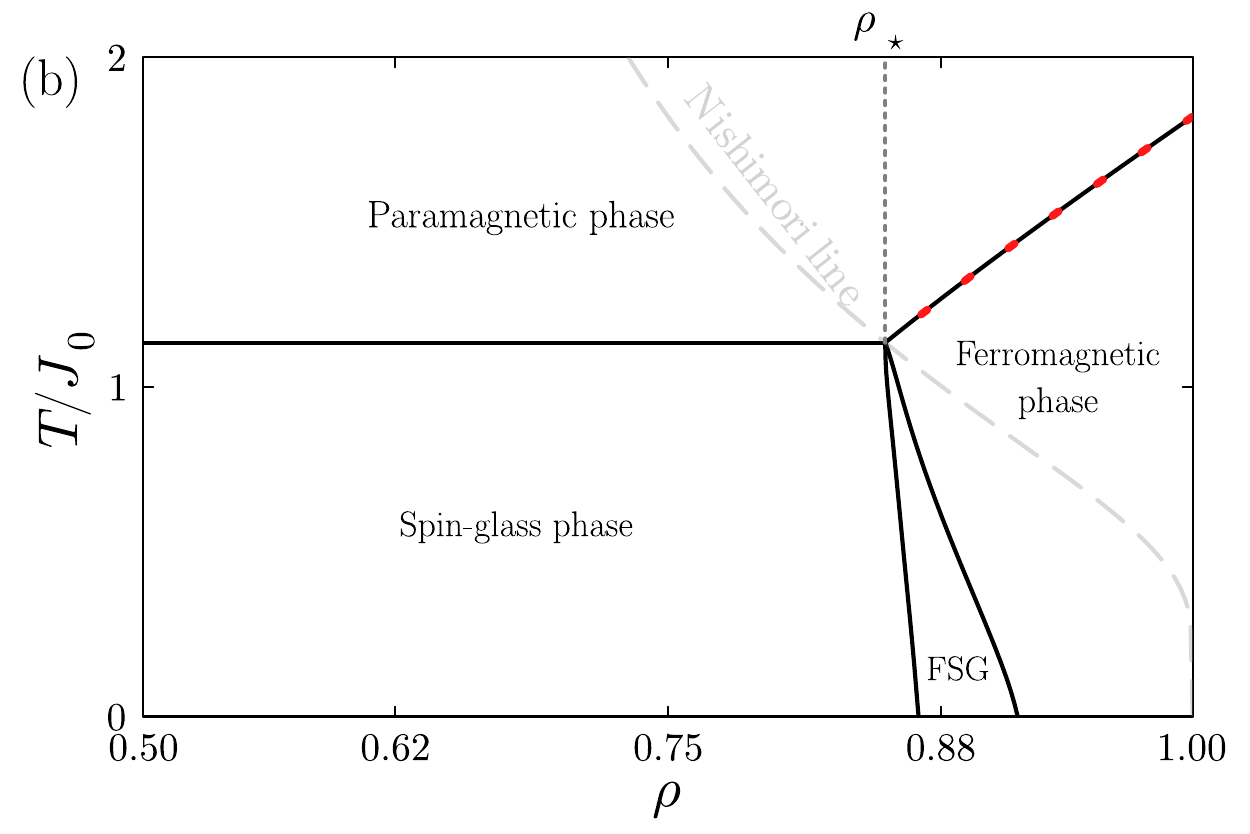}
    \caption{Phase diagram of (a) the ANNNI model on a Bethe lattice with $c+1=6$ adapted from Ref.~\onlinecite{zheng2022weakening} and (b) the frustrated RBIM on the Bethe lattice with $c+1=3$ (see Sec.~\ref{sec:rbim_phase}), both at zero external field. Phase transitions are determined as described in the text. The resulting multi-critical points are (a) $(\kappa_L, T_L/J) = (0.34(1), 1.02(2))$, and (b) $(\rho_\star,T_\star/J_0)=\left( \frac{1+\sqrt{2}}{2 \sqrt{2}}, \frac{1}{\atanh(1/\sqrt{2})}\right)$. In the Ising-like transition regime, the percolation temperature $T_p$ of the generalized FK--CK clusters (red dots) is consistent with $T_c$ to within one part in $10^{3}$. 
    In (b), the Nishimori line (gray dashed line) and the multi-critical point dilution, $\rho_\star$ (gray dotted line) are also included for reference.} 
    \label{fig:annni_rbim}
\end{figure*}
\label{sec:discussion}
Using the cavity expressions for the Bethe lattice derived in Sec.~\ref{sec:Bethe}, we here discuss the phase diagram results as well as the percolation properties of the three models considered in this work.

\subsection{Phase diagrams}
As noted in Sec.~\ref{sec:models}, the phase behavior of the three models presents various similarities. At weak frustration an Ising-like P-F transition is observed, and at strong frustration new phases emerges. In all cases, at zero external field these two regimes are separated by a multicritcal point. Important differences, however, are also noted. Most notably, for SALR interaction models the new phase is modulated, while for the RBIM it is a spin glass. In the RBIM phase diagram, a fourth phase also emerges, exhibiting spin-glass order and non-zero magnetization. Such ferromagnetic-spin-glass (FSG) phase is commonly found in mean-field models with (unbalanced) spin-glass--type frustration.

\begin{figure*}[t]
\centering
\includegraphics[width=\linewidth]{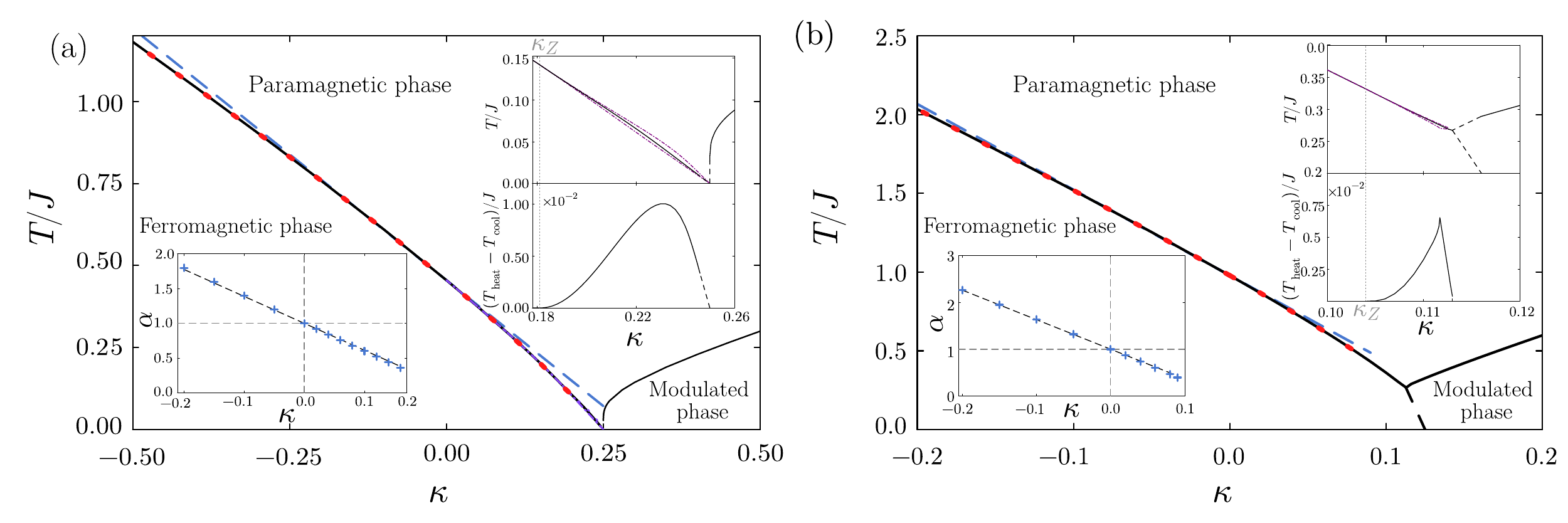}
\caption{Phase diagram of the isotropic SALR interaction model on a Bethe lattice with (a) $c+1 = 3$ and (b) $c+1 = 5$. At negative and small positive $\kappa$ the P-F phase transition is Ising-like, while beyond the Lifshitz point, $\kappa>\kappa_\mathrm{L}$, the paramagnetic-to-modulated phase transition is part of the XY universality class (solid black lines). The linear expansion of $T_c$ around $\kappa=0$ (blue dashed line) highlights the analytical continuity between the purely attractive and the SALR interaction regimes. In the Ising-like transition regime, the percolation temperature $T_p$ of the generalized FK--CK clusters (red dots) is consistent with $T_c$ to within one part in $10^{3}$. Numerical convergence problems in the vicinity of the multi-critical (Lifshitz) point require extrapolating the phase coexistence lines (dashed black lines), which gives (a) $(\kappa_L, T_L/J) = (0.2500(1),0.0000(1))$ for $c+1=3$ and (b) $(\kappa_L, T_L/J) = (0.113(1),0.267(7))$ for $c+1=5$. (Insets) Enlarged weakly first-order P-F transition regime, intermediate between the Ising-like P-F regime and the Lifshitz point, for $\kappa\in(\kappa_Z,\kappa_L)$, with (a) ($\kappa_Z,T_Z/J) = (0.181(6), 0.141(6))$ for $c+1=3$ and (b) $(\kappa_Z,T_Z/J)=(0.104(8),0.325(6))$ for $c+1=5$ (vertical dotted lines). The metastability limits, $T_\mathrm{heat}$ and $T_\mathrm{cool}$ (dash-dotted lines, top), bubble over that range, as captured by the hysteretic gap (solid line, bottom). In this regime, $T_p$ coincides with the metastability limit of the metastable phase considered: $T_\mathrm{cool}$ for the paramagnetic phase upon cooling and $T_\mathrm{heat}$ for the ferromagnetic phase upon heating.
}
\label{fig:salr_phase}
\end{figure*}

To various extents, the phase diagrams of the three models on Bethe lattices have been studied quantitatively in prior works. While for the ANNNI~\cite{yokoi1985strange,moreira1993modulated,zheng2022weakening} 
and the 
RBIM~\cite{dotsenko1994introduction,nishimori_statistical_2001} models detailed and accurate results exist, for the isotropic SALR interaction model the earlier analysis~\cite{charbonneau2021solution} suffers from various numerical deficiencies.  Figure~\ref{fig:annni_rbim} shows updated versions of the first two, and Fig.~\ref{fig:salr_phase} presents new results for the third. Because the RBIM results can be obtained somewhat straightforwardly and are by now fairly canonical (see Sec.~\ref{sec:rbim_phase}), the rest of this subsection mostly focuses on the SALR interaction models.

In general, using the cavity field equations, Eq.~\eqref{eq:spin_configs_iso} and Eqs.~\eqref{eq:spin_configs_a}-\eqref{eq:spin_configs_na}, phase diagrams for the SALR interaction models with various connectivity $c+1$ can be obtained. Such a brute-force approach, however, is rarely the most efficient. Various other schemes are employed instead, each tailored to a specific regime.

The P-F phase transition is determined by linear stability analysis (LSA), as described in Sec.~\ref{sec:spin_config_lsa}. In the Ising-like regime, the LSA results are (linearly) fitted to extract the temperature such that the leading eigenvalue is marginal, $\lambda_\mathrm{max}=1$, which is the onset of metastability. This approach provides $T_c$ results with markedly higher-accuracy than can be achieved by relying on the singularity in the derivatives of free energy.  This scheme can also efficiently approach $\kappa_L$, whereas the recursive equations grow numerically challenging to converge in the vicinity of the multicritical point. 

The accuracy of the LSA also brings into focus an unexpected feature of isotropic SALR interaction models with $3 \le c+1 \le 5$: In the vicinity of $\kappa_L$, both the homogeneous and heterogeneous phases are LSA (meta)stable. The transition is then first-order in nature, and determined---as described in Sec.~\ref{sec:spin_config_lsa}---by the free energy of the two phases crossing.  For $c+1=3$, for instance, the first-order transition region extends from the onset of hysteresis, ($\kappa_Z,T_Z/J) = (0.181(6), 0.141(6))$, up to the Lifshitz point. In between, a \emph{bubble} of metastability opens (see Fig.~\ref{fig:salr_phase} insets)~\footnote{The higher precision LSA results differ from those of Ref.~\onlinecite{charbonneau2021solution}, which had inaccurately found the two branches to grow monotonically more distant upon approaching $\kappa_L$; the connectivity range for the first-order transition regime is here also slightly wider than what was then reported.}. 

The paramagnetic-to-modulated phase transition is inaccessible by LSA because the configuration probability for the paramagnetic phase has an eigenvalue always smaller than unity, and therefore presents it as stable even when it is not. It is instead obtained from the strategy presented in Sec.~V of Ref.~\onlinecite{charbonneau2021solution} to account for the incommensurability of the finite-temperature modulated phase. Determining the finite-$T$ ferromagnetic-to-modulated transition would require properly sampling the modulated phase regime, and thus surmounting the same commensurability difficulties as for the paramagnetic-to-modulated phase transition. 
Because this regime is of limited interest for the current work and is numerically challenging to determine for models on Bethe lattices~\cite{moreira1993modulated,yokoi1985strange,charbonneau2021solution}, we here approximate it by linearly interpolating between the Lifshitz point and the $T=0$ result for the transition: $\kappa_0 = 1/(2c)$ for the isotropic case~\cite{charbonneau2021solution}, and $\kappa_0=1/2$ for the ANNNI case. For the latter, however, the modulated-side approach to the tricritical point  is then demonstrably unphysical, because all transition lines are expected to meet smoothly at that point~\cite{yokoi1985strange,moreira1993modulated,nascimento2014modulated,bak1980ising,selke1984mean,da1986ising,zhang2010monte,da2013time}.

Given this phenomenology, the Lifshitz point is identified via two different schemes. (i) For the isotropic SALR interaction model, it appears at the end point of the first-order region, where the hysteretic bubble closes (see Fig.~\ref{fig:salr_phase}): $(\kappa_L, T_L/J) = (0.2500(1),0.0000(1)$ for $c+1=3$, and $(0.113(1),0.267(7))$ for $c+1=5$. (For the former, the transition is indistinguishable from the $T=0$ change in ground state from ferromagnetic to modulated at $\kappa=1/4$, but a finite-$T$ Lifshitz point cannot be excluded either.)
(ii) For the ANNNI model with $c+1=6$, the Lifshitz point appears at the crossing point of the two second-order transition branches by (slightly) extrapolating them using a linear fit to neighboring points: $(\kappa_L, T_L/J) = (0.34(1), 1.02(2))$.

Interestingly, the putative existence of a first-order P-F transition regime and of an associated tricritical point at $T_Z$ has recently been debated in the context of the isotropic SALR interaction model on a two-dimensional honeycomb lattice~\cite{Bobak2016,Zukovic2021,Gessert2025}. The latest numerical estimate for this particular model finds that metastable states impede equilibration upon approaching the zero-temperature P-F phase transition, as is observed here for the same model on the Bethe lattice with $c+1=3$. Such atypical behavior might also underlie some of the challenges associated with determining the phase diagram of $d=2$ systems using transfer matrices in this same regime (see Sec.~\ref{sec:models})~\cite{hu2021numerical}. In any event, the physical origin of this phenomenon remains unclear. How could an Ising-like model give rise to a first-order P-F transition? The existing field-theoretic description identifies no such regime~\cite{Tarzia2006}, but some insight might be gained from a Bethe $M$-layer expansion around the infinite-connectivity limit, $c\rightarrow\infty$, as has recently been done for the standard Ising model~\cite{Altieri2017,Angelini2024}. However, given the technical difficulty involved in pursuing this calculation it is left as future work. 

\subsection{Percolation thresholds}
As introduced in previous sections and also discussed in Ref.~\cite{zheng2022weakening}, the generalized FK--CK cluster scheme is the thermodynamically appropriate description of any non-frustrated systems. Reference~\cite{zheng2022weakening} further suggests that the scheme can be generalized to frustrated models using a ``negative probability'' for bonding. In this subsection, we examine the generality of this proposal for both SALR interaction models and the frustrated RBIM.

\subsubsection{SALR models}
For SALR interaction models, we calculate $T_p$ from the percolation probabilities (as described in Sec.~\ref{sec:iso_nnn} and Sec.~\ref{sec:annni_nnn}). Because this approach relies strongly on the sampled cavity configuration probabilities, the homogeneous fixed point (as a forced solution) cannot be used to solve for $T_p$, and convergence problems are encountered upon approaching the transition point. The threshold is therefore identified as the point for which the percolation probability becomes nonzero (within a numerical tolerance of $10^{-4}$). Figures~\ref{fig:annni_rbim} and \ref{fig:salr_phase} show that the results for $T_p$ coincide with $T_c$ in the Ising-like regime. By contrast, in the first-order regime $\kappa\in (\kappa_Z,\kappa_L)$, the percolation temperature depends on the metastable branch considered. For the heating branch, the calculated transition is nearly coincident with the end of the metastability branch at $T_\mathrm{heat}$, and the opposite for the cooling branch. In the vicinity of $\kappa_L$, because numerical convergence is particularly challenging the putative correspondence between percolation and phase transition beyond the Lifshitz point, $\kappa>\kappa_L$, cannot be assessed directly. This difficulty, however, may reflect the physical impossibility of such correspondence. We get back to this point in the discussion of the RBIM below as well as in the conclusion, Sec.~\ref{sec:conclusion}. 

We next consider the percolation of the $\alpha$-parameter cluster model (defined in Sec.~\ref{sec:iso_revCK}). As discussed above, defining FK--CK clusters with negative probabilities is unphysical and hence cannot be constructively realized. Reducing the bonding probability between parallel nearest-neighbor spins---as in the $\alpha$-parameter cluster model---might then be leveraged to sidestep this issue in an effective sense. One simply needs to tune $\alpha$ using a simple search, until $T_p(\alpha) = T_c$ (within a numerical tolerance of $10^{-4}$).

\subsubsection{Frustrated RBIM model}
\begin{figure}
    \centering
    \includegraphics[width=\linewidth]{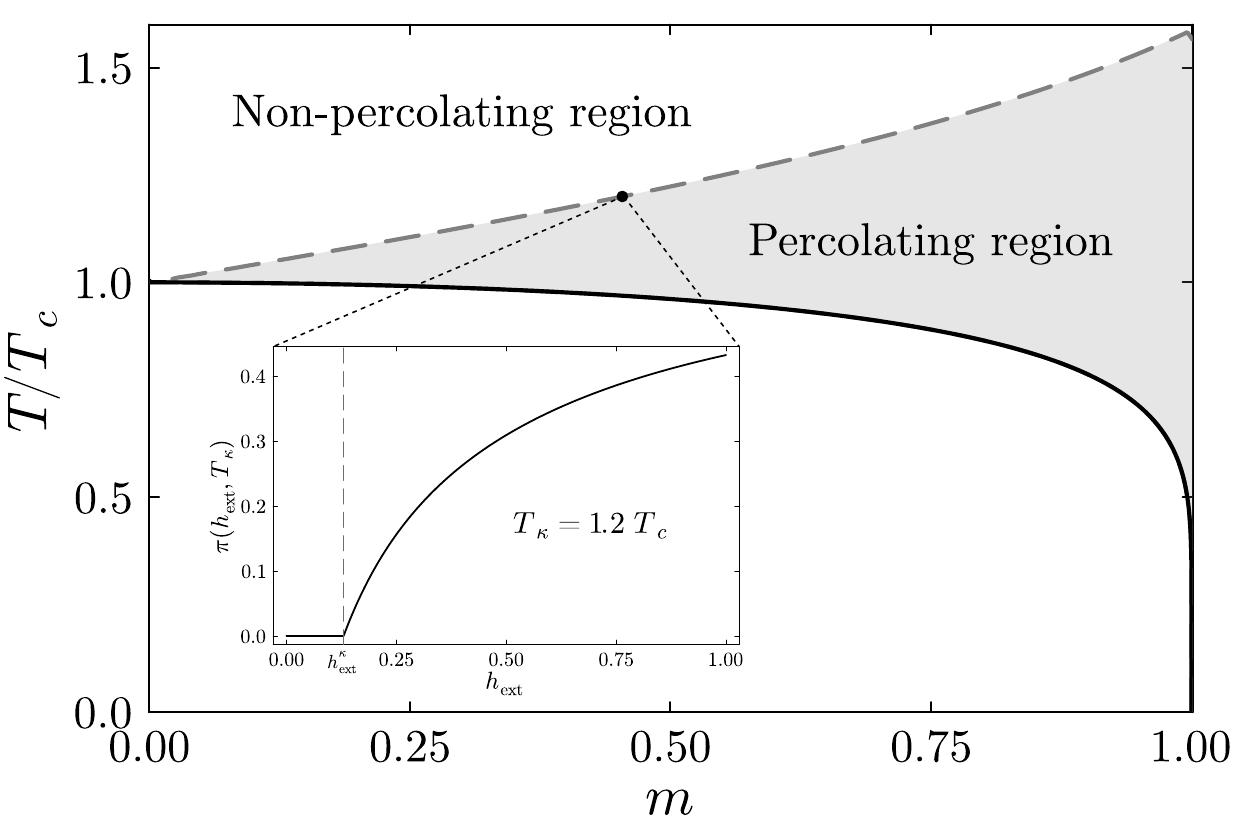}
    \caption{Percolation properties of the pure Ising model ($\rho = 1$) on the Bethe lattice with $c+1=3$. Kert\'esz line (dashed line) along with the reversed representation of the magnetization density with (normalized) temperature (solid line). In the Ising model the Kert\'esz line converges to the Ising critical point $T_c$ at $h_{\rm ext} = 0$, indicative of the matching between percolation and thermodynamic transitions. The inset shows the cavity percolation probability at $T_\kappa = 1.2\,T_c$.
    The onset $h^\kappa_{\rm ext}=0.1299(1)$, at which the curve detaches from zero, corresponds to a point on the Kert\'esz line.}\label{fig:ExampleKerteszPoint}
\end{figure}

For the frustrated RBIM---as for the models with SALR interactions---the generalized FK--CK clusters percolate at the critical transition for the Ising-like P-F transition regime (see Fig.~\ref{fig:annni_rbim}). This result therefore lends further credence to the claim that clusters defined with a ``negative probability'' for bonding capture the proper spin-spin correlations. 

Because tuning $h_\mathrm{ext}$ for the frustrated RBIM is not as algebraically onerous as for the SALR interaction models, additional possibilities also open up for examining the percolation-criticality correspondence. Although the presence of an external magnetic field renders the free energy density analytic, which means that no true thermodynamic transition occurs at $h_{\rm ext} \neq 0$, a percolation transition for FK--CK clusters can  still be defined. For a given temperature $T_\kappa$, there indeed exists an associated external magnetic field $h_{\rm ext}^\kappa$ for which the percolation probability starts to deviate from zero, and hence FK--CK clusters percolate. 

Figure~\ref{fig:ExampleKerteszPoint} illustrates the effect for the pure Ising model ($\rho=1$) on a Bethe lattice with $c+1=3$ at $T_\kappa = 1.2~T_c$, where $T_c$ the Ising critical temperature given by Eq.~\eqref{eq:critical_temps_RBIM}. The onset of percolation then takes place at $h_{\rm ext}^\kappa \simeq 0.1299(1)$. Repeating this procedure for all $T_\kappa \geq T_c$ identifies the points $(h^\kappa_{\rm ext}, T_{\kappa})$ that define the Kert\'esz line~\cite{Kertsz1989}, which identifies the percolation threshold for the FK--CK clusters in presence of an  external field. Alternatively, the external field can be parametrized in terms of the magnetization it induces, thus providing the Kert\'esz line in the $(m, T)$ plane along with the spontaneous magnetization at zero field. In this case, the Kert\'esz line separates a non-percolating region with high temperature and weak spin correlations, from a percolating region at low temperatures and spins correlated by the external field, thus resulting in the presence of a spanning cluster. By construction, in the zero external field limit the Kert\'esz line coincides with the Ising critical point, $(h_{\rm ext}^\kappa=0, T_{\kappa}=T_c)$. Therefore, the thermodynamic and percolating transitions then coincide. 

Note that the Kertész line does not have a direct physical interpretation in terms of clusters, because the CK bonding probability is not strictly valid when $h_{\rm ext} \neq 0$. In fact, there are two symmetric Kertész lines in the phase diagram. One is associated with parallel up spins, the other with parallel down spins, corresponding to positive and negative magnetic fields, respectively. These two lines merge at the critical point $(T_c, h_{\rm ext} = 0)$. Formally, it can be shown that in the presence of an external magnetic field, the bonding probability must be modified to account for the field---for instance, by introducing a ghost spin~\cite{Stauffer1994-qu, coniglio1989exact}. Tracing the Kertész line is nevertheless useful for our purpose as it highlights the fact that it terminates precisely at the critical point. This, in turn, implies that at $h_{\rm ext} = 0$, the clusters defined using the CK bonding probability percolate exactly at criticality.

It is important to note that the points lying within the region bounded by the spontaneous magnetization curve in Fig.~\ref{fig:ExampleKerteszPoint} are not physically realizable. If a system is prepared at a state $(m, T)$ within this out-of-equilibrium region, it will undergo phase separation and relax to one of the two equilibrium magnetizations at temperature $T$. Owing to the symmetry of the curve under $m \to -m$, these two stable magnetization values are equal in magnitude and opposite in sign.

\begin{figure}[h!]
    \centering
    \includegraphics[width=\linewidth]{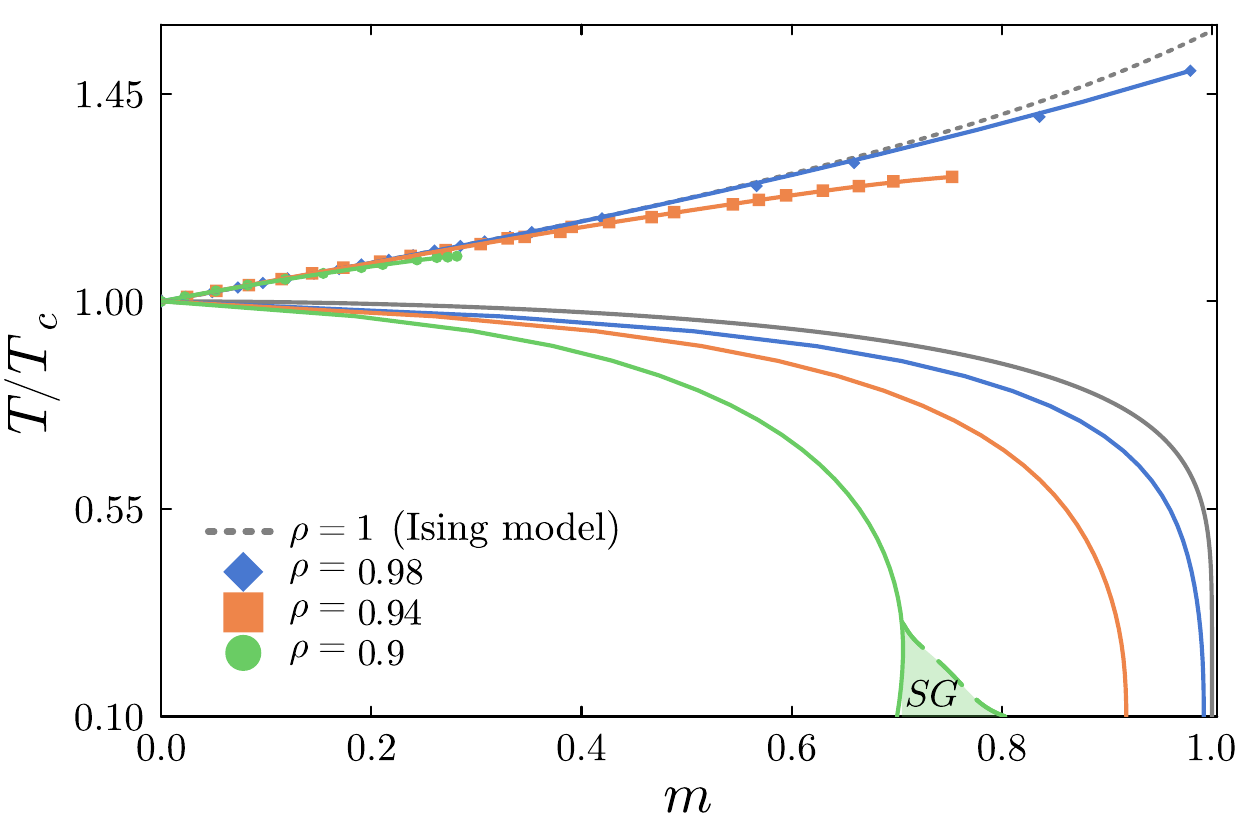}
    \caption{Kert\'esz lines (points; lines are guides for the eye) and spontaneous magnetization (solid lines) for the frustrated RBIM on the 
    Bethe lattice with $c+1=3$ for the temperature in the vertical axis normalized by the critical value at the corresponding $\rho$, Eq.~\eqref{eq:critical_temps_RBIM}. For $\rho = 0.9$ and at low temperatures, the spin-glass phase (SG) reappears due to the presence of the de Almeida--Thouless line.
    }
    \label{fig:KL-RBIM}
\end{figure}

Figure~\ref{fig:KL-RBIM} shows the Kert\'esz line in the frustrated RBIM for various $\rho \leq 1$. In the (Ising-like) P-F transition regime, $\rho > \rho_\star$, these lines clearly overlap upon approaching the critical temperature, nicely converging towards their respective $T_{c}$. The lines, however, shorten as $\rho$ decreases. In addition to their shortening due to magnetization decreasing---as the fraction of anti-ferromagnetic bonds grows---we note an unexpected systematic shortening as $\rho$ decreases. 

The numerical convergence of the percolation probability computed using the FK--CK cluster definition becomes increasingly unstable upon approaching the multicritical point $(T_\star, \rho_\star)$, where a different order---here, the spin-glass phase---emerges. An analogous loss of stability is observed in SALR interaction models near the Lifshitz point, suggesting that the onset of a competing order may hinder the stability of the random-cluster percolation equations. We get back to this point in the conclusion, Sec.~\ref{sec:conclusion}.

In the phase diagram at zero external field, $h_{\mathrm{ext}}=0$, a ferromagnetic spin-glass (FSG) phase appears for $\rho > \rho_\star$ (see Fig.~\ref{fig:annni_rbim}). This phase, characterized by a finite magnetization $m \neq 0$ and a nonzero Edwards--Anderson parameter $q_{\rm EA}>0$, persists up to $\rho \simeq 0.916$. The spontaneous magnetization curve at $\rho=0.9$ clearly displays the onset of this FSG phase at very low temperatures (Fig.~\ref{fig:KL-RBIM}). In the presence of an external field, the spin-glass order extends up to the de Almeida--Thouless line~\cite{Almeida1978, Matsuda2010, Pagnani2003}, smoothly connecting to the zero-field FSG region that emerges directly along the spontaneous magnetization curve. This spin-glass phase is accompanied by a reentrant behavior in the spontaneous magnetization curve. However, one should keep in mind that our results for $m$ at zero field are obtained within the replica-symmetric (RS) approximation. Because the FSG phase strictly requires the full replica symmetry breaking (RSB) solution, it remains unclear whether this reentrance is a genuine feature of the model or merely an artifact of the RS approximation.

The $\alpha$-parameter cluster model for the frustrated RBIM is also considered. For dilution $\rho = 0.9$, for instance, we find $\alpha \simeq 0.7845(7)$. 

\subsection{Percolation universality class}
In addition to the correspondence between percolation and thermodynamic critical points, the critical clusters are expected to belong to the Ising universality class. To test this hypothesis, we extract the Ising critical exponents from the FK--CK and the $\alpha$-parameter cluster model percolation probabilities. 

For $h_\mathrm{ext}=0$ near $T_c$, cluster percolation arises from spontaneous magnetization. As a result, the percolation probabilities on a Bethe lattice are expected to scale critically,
\begin{equation}
    P \propto |T - T_c|^{\beta},
    \label{eq:Pbeta}
\end{equation} 
with the Ising mean-field critical exponent $\beta =1/2$ (instead of $\beta=1$ for standard percolation~\cite{Stauffer1994-qu}). For the frustrated RBIM, we can further extract the critical exponent associated with the external magnetic field at $T_c$. The percolation probability is then expected to scale critically, 
\begin{equation}
    P \propto |h_{\rm ext}|^{1/\delta},
    \label{eq:Pdelta}
\end{equation} 
with the Ising mean-field critical exponent $\delta = 3$. 

\begin{figure}[t]
\centering
\includegraphics[width=\linewidth]{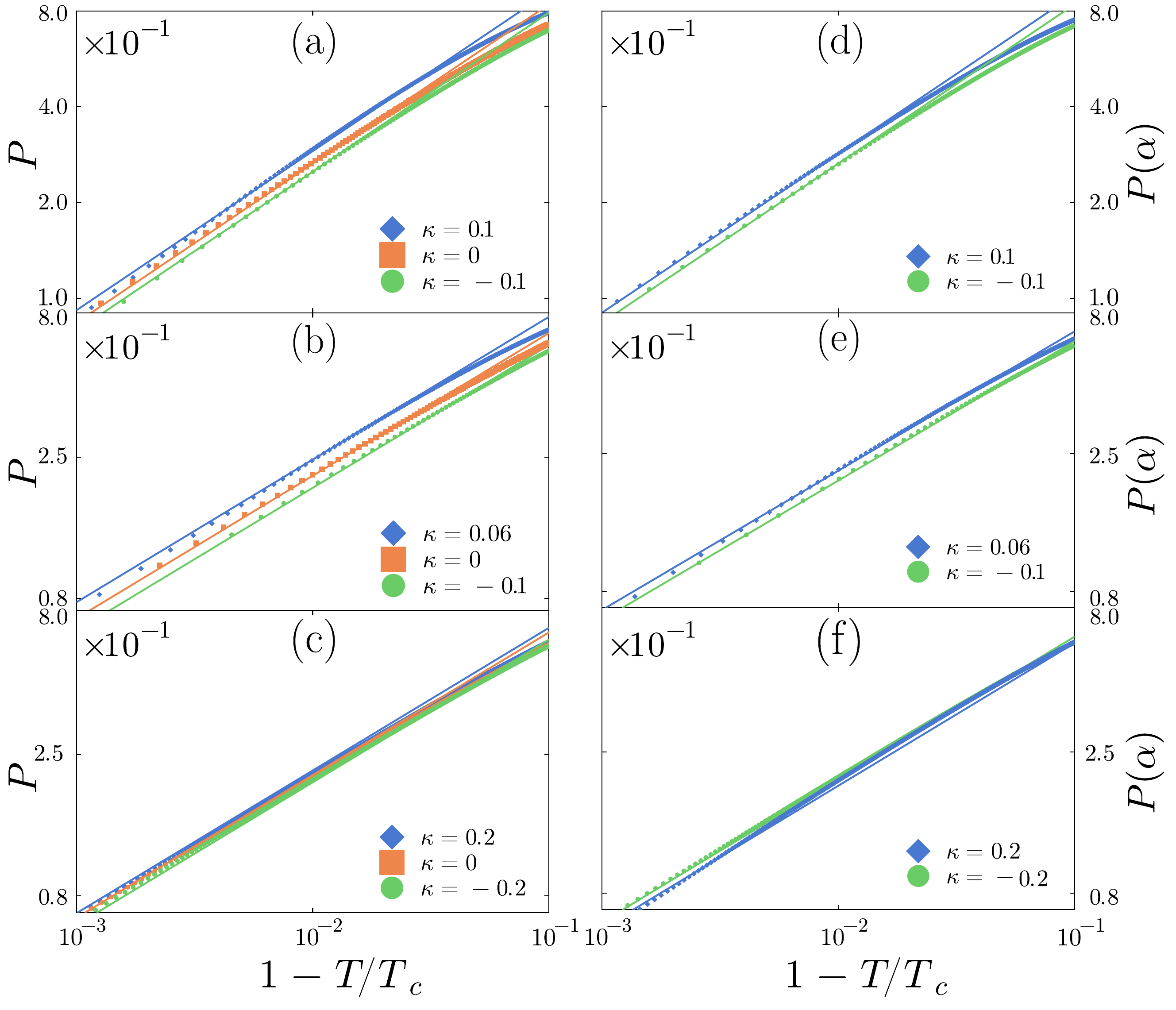}
\caption{Percolation probability of (left) generalized FK--CK clusters and (right) $\alpha$-parameter cluster model for the (frustrated and unfrustrated) isotropic model with (a)(d) $c+1=3$, (b)(e) $c+1=5$ and (c)(f) the ANNNI model with $c+1=6$. In all cases, upon approaching $T_c$ the critical scaling follows the mean-field Ising universality class with $\beta=1/2$ (left: solid lines; right: dash-dotted lines)}
\label{fig:universality_salr}
\end{figure}

\begin{figure}[t]
\centering
\includegraphics[width=\linewidth]{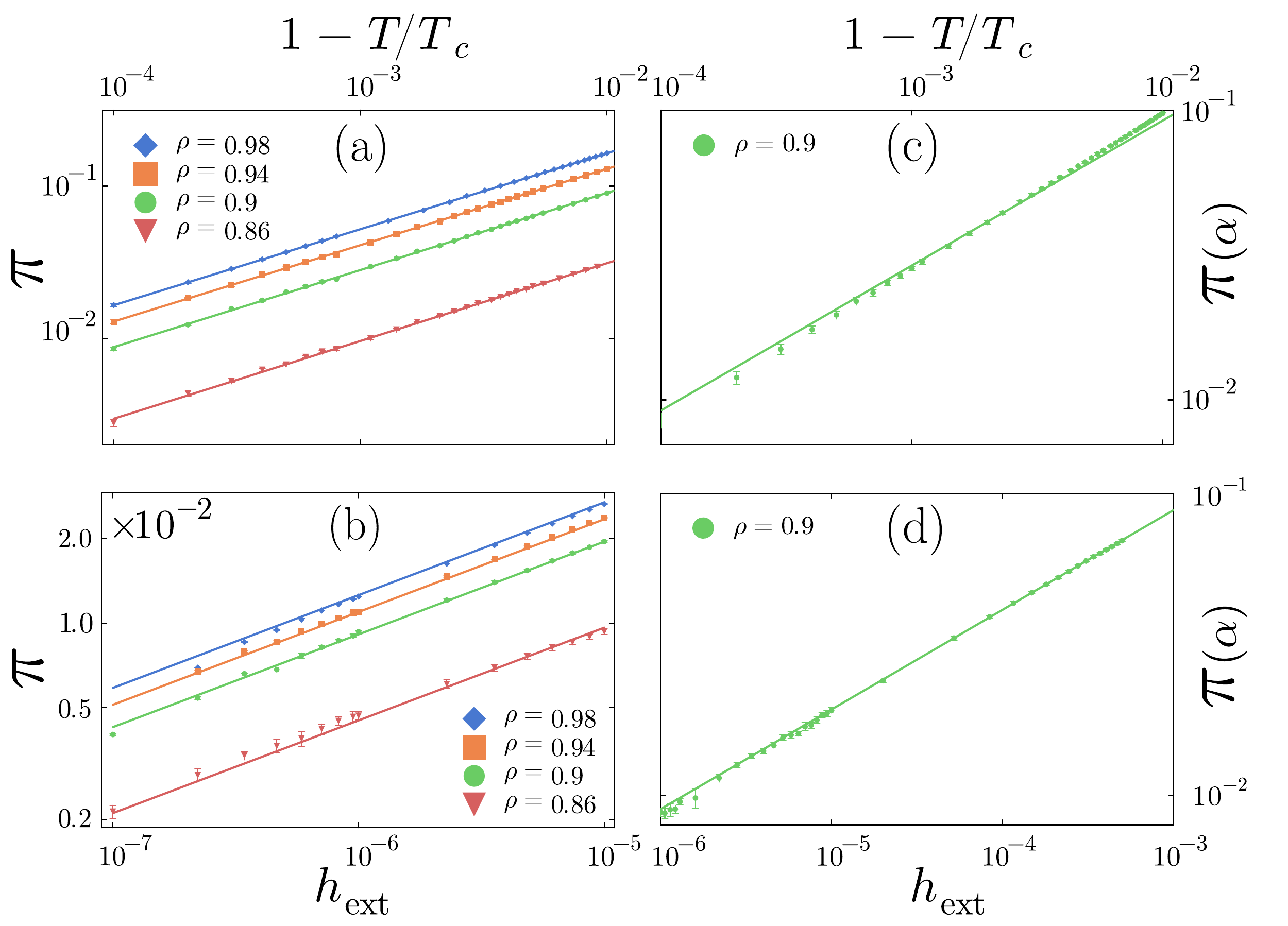}
\caption{Average cavity percolation probability of (left) FK--CK clusters (right) $\alpha$-parameter cluster model for the frustrated RBIM as a function of (a)(c) the distance to the critical temperature at zero field and (b)(d) the external field at the critical temperature $T_c$ of each respective $\rho$. In all cases, upon approaching $T_c$ or $h_{\rm ext} = 0$ the critical scaling follows the mean-field Ising universality class with $\beta=1/2$ and $\delta = 3$, respectively.}
\label{fig:universality_rbim}
\end{figure}

Figure~\ref{fig:universality_salr} considers the percolation  probability of generalized FK--CK clusters for various SALR interaction models as well as for the $\alpha$-parameter cluster model on the ANNNI model. Figure~\ref{fig:universality_rbim} does the same for the cavity percolation probability---which has the same critical scaling as the percolation probability $P$ (see Sec.~\ref{sec:FKCK-RBIM})---for both $h_\mathrm{ext}=0$ and $h_\mathrm{ext}>0$ for the frustrated RBIM. The former is obtained by the cavity method as in Secs.~\ref{sec:iso_nnn} and \ref{sec:annni_nnn}, while the latter is computed through the population dynamics algorithm, also used in the calculation of the K\'ertesz line in Fig.~\ref{fig:KL-RBIM}. In this case we have solved the percolation probability defined in Eq.~\eqref{eq:Probc2}, along with its corresponding expression for the $\alpha$-parameter cluster model, through iterations. Specifically, we evolved populations of percolation probabilities with sizes ranging from $\mathcal{M} = 6 \times 10^6$ to $\mathcal{M} = 5 \times 10^7$, iterating them for approximately $\mathcal{N} \simeq 5 \times 10^4$ steps to obtain multiple converged samples. 

In the Ising-like regime, the mean-field Ising scaling is observed for all cases. Surprisingly, the $\alpha$-parameter cluster model exhibits the same scaling behavior as the generalized FK--CK clusters. This suggests that, with sufficient numerical precision, it is possible to define positive bond probabilities that can be tuned to yield an effective model with the same critical temperature and critical exponents as the associated thermodynamic transition. Remember, however, that the clusters generated by the $\alpha$-parameter cluster model fail to reproduce the spin–spin correlations required to identify genuine critical clusters. This limitation arises because introducing a non-zero $\alpha$ necessarily modifies the expressions for the correlations $\langle \gamma_{ij} \rangle_W$, introducing an additional term that causes them to deviate from the actual spin–spin correlations $\langle s_i s_j \rangle$. In other words, the Ising critical scaling of the clustering probability is a necessary but insufficient condition for identifying the thermodynamically relevant clusters. 

\section{Conclusion}
\label{sec:conclusion}
In this work, we have generalized the definition of the random FK--CK clusters, which capture the physical correlations associated with the ferromagnetic critical point in the simple Ising model, to frustrated models with negative couplings, thus extending the result of Ref.~\cite{zheng2022weakening}. We first presented a more formal derivation of the cluster construction, extending the proof of Ref.~\cite{coniglio1980clusters} to systems with antiferromagnetic frustration. We then implemented the generalized FK--CK scheme for three exactly solvable models on the Bethe lattice (namely, the isotropic and anisotropic SALR interaction models as well as the frustrated RBIM), in which the clusters can be constructed by analytically continuing certain bond probabilities to negative values. In particular, we have explicitly verified that, near the line of P-F critical points, the critical properties of the clusters fall within the (mean-field) Ising universality class. 

In a sense, this paper provides a \emph{negative} result. Because the statistical weight of certain cluster configurations becomes negative in models with frustration, configurational sampling based on FK--CK clusters to accelerate the critical dynamics is clearly unfeasible.  Although these clusters properly encode thermodynamic correlations, they cannot be constructed in the first place. Our results further rule out numerous alternative proposals that have been put forward in recent decades. In so doing, this work settles a long-standing debate on the correct definition of clusters and discourages further pursuit of research directions that ultimately lead to dead ends. A similar evaluation of cluster schemes for other phase transitions could further dampen enthusiasm for the approach~\cite{Alfaro2025}.

The percolation–criticality correspondence that we have established exactly on Bethe lattices is expected to extend to finite-dimensional, more physical lattice geometries~\cite{Gomez2025}, such as square and cubic lattices in $d=2$ and $d=3$. Importantly, this correspondence is already rigorously confirmed in $d=1$ for the isotropic SALR interaction model (Sec.~\ref{sec:iso_nnn}). In that case, the correspondence holds for arbitrary temperatures and system sizes. Because the same picture emerges both in $d=1$---where fluctuations are maximal---and at the mean-field level represented by the Bethe lattice, there is no indication that fluctuations in intermediate dimensions would qualitatively destroy the correspondence. 
The structural relation between percolation and thermodynamic criticality is therefore expected to remain intact in all $d$, including $d=2$ and $d=3$. Future work might consider these two cases explicitly: in $d=2$ via transfer-matrix calculations~\cite{hu2021resolving}, albeit with size limitations due to computational cost, and in $d=3$ by examining whether the signed-probability formulation developed here could be embedded into a renormalization-group (RG) framework.

From a different viewpoint, this paper offers some hope for cluster-based sampling schemes. The fact that physically relevant clusters cannot be generated through standard constructive schemes does not necessarily mean that they cannot be generated {\it at all}. Alternative generation methods remain possible. An appealing prospect entails the use of AI-based generative models or other machine learning-based schemes to \emph{learn rather than construct} clusters. We emphasize, however, that this possibility remains highly speculative at present. ``Learning clusters'' would entail training generative models---such as normalizing flows~\cite{Gabri2022, Albergo2022}, diffusion models~\cite{Biroli2023}, or autoregressive neural networks~\cite{Biazzo2023, Biazzo2024}---to sample cluster configurations directly from the appropriate probability distribution, thereby bypassing the need for constructive cluster algorithms altogether. Recent advances in machine learning-based sampling of complex probability distributions in statistical physics~\cite{DelBono2025, DelBono2025b,DelBono2026} suggest this approach may be feasible, but significant technical challenges remain to be overcome~\cite{Ciarella2023}. In this context, exactly solvable models on Bethe lattices offer particularly interesting benchmarks. 

Even in this speculative scenario, however, there is a---somewhat more subtle---hurdle to consider. Our work demonstrates that critical clusters may only form between parallel spins. However, in systems with antiferromagnetic couplings, configurations with parallel spins connected by such couplings are strongly suppressed by their Boltzmann weight at low temperatures, and hence contribute little to spin–spin correlation functions. Nevertheless, these same configurations may contribute significantly to cluster correlation functions. If the bond is present, the cluster weight may be negative; if the bond is absent, the ``probability'' of bonding may exceed unity, thus corresponding to a large weight. These configurations therefore cannot be neglected in the cluster representation. In other words, correctly generating physical clusters then requires sampling rare spin configurations with very low statistical weight.

To illustrate this phenomenon, consider a simple $d=1$ chain of four spins with three ferromagnetic bonds $J > 0$, and a single antiferromagnetic coupling, $-J$, in the low-temperature limit $\beta \to \infty$, i.e., near the zero-temperature critical point of the extended $d=1$ chain. 
    \begin{figure*}[htb]
        \centering
    \includegraphics[width=\textwidth]{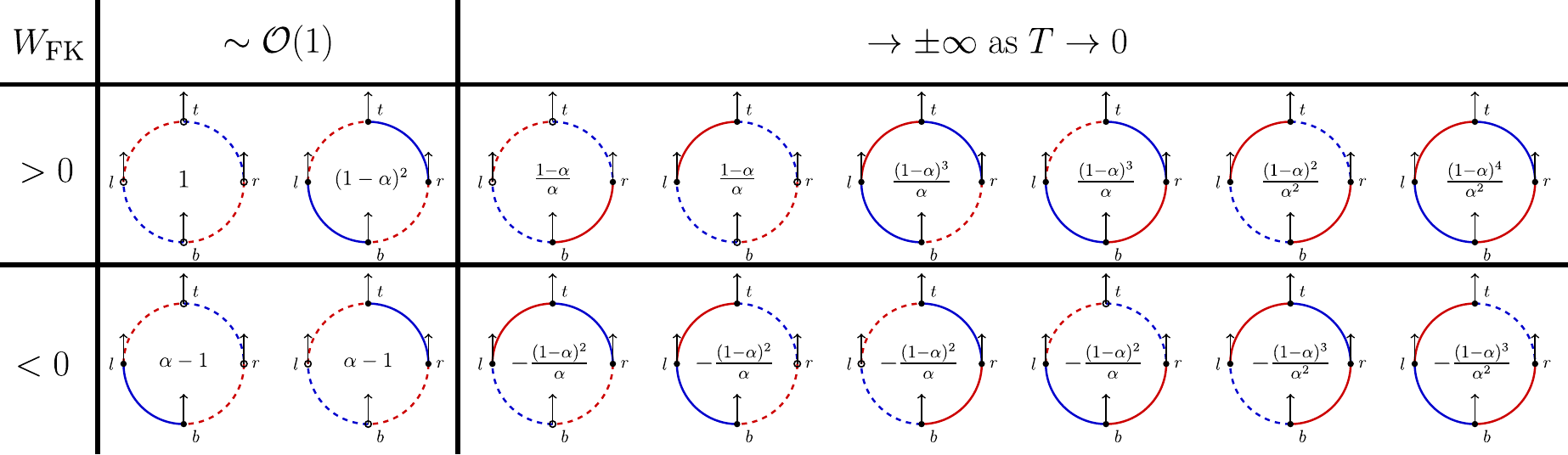}
        \caption{Possible clusters for the spin chain configuration $(s_t, s_r,s_b,s_l)=(\uparrow, \uparrow, \uparrow, \uparrow)$, with periodic boundary conditions, and fixed realization of the random bonds $(J_{tr} = -J, J_{rb} = +J, J_{bl} = -J, J_{lt} = +J)$. Ferromagnetic bonds, $J>0$, are colored red while anti-ferromagnetic bonds, $-J$, are colored blue. Spins have a filled black circle at their tails if they belong to a random cluster, and the circle is left unfilled otherwise. Despite the associated Boltzmann weight being vanishingly small at low temperatures, the possible clusters reveal that non-negligible contributions to the spin-spin correlation functions arise from the weights of the clusters $W_{\rm FK}$ in the correlated percolation problem. These weights are given in the center of their respective cluster, with $\alpha = e^{-2 \beta J}$.
       Although some contributions are of order one (first column), most contributions grow exceedingly large---in absolute value---as temperature decreases (second column). However, the positive and negative contributions, which appear in matching numbers, effectively balance out to yield the physical thermodynamic weight of this fully ordered spin configuration.}
        \label{fig:cluster-configurations}
    \end{figure*}
Spin configurations in which the two spins connected by the antiferromagnetic bond are parallel are then strongly suppressed by a factor $e^{-2\beta J}$. Nevertheless, certain cluster configurations associated with these suppressed spin states contribute significantly to the cluster correlation functions---as can be seen in Fig.~\ref{fig:cluster-configurations}---because the cluster weights $W_{\rm FK}$ explode in the limit $\beta \to \infty$.

The relationship between physical and geometrical correlations may therefore be more intricate in the presence of frustration than in unfrustrated systems. Going back to the model on the Bethe lattice, as the system approaches a different kind of order---one that deviates from the standard ferromagnetic phase---we observed that the cavity equations governing percolation probabilities fail to converge. For instance, near the Lifshitz point in SALR interaction models or close to the multicritical point for the frustrated RBIM, we are unable to find a fixed point of the recursion.

This difficulty also arises far from the critical point. For instance, it is observed at higher temperatures and stronger magnetic fields in the RBIM, as evidenced by the shortening of the K\'ertesz line for values of $\rho$ near the multicritical point $\rho_\star$. In these regions, the system approaches the low-temperature spin-glass phase that emerges along the coexistence line. These observations suggest that the appearance of an additional local minimum in the free-energy landscape---associated with a competing form of order---undermines the stability of the equations governing the percolation of physical clusters. This effect might therefore limit the regime over which relevant clusters can be identified and play a significant physical role.

Understanding whether a deeper conceptual mechanism underlies this phenomenon therefore remains an important direction for future research.

\paragraph*{Acknowledgments.}
We acknowledge partial support from the Simons Foundation (Grant No. 454937), 
the National Science Foundation (Grant No. DMR-1749374), and the ANR-24-CE30-5851 ManyBodyNet grant. Computations were carried out on the Duke Compute Cluster and on the SACADO MeSU platform at Sorbonne Universit\'e.

\paragraph*{Data availability statement.}
The data that support the findings of this article are openly available~\cite{RDR}.

\appendix

\section{Percolation and stability analysis of the $\alpha$-parameter cluster model}
\label{sec:appendix_iso_revCK}

In this appendix, we obtain a few analytical results for the small $\kappa$ regime of the $\alpha$-parameter cluster model defined by Eq.~\eqref{eq:alphaparameterdef}. In order to determine $T_p$ for this model, we follow the strategy of Ref.~\onlinecite{charbonneau2021solution} for the cavity method, designing recursive relations to calculate the \textit{percolation probability} that a spin belongs to the percolating cluster $C_\infty$. For simplicity, consider clusters with all up spins. (The results for clusters of down spins are the same by symmetry.) We define the probabilities that a spin points up and belongs to $C_\infty$ in terms of the cavity field
\begin{align}
\label{eq:ck_pr_def_appendix}
    \pi_R &\equiv \textrm{Pr} (s_i=\; \uparrow \wedge \; s_j= \; \uparrow \wedge\, i \in {\cal C}_\infty) \\
    q_R &\equiv \textrm{Pr}(s_i= \; \uparrow \wedge \; s_j= \; \uparrow \wedge\, i \not\in {\cal C}_\infty)
\end{align}
where $\pi_R + q_R = R$. Taking the bonding probability into account, we define the auxiliary quantities
\begin{equation}
\label{eq:revised_pr_appendix}
        \hat{\pi} = \pi_R \, p_{B}^\mathrm{rev}, \qquad \hat{q} = q_R + (1-p_{B}^\mathrm{rev})\pi_R.
\end{equation}
which denote the probability that an up spin is connected or not to the parallel and percolated neighbor spin by a bond, respectively.

Recursive equations can then be obtained,
\begin{widetext}
\begin{equation}
\label{eq:ck_pr_appendix}
    \pi_R = Z_R^{-1} e^{\beta h} \sum\limits_{l=1}^{c} \binom{c}{l}e^{\beta J(2l-c)} e^{-\beta \kappa J(2l-c)} F^{c-l} \sum\limits_{k=1}^{l}\binom{l}{k}\hat{\pi}^k \, \hat{q}^{l-k} e^{-\beta \kappa J\frac{l(l-1)+(c+1-l)(c-l)-2l(c+1-l)}{2}},
\end{equation}
\begin{equation}
\label{eq:ck_qr}
    q_R = Z_R^{-1} e^{\beta h} \sum\limits_{l=0}^{c} \binom{c}{l}e^{\beta J(2l-c)} e^{-\beta \kappa J(2l-c)} F^{c-l} \hat{q}^l e^{-\beta \kappa J\frac{l(l-1)+(c+1-l)(c-l)-2l(c+1-l)}{2}},
\end{equation}
\end{widetext}
where  $Z_R$ is such that $\pi_R + q_R = R$.

Given converged $\pi_R$ and $q_R$, the percolation probability is
\begin{eqnarray}
\label{eq:salr_nnn_p}
    P &= Z_\mathrm{site}^{-1} \sum\limits_{l=1}^{c+1} \binom{c+1}{l}e^{\beta\left[\mu+l(\varepsilon-\frac{l-1}{2}\kappa)\right]} \\
    &\times F^{c+1-l} \sum\limits_{k=1}^{l}\binom{l}{k}\hat{\pi}^k \, \hat{q}^{l-k},
    \end{eqnarray}
    \begin{eqnarray}
    Q &= Z_\mathrm{site}^{-1} \sum\limits_{l=0}^{c+1} \binom{c+1}{l}e^{\beta\left[\mu+l(\varepsilon-\frac{l-1}{2}\kappa)\right]} F^{c+1-l} \hat{q}^l,
\end{eqnarray}
where $Z_\mathrm{site}$ is such that $P + Q = \rho$.

In order to gain some physical intuition for $\alpha$, we first consider the small $\kappa$ regime. To linear order, we expect $\alpha = 1 - f(c)\kappa$, with $f(c) > 0$. To obtain $f(c)$, we consider the linear stability of the CK percolation probability, which has leading eigenvalue
\begin{align}
\label{eq:spin_revck_eigs}
    \lambda &= Z_R^{-1} e^{\beta h} \, p_B^\mathrm{rev} \, \sum\limits_{l=1}^{c} \binom{c}{l}F^{c-l} \, R^{l-1} \, l \, \\
    &\times e^{\beta J (2l-c)} e^{-\beta \kappa J(2l-c)} e^{-\beta \kappa J\frac{l(l-1)+(c-l)(c-2l-1)}{2}}\nonumber.
\end{align}
For a system with a fixed connectivity $c+1$, $p_B^{\mathrm{rev}}$ depends only on $\beta$ and $\kappa J$. Given the configuration probabilities at leading order in $\kappa$ (see Sec.~\ref{sec:spin_config_ls}), we can rewrite Eq.~\eqref{eq:spin_revck_eigs} as
\begin{equation}
\label{eq:spin_revck_eigs2}
\lambda = \mathcal{F}(\beta, \kappa)\, p_B^{\mathrm{rev}}(\beta, \kappa).
\end{equation}
where $\mathcal{F}(\beta, \kappa)$ includes configuration probabilities and Boltzmann weights. At the critical (or percolation) point, we expect $p_B^{\mathrm{rev}}$ (or rather $f(c)$ in $p_B^{\mathrm{rev}}$) to be such that $\lambda = 1$, and the resulting $\partial_{\kappa J} \beta$ equal to that of the linear stability analysis of the configuration probabilities (see Sec.~\ref{sec:spin_config_ls}). We can therefore solve for the derivative of Eq.~\eqref{eq:spin_revck_eigs2} about $\kappa$ around the Ising transition point $(\kappa_0,\beta_0) = (0,\frac{1}{2}\ln{(\frac{c+1}{c-1})})$,
\begin{align}
\label{eq:Tp_slope}
    0 &= (\frac{\partial \mathcal{F}}{\partial \beta} \, \frac{\partial\beta}{\partial \kappa J}+ 
    \frac{\partial \mathcal{F}}{\partial \kappa J})p_B^{\mathrm{rev}}
    \nonumber
    \\
    &+ \mathcal{F} \, (\frac{\partial p_B^{\mathrm{rev}}}{\partial\beta} \, 
    \frac{\partial\beta}{\partial \kappa J} + \frac{\partial p_B^{\mathrm{rev}}}{\partial \kappa J})\nonumber\\
    &= (\frac{\partial \mathcal{F}}{\partial \beta}\, \frac{\partial\beta}{\partial \kappa J} + \frac{\partial \mathcal{F}}{\partial \kappa J})p_B^{\mathrm{rev}}
    \nonumber \\
    &+ \mathcal{F}\, (2J e^{-2\beta J}\, \frac{\partial\beta}{\partial \kappa J} - e^{-2\beta J}\, 2\beta f(c)),
\end{align}
where all derivatives and variables can be evaluated at linear order in $\kappa$.

Solving for $f(c)$ from Eq.~\eqref{eq:Tp_slope} is challenging, because the expression involves summing over an increasing number of terms with respect to $c$. We here only evaluate $f(c)$ for specific $c=2, \ldots 6$ (Table~\ref{table:salr_revck}). As expected, these (exact) results lead to equal slopes, $dT_p/d\kappa=dT_c/d\kappa$, around $\kappa=0$ at linear order in $\kappa$. 
Because obtaining these values is tedious even with symbolic software, we extend their range by (empirically) inferring the generic expression, 
\begin{equation}
\label{eq:f(c)}
    f(c) = (2+c) + \frac{3c-7}{4c} = \frac{4c^2+11c-7}{4c}.
\end{equation}
Although Eq.~\eqref{eq:f(c)} is not formally derived, we nevertheless expect it to remain valid for arbitrary $c$. Checking that result, however, is left as future work.

From a physical standpoint a couple of observations are important. First, the change of $p_B^{rev}$ around $\kappa = 0$ is continuous and symmetric for attractive and repulsive next-nearest-neighbor interactions. Second, a consideration of higher-order terms in the configuration probabilities to capture its behavior at larger $\kappa$ is expected to generate similarly analytic terms in powers of $\kappa$ for $p_B^\mathrm{rev}$. (Such work is not attempted here.) The resulting expansion is therefore seemingly oblivious to the discontinuity in physical properties around the multi-critical point. 

\begin{table}[h]
\vspace{0.5cm}
\centering
\begin{tabular}{ m{2cm}<{\centering} m{2cm}<{\centering} } 
\hline
$c$ & $2f(c)$  \\
\hline
2 & $8 - 1/4$ \\ 
3 & $10 + 2/6$ \\ 
4 & $12 + 5/8$ \\ 
5 & $14 + 8/10$ \\ 
6 & $16 + 11/12$ \\ 
\hline
\end{tabular}
\caption{Values of $f(c)$ in $p_B^{\mathrm{rev}} = 1-\exp\{-2\beta J[1-f(c)\kappa]\}$ for the revised CK percolation criterion.}
\label{table:salr_revck}
\end{table}

\section{Connection probabilities of generalized FK--CK clusters for a $d=1$ chain (case $c+1=2$)}
\label{sec:appendix_tm}

In this appendix, we calculate the probability that two spins---one at site $i$ and the other at site $j$---are parallel and belong to a same cluster, i.e., are connected. If we set $i=1$ and $j=1+r$ in the $d=1$ chain, then $\langle\gamma_{ij}^\parallel\rangle$, or the distribution of the connection probability $P(1+r)$ for the spin at site $1+r$, can recursively be obtained from that of its backward site $P(r)$ through a transfer matrix $M$,
\begin{equation}
    P(1+r)=MP(r)=M^rP(1).
\end{equation}
The recursive equations for percolation, Eqs.~\eqref{eq:perc_1d_1}--\eqref{eq:perc_1d_9}, describe the connections between spins, and the transfer matrix can be written as  
\pagebreak
\begin{widetext}
\begin{table}[h]
    \centering
    \begin{tabular}{ccc}
       &  & 
      \begin{minipage}[b]{1.6\columnwidth}
      \centering
      \raisebox{-.5\height}{
      \begin{minipage}[b]{0.18\columnwidth}
      {\includegraphics[scale=0.35]{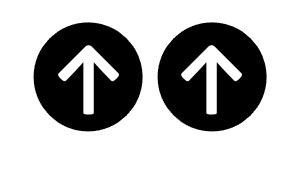}}
      \end{minipage}
      \begin{minipage}[b]{0.19\columnwidth}
      {\includegraphics[scale=0.35]{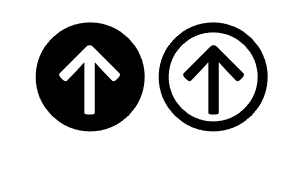}\qquad}
      \end{minipage}
      \begin{minipage}[b]{0.16\columnwidth}
      {\includegraphics[scale=0.35]{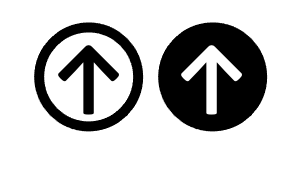}}
      \end{minipage}
      \begin{minipage}[b]{0.15\columnwidth}
      {\includegraphics[scale=0.35]{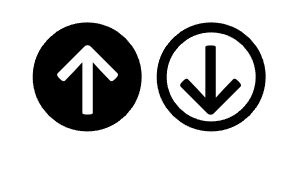}}
      \end{minipage}
      \begin{minipage}[b]{0.18\columnwidth}
      {\includegraphics[scale=0.35]{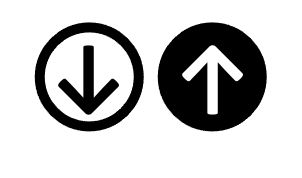}\qquad\qquad}
      \end{minipage}
      }
      \end{minipage}
      \\
     \qquad\qquad$M=$ & 
     \begin{minipage}[b]{0.1\columnwidth}
      \centering
      \raisebox{-.5\height}{
      \includegraphics[scale=0.35]{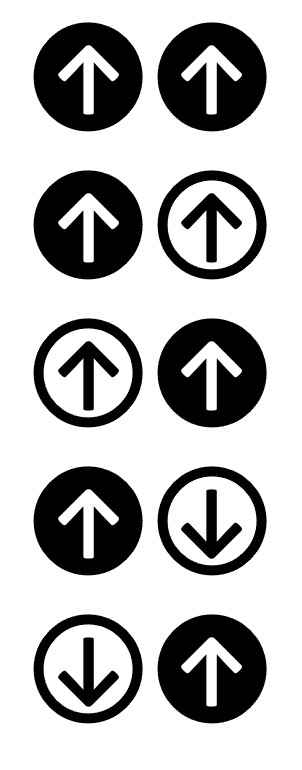}
      }
      \end{minipage}
     & 
     \begin{minipage}[b]{1.6\columnwidth}
      \centering
      \begin{equation}
      \left( \begin{array}{ccccc}
    p_2e^{\beta J(1-\kappa)} & (1-p_2)e^{\beta J(1-\kappa)} & 0 & e^{\beta J(1+\kappa)} & 0 \\
    p_1p_2e^{\beta J(1-\kappa)} & p_1(1-p_2)e^{\beta J(1-\kappa)} & (1-p_1)p_2e^{\beta J(1-\kappa)} & p_1e^{\beta J(1+\kappa)} & 0 \\
    p_1p_2e^{\beta J(1-\kappa)} & (1-p_1p_2)e^{\beta J(1-\kappa)} & 0 & e^{\beta J(1+\kappa)} & 0 \\
    0 & 0 & 0 & 0 & p_2e^{-\beta J(1+\kappa)} \\
    0 & 0 & 0 & e^{-\beta J(1+\kappa)} & 0
     \end{array} \right).
     \label{eq:NNNA_M}
    \end{equation}
      \end{minipage}
    \end{tabular}
\end{table}
\end{widetext}
Note that we neglect cases in which two consecutive spins do not belong to the same cluster as spin $1$, because the cluster then cannot grow further.

We next consider the boundary conditions at the beginning and at the end of the cluster, and construct

\begin{widetext}
\begin{table}[b!]
    \centering
    \begin{tabular}{ccc}
       &  & 
      \begin{minipage}[b]{1.4\columnwidth}
      \centering
      \raisebox{-.5\height}{
      \begin{minipage}[b]{0.22\columnwidth}
      {\quad\includegraphics[scale=0.35]{Figures_Mingyuan/matrix_configs_1d/1D_configs1.png}}
      \end{minipage}
      \begin{minipage}[b]{0.21\columnwidth}
      {\quad\includegraphics[scale=0.35]{Figures_Mingyuan/matrix_configs_1d/1D_configs2.png}\qquad}
      \end{minipage}
      \begin{minipage}[b]{0.13\columnwidth}
      {\includegraphics[scale=0.35]{Figures_Mingyuan/matrix_configs_1d/1D_configs3.png}}
      \end{minipage}
      \begin{minipage}[b]{0.14\columnwidth}
      {\includegraphics[scale=0.35]{Figures_Mingyuan/matrix_configs_1d/1D_configs4.png}}
      \end{minipage}
      \begin{minipage}[b]{0.18\columnwidth}
      {\includegraphics[scale=0.35]{Figures_Mingyuan/matrix_configs_1d/1D_configs5.png}\qquad\quad}
      \end{minipage}
      }
      \end{minipage}
      \\
     \qquad\qquad$M_0=$ & 
     \begin{minipage}[b]{0.1\columnwidth}
      \centering
      \raisebox{-.5\height}{
      \includegraphics[scale=0.34]{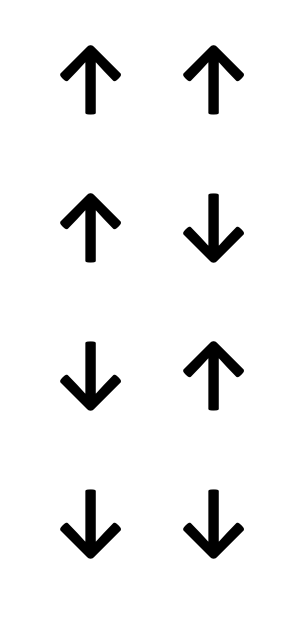}
      }
      \end{minipage}
     & 
     \begin{minipage}[b]{1.4\columnwidth}
      \centering
      \begin{equation}
      \left( \begin{array}{ccp{1.5cm}cp{1cm}}
    p_1p_2e^{\beta J(1-\kappa)} & (1-p_1p_2)e^{\beta J(1-\kappa)} & \qquad 0 & e^{\beta J(1+\kappa)} & \qquad 0 \\
    0 & 0 & \qquad 0 & 0 & \qquad 0 \\
    0 & e^{-\beta J(1-\kappa)} & \qquad 0 & e^{-\beta J(1+\kappa)} & \qquad 0 \\
    0 & 0 & \qquad 0 & 0 & \qquad 0 
     \end{array} \right),
     \label{eq:NNNA_M0}
    \end{equation}
      \end{minipage}
    \end{tabular}
\end{table}
%
\begin{table}[b!]
    \centering
    \begin{tabular}{ccc}
       &  & 
      \begin{minipage}[b]{1.05\columnwidth}
      \centering
      \raisebox{-.5\height}{
      \begin{minipage}[b]{0.16\columnwidth}
      {\includegraphics[scale=0.35]{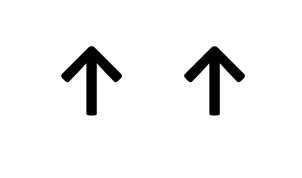}}
      \end{minipage}
      \begin{minipage}[b]{0.18\columnwidth}
      {\includegraphics[scale=0.35]{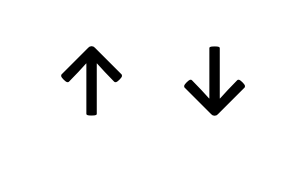}}
      \end{minipage}
      \begin{minipage}[b]{0.16\columnwidth}
      {\includegraphics[scale=0.35]{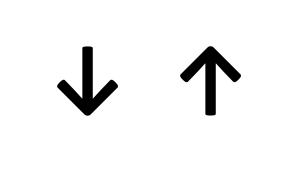}}
      \end{minipage}
      \begin{minipage}[b]{0.2\columnwidth}
      {\includegraphics[scale=0.35]{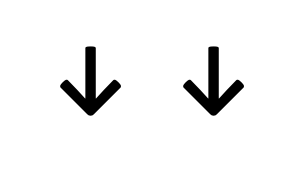}}
      \end{minipage}
      }
      \end{minipage}
      \\
     \qquad\qquad$M_1=$ & 
     \begin{minipage}[b]{0.1\columnwidth}
      \centering
      \raisebox{-.5\height}{
      \includegraphics[scale=0.35]{Figures_Mingyuan/matrix_configs_1d/1D_configs_v.png}
      }
      \end{minipage}
     & 
     \begin{minipage}[b]{1.05\columnwidth}
      \centering
      \begin{equation}
      \left( \begin{array}{ccp{1.5cm}p{1cm}}
    e^{\beta J(1-\kappa)} & e^{\beta J(1+\kappa)} & \qquad 0 & \qquad 0 \\
    p_1e^{\beta J(1-\kappa)} & p_1e^{\beta J(1+\kappa)} & \qquad 0 & \qquad 0 \\
    e^{\beta J(1-\kappa)} & e^{\beta J(1+\kappa)} & \qquad 0 & \qquad 0 \\
    0 & 0 & \qquad 0 & \qquad 0 \\
    e^{-\beta J(1-\kappa)} & e^{-\beta J(1+\kappa)} & \qquad 0 & \qquad 0
     \end{array} \right).
     \label{eq:NNNA_M1}
    \end{equation}
      \end{minipage}
    \end{tabular}
\end{table}
\end{widetext}
The connection probability of two spins a distance $r$ apart can then be expressed as
\begin{equation}
    \langle \gamma_{ij}^\parallel \rangle = 
    P(1+r) = \frac{2}{Z}\mathrm{Tr}[M_0 M^{r-1}M_1T^{N-r-1}],
\end{equation}
where $\mathbf{T}=\exp{(s_i (s_{i+1} -\kappa s_{i+2})\beta J)}$ is a $4\times 4$ transfer matrix for the configuration distribution, and $Z = \mathrm{Tr}[T^N]$ is the partition function. In order to be consistent with the Bethe lattice construction, we take the thermodynamic limit, $N\rightarrow\infty$. More details can be found in Ref.~\cite{zheng2022weakening}, but note that the cavity method here provides a different definition of bonding relations and results in a more compact transfer matrix expression.

\section{Percolation probabilities of generalized FK--CK clusters for SALR interaction model}
\label{sec:appendix_iso_perc}
In this appendix,  we obtain expressions for the percolation probabilities of generalized FK--CK clusters in the isotropic SALR interaction model. Recall that the nine variables and the auxiliary functions ($\Phi(k,m)$, $\Psi(k, m)$ and $\Theta(k,m,n)$) are defined in Sec.~\ref{sec:iso_nnn}, and that we have defined $x = \exp[-\beta\kappa\frac{l(l-1)+(c-l)(c-3l-1)}{2}]$.

\begin{widetext}
For the configuration ($\uparrow, \uparrow$), denoted $E$ in Fig.~\ref{fig:bethe_configs}, one has $\eta_\mathrm{cav}(\uparrow,\uparrow) = \pi(\uparrow,1)+\pi(\uparrow,0)+q(\uparrow,1)+q(\uparrow,0)$ with 
\begin{equation}
\label{eq:SALRperc1}
\begin{split}
    \pi(\uparrow,1) &= Z_\mathrm{cav}^{-1}\left\{\sum_{l=0}^{c}\sum_{k=0}^{l}\sum_{a+b+d>0}\binom{c}{l}\binom{l}{k}e^{\beta J (2l-c)} e^{-\beta \kappa J(2l-c)} x \, w(\uparrow,0)^{c-l-a} w(\uparrow,1)^a q(\uparrow,0)^{l-k-b} q(\uparrow,1)^b \pi(\uparrow,0)^{k-d} \pi(\uparrow,1)^d\right.\\ &\times\sum\limits_{m=0}^{l-k}\binom{l-k}{m}\Psi(k,m) [(1-p_1)(1-p_2)^{k+m}]^{l-k-m}[1-(1-p_2)^{k+m}]\\
    &+\sum\limits_{l=0}^{c}\binom{c}{l}e^{\beta J (2l-c)} e^{-\beta \kappa J(2l-c)} x \, w(\uparrow,0)^{c-l} \sum\limits_{k=0}^{l}\binom{l}{k}q(\uparrow,0)^{l-k}\pi(\uparrow,0)^{k}\\
    &\left.\times \sum\limits_{m=0}^{l-k}\sum\limits_{n=0}^{l-k-m} \binom{l-k}{m}\binom{l-k-m}{n}\Theta(k,m,n)[(1-p_1)(1-p_2)^{k+m+n}]^{l-k-m-n}\, [1-(1-p_2)^{k+m+n}]\right\}.
\end{split}
\end{equation}
(Notice that the exponents involving $a$, $b$, and $d$ should be greater than or equal to zero.)

\begin{equation}
\begin{split}
    \pi(\uparrow,0) &= Z_\mathrm{cav}^{-1}\left\{\sum_{l=0}^{c}\sum_{k=0}^{l}\sum_{a+b+d>0}\binom{c}{l}\binom{l}{k} e^{\beta J (2l-c)} e^{-\beta \kappa J(2l-c)} x \, w(\uparrow,0)^{c-l-a} w(\uparrow,1)^a q(\uparrow,0)^{l-k-b} q(\uparrow,1)^b \pi(\uparrow,0)^{k-d} \pi(\uparrow,1)^d\right.\\ &\times\sum\limits_{m=0}^{l-k}\binom{l-k}{m}\Psi(k,m) [(1-p_1)(1-p_2)^{k+m}]^{l-k-m}(1-p_2)^{k+m}\\
    &+\sum\limits_{l=0}^{c}\binom{c}{l}e^{\beta J (2l-c)} e^{-\beta \kappa J(2l-c)} x \, w(\uparrow,0)^{c-l} \sum\limits_{k=0}^{l}\binom{l}{k}q(\uparrow,0)^{l-k}\pi(\uparrow,0)^{k}\\
    &\left.\times \sum\limits_{m=0}^{l-k}\sum\limits_{n=0}^{l-k-m} \binom{l-k}{m}\binom{l-k-m}{n}\Theta(k,m,n)[(1-p_1)(1-p_2)^{k+m+n}]^{l-k-m-n}\, (1-p_2)^{k+m+n}\right\}.
\end{split}
\end{equation}

\begin{equation}
\begin{split}
    q(\uparrow,1) &= Z_\mathrm{cav}^{-1} \sum\limits_{l=0}^{c}\binom{c}{l}e^{\beta J (2l-c)} e^{-\beta \kappa J(2l-c)} x \, w(\uparrow,0)^{c-l} \sum\limits_{k=0}^{l}\binom{l}{k}q(\uparrow,0)^{l-k}\pi(\uparrow,0)^{k}\\
    &\times \sum\limits_{m=0}^{l-k} \binom{l-k}{m}\Phi(k,m)(1-p_2)^{(k+m)(l-k-m)}(1-p_1)^{k+m}[1-(1-p_2)^{k+m}].
\end{split}
\end{equation}

\begin{equation}
\begin{split}
    q(\uparrow,0) &= Z_\mathrm{cav}^{-1} \sum\limits_{l=0}^{c}\binom{c}{l}e^{\beta J (2l-c)} e^{-\beta \kappa J(2l-c)} x \, w(\uparrow,0)^{c-l} \sum\limits_{k=0}^{l}\binom{l}{k}q(\uparrow,0)^{l-k}\pi(\uparrow,0)^{k}\\
    &\times \sum\limits_{m=0}^{l-k} \binom{l-k}{m}\Phi(k,m)(1-p_2)^{(k+m)(l-k-m)}(1-p_1)^{k+m}(1-p_2)^{k+m}.
\end{split}
\end{equation}

For the configuration ($\uparrow, \downarrow$) denoted $O$ in Fig.~\ref{fig:bethe_configs}, one has 
$\eta_\mathrm{cav}(\uparrow,\downarrow) = \pi(\downarrow,0)+q(\downarrow,0)$ with
\begin{equation}
\begin{split}
    \pi(\downarrow,0) &= Z_\mathrm{cav}^{-1}\left\{\sum_{l=0}^{c}\sum_{k=0}^{l}\sum_{a+b+d>0}\binom{c}{l}\binom{l}{k} e^{\beta J (2l-c)} e^{\beta \kappa J(2l-c)} x \, w(\uparrow,0)^{c-l-a} w(\uparrow,1)^a q(\uparrow,0)^{l-k-b} q(\uparrow,1)^b \pi(\uparrow,0)^{k-d} \pi(\uparrow,1)^d\right.\\
    &+\sum\limits_{l=0}^{c}\binom{c}{l}e^{\beta J (2l-c)} e^{\beta \kappa J(2l-c)} x \, w(\uparrow,0)^{c-l} \sum\limits_{k=0}^{l}\binom{l}{k}q(\uparrow,0)^{l-k}\pi(\uparrow,0)^{k}\\
    &\left.\times\sum\limits_{m=0}^{l-k} \binom{l-k}{m}\Phi(k,m)(1-p_2)^{(k+m)(l-k-m)}[1-(1-p_1)^{k+m}]\right\}.
\end{split}
\end{equation}

\begin{equation}
\begin{split}
    q(\downarrow,0) &= Z_\mathrm{cav}^{-1} \sum\limits_{l=0}^{c}\binom{c}{l}e^{\beta J (2l-c)} e^{\beta \kappa J(2l-c)} x \, w(\uparrow,0)^{c-l} \sum\limits_{k=0}^{l}\binom{l}{k}q(\uparrow,0)^{l-k}\pi(\uparrow,0)^{k}\\
    &\times \sum\limits_{m=0}^{l-k} \binom{l-k}{m}\Phi(k,m)(1-p_2)^{(k+m)(l-k-m)}(1-p_1)^{k+m}.
\end{split}
\end{equation}

For the configuration ($\downarrow, \uparrow$) denoted $F$ in Fig.~\ref{fig:bethe_configs}, one has 
$\eta_\mathrm{cav}(\downarrow,\uparrow) = w(\uparrow,1)+w(\uparrow,0)$ with
\begin{equation}
\begin{split}
    w(\uparrow,1) &= Z_\mathrm{cav}^{-1} \sum\limits_{l=0}^{c}\binom{c}{l}e^{-\beta J (2l-c)} e^{-\beta \kappa J(2l-c)} x \, w(\downarrow,0)^{c-l} \sum\limits_{k=0}^{l}\binom{l}{k}q(\downarrow,0)^{l-k}\pi(\downarrow,0)^{k}\\
    &\times \sum\limits_{m=0}^{l-k} \binom{l-k}{m}\Phi(k,m)(1-p_2)^{(k+m)(l-k-m)}[1-(1-p_2)^{k+m}].
\end{split}
\end{equation}

\begin{equation}
\begin{split}
    w(\uparrow,0) &= Z_\mathrm{cav}^{-1} \sum\limits_{l=0}^{c}\binom{c}{l}e^{-\beta J (2l-c)} e^{-\beta \kappa J(2l-c)} x \, w(\downarrow,0)^{c-l} \sum\limits_{k=0}^{l}\binom{l}{k}q(\downarrow,0)^{l-k}\pi(\downarrow,0)^{k}\\
    &\times \sum\limits_{m=0}^{l-k} \binom{l-k}{m}\Phi(k,m)(1-p_2)^{(k+m)(l-k-m)}(1-p_2)^{k+m}.
\end{split}
\end{equation}

Finally, for the configuration ($\downarrow, \downarrow$) denoted $E$ in Fig.~\ref{fig:bethe_configs}, one has 
$\eta_\mathrm{cav}(\downarrow,\downarrow) = w(\downarrow,0)$ with
\begin{equation}
\label{eq:SALRperc9}
\begin{split}
    w(\downarrow,0) &= Z_\mathrm{cav}^{-1} \sum\limits_{l=0}^{c}\binom{c}{l}e^{-\beta J (2l-c)} e^{\beta \kappa J(2l-c)} x \, w(\downarrow,0)^{c-l} \sum\limits_{k=0}^{l}\binom{l}{k}q(\downarrow,0)^{l-k}\pi(\downarrow,0)^{k}.
\end{split}
\end{equation}
\end{widetext}

%

\end{document}